\documentclass[11pt]{ucthesis}  %[12pt,draft]
\usepackage{psfig,natbib} 
\def\fdg{\hbox{$.\!\!^\circ$}}
\def\mycaption#1{\begin{quote}\noindent{#1}\end{quote}}
\def\ssp{\def\baselinestretch{1.0}\large\normalsize}

\begin{document}

% Declarations for Front Matter

\title{Big Bang Leftovers in the Microwave:\\  
Cosmology with the Cosmic Microwave Background Radiation}
\author{Eric Joseph Gawiser}
\degreeyear{Spring 1999}
\degree{Doctor of Philosophy}
\chair{Professor Joseph Silk}
\othermembers{Professor George F. Smoot\\ 
Professor Thomas Broadhurst}
\numberofmembers{3}
\prevdegrees{A.B. (Princeton University)
1994\\
M.A. (University of California, Berkeley) 
1996}
\field{Physics}
\campus{Berkeley}

\maketitle
%\approvalpage
\copyrightpage

\begin{abstract}
%Not as abstract as it could be...  

	We combine detections of anisotropy in the Cosmic Microwave Background 
radiation with observations of inhomogeneity in 
the large-scale distribution of galaxies to 
test the predictions of models of cosmological structure formation.
This combination probes spatial
scales varying by three orders of magnitude, including a significant region
where the two types of data overlap.  We examine Cold Dark Matter models 
with 
adiabatic density perturbations, 
isocurvature models, and a topological defects model.
% and give a quantitative measure of their success.  
We set upper limits on the neutrino mass and find the primordial 
power spectrum needed 
%The possibility of a non-scale-invariant 
%primordial power spectrum is considered and is used 
to reconcile an  
%can cut apparent if needed, then switch an to a 
apparent disagreement between structure formation observations and 
direct observations 
of cosmological parameters.  

	Present and future observations of Cosmic Microwave Background 
anisotropy suffer from foreground contamination.  We develop detailed 
predictions for microwave emission from radio and infrared-bright galaxies and the Sunyaev-Zeldovich effect from clusters.  
We present realistic simulations of the microwave sky, produced as part of the 
``WOMBAT Challenge'' exercise, and 
introduce a pixel-space method for subtracting foreground contamination 
which can be tested on these simulations.

%\abstractsignature
\end{abstract}

\begin{frontmatter}

\begin{dedication}
\null\vfil
{\large
\begin{center}
To everyone who has ever gazed at the sky and wondered how it all began:    
Do not fear.  
This dissertation will leave you with plenty still to wonder about!  
\end{center}}
\vfil\null
\end{dedication}

\tableofcontents
\listoffigures
\listoftables
\begin{acknowledgements}

This research would not have been possible without the tireless efforts of hundreds of observers who gathered and analyzed the raw data on cosmological 
inhomogeneity and anisotropy.  I have benefitted from discussions with 
dozens of researchers in person and by email and have tried to acknowledge 
their help in my published work.  In particular, I want to thank Martin White, 
Helen Tadros, and Carlton Baugh for their patient and thorough responses 
to my email questions.  Kris Gorski, Eric Hivon, and Ben Wandelt (HEALPIX) and 
Uros Seljak and Matias Zaldarriaga (CMBFAST) deserve tremendous credit for 
making their high-tech software packages available to the cosmology 
community.  

I am grateful for research support from 
a variety of sources:  an NSF Graduate Fellowship, the NASA AISRP grant
(NAG-3941), 
a Phi Beta Kappa fellowship, a Department of Education-GANN fellowship, and 
a summer NASA grant.  I want to acknowledge my co-authors (Joe 
Silk, George Smoot, Aaron Sokasian, Andrew Jaffe, and the WOMBAT Team) since 
much of the material contained herein is a mild modification of multiple-author
works submitted for publication.  I appreciate comments from Doug Finkbeiner 
and Andrew Jaffe that improved the material in Chapter \ref{chap:pixel}.  
I thank Ted Bunn, Luis Tenorio, and Philip Stark for being my resources 
on statistics issues.  

My advisor, Joe Silk, has been a tremendous
supporter of my research efforts and has given me opportunities to travel 
widely through various fields of cosmology, both literally and figuratively.  
George Smoot got me started in the microwave 
foregrounds business the summer before 
graduate school; little did I know how much that field would grow!  
I owe a great debt to my officemates, 
Doug Finkbeiner, Lexi Moustakas, and Jeff Newman, for putting up with my 
whistling and yelling at the computer screen (it's all part of the creative
process!) and for giving me so much remedial help with UNIX and IDL.  
I thank my parents for understanding that I really {\bf was} working hard on 
my dissertation, but I didn't necessarily feel like discussing it.  And 
for my sweetie:  Sevil \c{c}ok ama en \c{c}ok seviyorum!  

\end{acknowledgements}

\end{frontmatter}

\part{Testing Models of Cosmological Structure Formation}

\chapter{Introduction:  An Overview of Cosmic Microwave Background Anisotropy}
\label{chap:overview}

\section{Motivation of this Thesis}

The Cosmological Principle holds that the universe is homogeneous 
and isotropic, i.e. that it looks roughly the same at every point 
and in every direction.  We know that this holds true for the Cosmic 
Microwave Background (CMB)
radiation, which has uniform intensity to a level of one part in one hundred 
thousand in every direction on the sky and is described in detail 
in the rest of this chapter.  We believe that this is 
true for the distribution of galaxies as well, with each large-scale 
region of the universe containing a similar mix of galaxies and 
galaxy clusters.  
However, we have abundant evidence that the 
Cosmological Principle is not absolute:  our very existence requires 
a highly inhomogeneous distribution of matter on small scales, in order 
to form galaxies and then stars and then planets containing the 
relatively heavy elements such as 
carbon and oxygen which are so critical for life 
as we know it.  The average density of the universe is enough to have 
only a few hydrogen atoms per cubic meter, so everything that makes up 
our daily world is an aberration.  
On the larger scale, we can readily observe that galaxies 
are clustered at a level much greater than a Poissonian distribution would 
suggest and that these clusters are themselves clustered into 
superclusters of galaxies.  This implies that density 
variations on very large scales existed in the early universe.

Such density variations are most directly probed using the Cosmic 
Microwave Background radiation.  Indeed, anisotropy (variation from 
one position on the sky to another) has now been seen in the CMB on 
scales corresponding to present-day superclusters.  We believe that the 
formation of structure in the universe is described by the paradigm of 
gravitational instability, which states that sufficiently 
overdense regions 
overwhelm the universal expansion and
 collapse to form galaxies and clusters while underdense regions 
expand to form voids.  Given the approximately 15 billion years since  
the Big Bang, gravity has been able to magnify density variations seen 
in the CMB at one part in a hundred thousand to the level needed to form 
galaxies today.  

This connection between slight inhomogeneities in the early universe, seen 
as slight anisotropies in the CMB, and the clear inhomogeneity in 
the galaxy distribution today is the motivation for Part I of this thesis.  
We compile the current set of observations of CMB 
anisotropy and the large-scale structure of the 
galaxy distribution (Chapter \ref{chap:obs}).  
A careful method is presented for 
comparing these observations with the predictions of various models 
of cosmological structure formation (Chapter \ref{chap:method}).  
The data are improving rapidly but the 
list of models which can be considered is growing even faster.  For this 
reason, we do not attempt an exhaustive search of a particular 
parameter space 
but rather examine the most popular models and a few exotic ones 
in order to gain 
insight into what aspects are well constrained by observation
(Chapter \ref{chap:results}, see also \citealp{gawisers98}).  
This 
allows us to identify two aspects for further analysis, the 
possibility of cosmologically significant neutrino mass (Chapter \ref{chap:nu})
and the possibility 
of a non-scale-free primordial power spectrum (Chapter \ref{chap:ppk}).

While we attempt to correct for known systematic issues in the large-scale 
structure observations in Part I, the field of CMB anisotropy observation is 
still in its infancy, and so far there is no proof of systematic corrections 
needed for the data.  The most likely source of such difficulties is 
foreground contamination due to Galactic and extragalactic microwave emission. 
Foreground contamination has been subtracted in a very primitive manner by 
most observing teams, and for the low level of precision of current 
observations this may be sufficient.  However, the 1\% level 
of precision in cosmological parameter estimation which is desired for 
future CMB observations will not allow for much error in foreground 
subtraction.   
In Part II, we describe the 
``WOMBAT Challenge'' exercise (Chapter \ref{chap:wombat}, see also 
\citealp{gawiseretal98}), 
which seeks to make realistic simulations 
of the microwave sky and test methods of foreground subtraction.   
We describe detailed studies of how to predict the contribution to 
microwave anisotropy from low-redshift infrared-bright galaxies 
(Chapter \ref{chap:ir}, see also \citealp{gawisers97}) and 
radio galaxies (Chapter \ref{chap:radio}, see also \citealp{sokasiangs98}) 
and set limits on the level of foreground contamination which can 
be caused by high-redshift infrared-bright galaxies, dim radio sources, and 
undiscovered families of point sources (Chapter \ref{chap:sources}, see 
also \citealp{gawiserjs98}).  We also give predictions for microwave 
anisotropy caused by the Sunyaev-Zeldovich effect in galaxy clusters 
and the filaments connecting them (Chapter \ref{chap:sz}) and present a 
pixel-space method for foreground subtraction which can be tested on 
the WOMBAT simulations (Chapter \ref{chap:pixel}).  As the quantity and 
quality of data are expected to improve rapidly in the 
coming decade, we will examine the prospects for detecting dusty 
high-redshift galaxies with forthcoming instruments and for setting 
improved constraints on models of structure formation and then conclude 
(Chapter \ref{chap:conclusion}).  The rest of this chapter sets the 
stage by providing an overview of Cosmic Microwave Background anisotropy.

%for Review, add plot of window functions of all expts
%for Review, add plots of 2ndary and primary anisotropies 

\section{Origin of the Cosmic Background Radiation}

Our present understanding of the beginning of the universe is based upon the 
remarkably successful theory of the Hot Big Bang.  We believe that our universe began about 15 billion years ago at 
a minute fraction of its present size (formally an infinitesimal
singularity) as a hot, dense, nearly uniform sea of radiation.  
If inflation occurred in the first fraction of a second, the universe became 
matter dominated while expanding exponentially and then returned to 
radiation domination by the reheating caused by the decay of the inflaton.  
Baryonic
matter formed within the first second, 
and the nucleosynthesis of the lightest elements 
took only a few minutes as the universe expanded and cooled.  
The majority of the baryonic matter in the universe was in the form of plasma 
until about 300,000 years after the Big Bang, when the universe had 
cooled to a temperature of about 3000 K, sufficiently cool for 
protons to capture free electrons and form atomic hydrogen; this process 
is referred to as recombination.  The recombination epoch 
occurred at an observed redshift of about 1100, meaning the universe is 
now over a thousand times larger than it was then.  
The ionization energy of a hydrogen atom is 13.6 eV, but 
recombination did not 
occur until the universe had cooled to a characteristic temperature (kT) of
0.3 eV \citep{padmanabhan93}.  
This delay had several causes.  The high entropy of the universe
makes the rate of electron capture only marginally faster than the rate of 
photodissociation.  Moreover, each electron captured directly into the ground
state emits a photon capable of ionizing another newly formed atom, so it 
was through 
recombination into excited states and the cooling of the universe to
temperatures below the ionization energy of hydrogen
that neutral matter finally condensed out of the plasma.  
Until recombination,
the universe was opaque to electromagnetic radiation due to scattering 
of the photons by free electrons.   As recombination occurred, the 
density of free electrons diminished greatly, leading to the decoupling of
matter and radiation as the universe became transparent to light.  

The Cosmic Background Radiation (CBR) 
released during this era of decoupling had a mean free path long 
enough to travel almost unperturbed until the present day, where we 
observe it peaked in the microwave region of the spectrum as the 
Cosmic Microwave Background (CMB).  
We see this radiation today coming from the surface of last 
scattering (which is really a spherical shell of finite thickness) 
at a distance of 
 15 billion light years.  
This Cosmic Background Radiation was predicted by the Hot Big Bang theory
and discovered with an antenna temperature of 3K 
in 1964 by \citet{penziasw65}.  
Because of the expansion of the universe, each epoch in its history is observed
today at a redshift which increases with the age of the epoch.
The number density of photons in the universe at a redshift $z$ is given by \citep{peebles93} 

\begin{equation}
 n_{\gamma} = 420 (1 + z)^{3} cm ^{-3}  
\end{equation}

\noindent 
where $(1 + z)$ is the factor by which the linear scale of the 
universe has expanded since then.  
The radiation temperature of the universe is given by $T = T_{0} (1 + z)$ so it
is easy to see how the conditions in the early universe 
at high redshifts were hot and dense.  

The CBR is our best probe into the conditions of the early universe.  Theories
of the formation of large-scale structure 
predict the existence of slight inhomogeneities in the distribution of 
matter in the early universe which eventually underwent gravitational 
collapse to form galaxies, galaxy clusters, and superclusters.  These density 
inhomogeneities lead to temperature anisotropies in the CBR because the
radiation leaving a dense area of the last scattering surface is 
gravitationally redshifted to a lower apparent temperature and vice versa
for an underdense region.  
The DMR (Differential Microwave Radiometer) 
instrument of the Cosmic Background Explorer (COBE) satellite 
discovered primordial temperature 
fluctuations on angular scales larger than $7^\circ$
of order $\Delta T/T = 10^{-5}$ \citep{smootetal92}.  
The subsequent observations of the CMB 
appear to reveal temperature anisotropies on smaller 
angular scales which correspond to the physical scale of 
observed structures such as galaxies 
and clusters of galaxies.

\subsection{Thermalization}

There were three main processes by which this radiation interacted with matter 
in the first few hundred thousand years, Compton scattering, double Compton 
scattering, and thermal bremsstrahlung.
The simplest interaction of matter and radiation is Compton
scattering of a single photon off a free electron, 
$ \gamma + e^{-} \rightarrow \gamma + e^{-}$. 
The photon will transfer
momentum and energy to the electron if it has significant energy in the
electron's rest frame.  However, the scattering will be
well approximated by the Thomson cross section if the photon's energy in 
the rest frame of the electron is significantly less than the rest mass, 
$h \nu \ll m_{e}c^{2}$.
When the electron is relativistic, the photon is blueshifted by 
roughly a factor 
$\gamma$ in energy when viewed from the 
electron rest frame, is then emitted at almost the same energy in the 
electron rest frame, and is blueshifted by another factor of $\gamma$
when retransformed to the observer's frame.  Thus, energetic 
electrons can efficiently transfer energy 
to the photon background of the universe.  
This process is referred to as Inverse Compton scattering.
The combination of cases where the photon gives energy to the electron 
and vice versa allows Compton scattering to generate thermal equilibrium 
(which is impossible in the Thomson limit of elastic scattering).
Compton scattering conserves the number of photons.
There exists a similar process, double Compton scattering,
 which produces (or absorbs)
photons, $e^- + \gamma \leftrightarrow e^{-} + \gamma + \gamma $.

Another electromagnetic interaction which occurs in the plasma of the early
universe is Coulomb scattering.  Coulomb scattering itself serves to establish
and maintain thermal equilibrium within the photon-baryon fluid
 but does not affect the
photons.  However, when electrons encounter ions they experience an 
acceleration and therefore emit electromagnetic radiation.  This is called 
thermal bremsstrahlung or free-free emission.  For an ion $X$, 
we have $e^{-} + X \leftrightarrow e^{-} + X + \gamma$.  The interaction
can occur in reverse because of the ability of the charged particles
to absorb incoming photons; this is called free-free absorption.  Each charged
particle emits radiation, but the acceleration is proportional to the mass,
so we can usually view the electron as being accelerated in the fixed Coulomb
field of the much heavier ion.  
Bremsstrahlung is dominated by electric-dipole 
radiation \citep{shu91} and can also 
produce and absorb photons.  

The net effect is that Compton scattering is dominant
for temperatures above 90 eV whereas bremsstrahlung is the primary process
between 90 eV and 1 eV.  At temperatures above 1 keV, double Compton 
is more efficient than bremsstrahlung.  All three processes occur faster than 
the expansion of the universe and therefore have an impact until decoupling.
A static solution for Compton scattering 
is the Bose-Einstein distribution, 

\begin{equation}
 f_{BE} = \frac {1} {e^{x + \mu} - 1} 
\end{equation}

\noindent 
where $\mu$ is a dimensionless chemical potential \citep{hu95}.  
At high optical depths, Compton scattering can exchange enough energy to 
bring the photons to this Bose-Einstein equilibrium distribution.  A Planckian
spectrum corresponds to zero chemical potential, which will occur only when 
the number of photons and total energy are in the same proportion as they 
would be for a blackbody.  Thus, unless the photon number starts out exactly
right in comparison to the total energy in radiation in the universe, Compton
scattering will only produce a Bose-Einstein distribution
and not a blackbody spectrum.  It is important to note, however, that 
Compton scattering can preserve a 
Planck distribution, which is given by 

\begin{equation}
 f_{P} = \frac {1}{e^{x} - 1 }. 
\end{equation}

All three interactions
will preserve a thermal spectrum if one is achieved at any point.  It has
long been known that the expansion of the universe serves to decrease
the temperature of a blackbody spectrum, 

\begin{equation}
B_{\nu} = \frac{2 h \nu^{3} / c^{2}} {e^{h \nu / k T} - 1}, 
\end{equation}

\noindent
 but keeps it thermal 
\citep{tolman34}.  This
occurs because both the frequency and temperature decrease as $(1 + z)$
leaving the scaled variable $x$ unchanged during expansion.  Although
Compton scattering alone cannot produce a Planck distribution, such a 
distribution will remain unaffected by electromagnetic interactions or the 
universal expansion once it is achieved.  
A non-zero
chemical potential will be reduced to zero by double Compton scattering
and, later, bremsstrahlung which will create and absorb photons until the 
number density matches the energy and a thermal distribution of zero
chemical potential is achieved. 
 This results in the thermalization 
of the CBR at redshifts much greater than that of recombination.

Thermalization, of course, should only be able to create an 
equilibrium temperature over regions that are in causal contact.  
The causal horizon at the time of last scattering was relatively small, 
corresponding to a scale today of about 200 Mpc, or a region of 
angular extent of one degree on the sky.  However, observations of the 
CMB show that it has an isotropic temperature on the sky to the 
level of one part in one hundred thousand!  This is the origin of the 
Horizon Problem, which is that there is no physical mechanism expected 
in the early universe which can produce thermodynamic equilibrium on 
superhorizon scales.  The inflationary universe paradigm 
\citep{guth81,linde82,albrechts82}
solves the Horizon 
Problem by postulating that the universe underwent a brief phase of 
exponential expansion during the first second after the Big Bang, during 
which our entire visible Universe expanded out of a region small 
enough to have already achieved thermal equilibrium.

\section{CMB Spectrum}

\begin{figure}
\centerline{\psfig{file=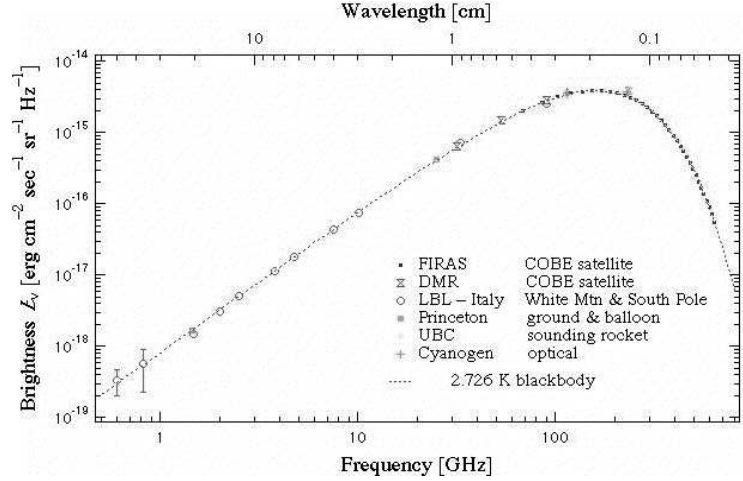,width=4in,angle=-90}}
\caption{Measurements of the CMB spectrum.}
\label{fig:spectrum}
\end{figure}

The CBR 
is the most perfect blackbody ever
seen, according to the FIRAS (Far InfraRed Absolute 
Spectrometer) instrument of COBE, which measured a temperature 
of $T_0 = 2.726 \pm 0.010 K$ \citep{matheretal94}.
%as shown in Figure 1.  
The theoretical prediction that the CBR will have a blackbody spectrum 
appears to be confirmed by the FIRAS observation
 (see Figure \ref{fig:spectrum}).  
But this is not the 
end of the story.  FIRAS only observed the peak of the blackbody.
%, as shown in Figure 2.  
Other experiments have mapped out the Rayleigh-Jeans part of the
spectrum at low frequency.  Most are consistent with a 2.73 K blackbody, but
some are not.  It is in the low-frequency limit that the greatest spectral
distortions might occur because a Bose-Einstein distribution differs from 
a Planck distribution there.  However, double Compton and 
bremsstrahlung are most effective at low frequencies so 
strong deviations
from a blackbody spectrum are not generally expected.
Possible spectral distortions in the Wien tail of the
spectrum would need to be very significant to be observed above the foreground
signal from interstellar dust at those high frequencies.  For example, 
broad emission lines from electron capture at recombination are predicted
in the Wien tail but cannot be distinguished due to foreground 
contamination \citep{whitess94}.  

Although Compton, double Compton, and bremsstrahlung interactions occur 
frequently until decoupling, the complex interplay between them required
to thermalize the CBR spectrum is ineffective at redshifts below 
$10^7$.  
This means that any process after that time
which adds a significant portion of 
energy to the universe  
will lead to a spectral distortion today.  Neutrino decays during this 
epoch should lead to a Bose-Einstein rather than a Planck distribution, and 
this allows the FIRAS observations to set constraints on the decay of 
neutrinos and other particles 
in the early universe \citep{kolbt90}.  The apparent
impossibility of thermalizing 
radiation at low redshift makes the blackbody nature of the CBR strong 
evidence that it did originate in the early universe and as a result serves
to support the Big Bang theory.  

The process of Compton scattering can cause spectral distortions if it
is too late for double Compton and bremsstrahlung to be effective.  In general,
low-frequency photons will be shifted to higher frequencies, thereby 
decreasing the number of photons in the Rayleigh-Jeans region and enhancing
the Wien tail.  This is referred to as a Compton-{\it y} distortion and it
is described by the parameter

\begin{equation}
 y = \int \frac {T_{e}(t)} {m_{e}} \sigma n_{e}(t) dt. 
\end{equation}

\noindent
The apparent temperature drop in the long-wavelength limit is

\begin{equation}
 \frac {\delta T}{T} = - 2 y. 
\end{equation}

\noindent
The most important example of this is Compton scattering of photons off 
hot electrons in galaxy clusters, called the Sunyaev-Zeldovich (SZ) 
effect.  The electrons transfer energy to the photons, and the spectral
distortion results from the sum of all of the scatterings off 
electrons in 
thermal motion, each of which has a Doppler shift.  The 
SZ effect from clusters can yield a distortion of $y \simeq 10^{-5} - 
10^{-3}$ and these distortions have been observed in several
rich clusters of galaxies.  The FIRAS observations place 
a constraint on any full-sky Comptonization by limiting the average
$y$-distortion to $y < 2.5 \times 10^{-5}$ \citep{hu95}.  The 
integrated $y$-distortion predicted from the SZ effect of 
galaxy clusters and large-scale structure is over a factor of ten lower 
than this observational constraint \citep{refregiersh98} 
but that from ``cocoons'' of radio galaxies \citep{yamadass99}
 is of the same order. 

%add description of spectral distortions due to reionization - hu95

\section{CMB Anisotropy}

The temperature anisotropy at a point on the sky $(\theta,\phi)$ can be 
expressed in the basis of spherical harmonics as 
\begin{equation}
 \frac{\Delta T}{T} (\theta, \phi) = 
\sum_{\ell m} a_{\ell m} Y_{\ell m}(\theta, \phi).
\end{equation}

\noindent
A cosmological model predicts the amplitude 
of the $a_{\ell m}$ coefficients 
over an ensemble of universes (or an ensemble of observational 
points within one universe, if the universe is ergodic).  The
assumptions of rotational symmetry and Gaussianity allow
 us to express this ensemble 
average in terms of the multipoles $C_{\ell}$ as 
\begin{equation}
 \langle a^{*}_{\ell m} a_{\ell' m'} \rangle \equiv 
C_{\ell} \delta_{\ell' \ell} \delta_{m' m}.
\end{equation}

\noindent
The predictions 
of a cosmological model can be expressed in terms of $C_{\ell}$ alone if 
that model predicts a Gaussian distribution of density perturbations, 
in which case the 
$a_{\ell m}$ will have mean zero  
and variance $C_\ell$.

%[some basic equations will be added here to show acoustic oscillations]
%explain physical origins of peak and damping tail, include $\Lambda$
%use angular size-redshift relation and derive formulae for LS
%theoretical implications of neutrinos - more detail 

The conventional wisdom in cosmology holds that the temperature anisotropies
detected by COBE are the result of inhomogeneities in the distribution of 
matter at the epoch of decoupling.  Because Compton scattering is an 
isotropic process in the electron rest frame, any primordial anisotropies (as 
opposed to inhomogeneities) should have been smoothed out before
decoupling.  This lends credence
to the interpretation of the observed 
anisotropies as the result of density perturbations
which seeded the formation of galaxies and clusters.  
The discovery of temperature anisotropies by 
COBE provides evidence that 
such density inhomogeneities existed in the early
universe due to quantum fluctuations in the scalar field of inflation
or to topological defects caused by a phase transition.  
Gravitational collapse of 
these primordial density inhomogeneities 
appears to have formed
the large-scale structures of galaxies,
clusters, and superclusters that we observe today.  

	There are now a plethora of theoretical 
models which predict the development of a primordial 
power spectrum of density perturbations into the  
radiation power spectrum of microwave 
background anisotropies.  These models differ in their explanation 
of the origin of density inhomogeneities 
(inflation or topological defects), the nature of the dark matter (hot,
cold, baryonic, or a mixture of the three), 
the curvature of the universe ($\Omega$), 
the value of the cosmological constant ($\Lambda$), 
the value of Hubble's constant ($H = 100 h$ km/s/Mpc), and 
the possibility of reionization at some redshift $z$ which 
wholly or partially erased temperature anisotropies in the CMB on 
scales smaller than the horizon size.  Available data does not allow 
us to constrain all (or even most) of these parameters, so it is
clear that in analyzing new CMB anisotropy data we should seek 
a model-independent approach.  

Anistropy measurements on 
small angular scales ($0\fdg1$ to $1^{\circ}$) 
are expected to reveal the so-called
first acoustic
 peak of the CMB power spectrum.  This acoustic peak corresponds
to the scale where acoustic oscillations of the photon-baryon fluid caused
by primordial density inhomogeneities are just reaching their maximum 
amplitude at the surface of last scattering.  Further acoustic  
peaks occur at scales that are reaching their second, third, fourth, etc.
antinodes of oscillation.  For a given model,  
the size and location of the 
first acoustic peak can yield information about $\Omega$, the ratio of 
the density of the universe to the critical density needed to stop
its expansion, and $\Omega_b$, the 
fraction of this critical density which is contained in baryonic matter.   
A precise measurement of all three acoustic peaks can reveal information 
on the fraction of hot dark matter and even potentially the number 
of neutrino species \citep{dodelsongs96}.  
It seems
reasonable to view the mapping of the acoustic peaks as
a means of determining the nature of parameter space
before going on to fitting cosmological parameters directly.  
The CMB anisotropy damping tail on arcminute scales, 
where the fluctuations are decreased due to photon diffusion 
\citep{silk67}
and the finite thickness of the last-scattering surface,
 is a sensitive probe 
of cosmological parameters and has the potential to 
break degeneracies between models which explain the larger-scale anisotropies
\citep{huw97a,metcalfs98}.  

%expand ``damping envelope'', give equations 

\subsection{Reionization}

The possibility 
that post-decoupling interactions between ionized matter and the CBR
have affected the anisotropies on scales 
smaller than those measured by COBE is of great significance
for current experiments.  
Reionization is inevitable but its effect on anisotropies 
depends significantly on when it occurs
(see \citealp{haimank99} for a review).  
  Early reionization leads to 
a larger optical 
depth and therefore a greater damping of the anisotropy 
power spectrum due to the secondary scattering of CMB photons 
off of the newly free electrons.    
Attempts to measure
the temperature anisotropy on angular scales of less than a degree which 
correspond to the size of galaxies could lead to a surprise; 
if the universe was reionized after recombination to the extent
that the CBR was significantly scattered 
at redshifts less than 1100, the small-scale 
primordial anisotropies will have been washed out.
On arc-minute scales, the 
interaction of photons with reionized matter is expected to have eliminated
the primordial anisotropies and replaced them with smaller secondary 
anisotropies from this new surface of last scattering (the 
Ostriker-Vishniac effect and patchy reionization, see next section 
and \citealt{whitess94}).
To have an appreciable optical depth 
for photon-matter interaction, reionization cannot have occurred 
much later than a redshift of 20 \citep{padmanabhan93}.  
Large-scale anisotropies such as those
seen by COBE are not expected to be affected by reionization because they 
encompass regions of the universe which were not yet in causal contact
even at the proposed time of reionization.

\subsection{Secondary Anisotropies}

%ISW is a secondary anisotropy but matters on large scales, distinguishes 
%open, $\Lambda$, explain, give equations

Secondary CMB anisotropies 
occur when the photons of the Cosmic Microwave 
Background radiation are scattered after the original last-scattering 
surface (see \citealp{refregier99} for a review).  
  The shape of the blackbody 
spectrum can be altered through 
inverse Compton scattering by the thermal Sunyaev-Zeldovich (SZ) effect
 \citep{sunyaevz72}, which 
will be discussed in more detail in Chapter \ref{chap:sz}.
  The effective temperature of 
the blackbody can be shifted locally by 
a doppler shift from the peculiar velocity of the scattering medium (the 
kinetic SZ and Ostriker-Vishniac effects) as well as by passage through
the changing gravitational potential caused by the 
collapse of nonlinear structure (the Rees-Sciama effect) or 
the onset of curvature or cosmological constant domination (the Integrated 
Sachs-Wolfe effect).  
Simulations have been made of the impact of 
patchy reionization 
\citep{aghanimetal96, knoxsd98, gruzinovh98, peeblesj98}.
%(Aghanim et al. 1996, 
%Knox, Scoccimarro, \& Dodelson 1998, Gruzinov \& Hu 1998, Peebles \& 
%Juskiewicz 1998).  
The SZ effect itself is independent of redshift, so it can yield 
information on clusters at much higher redshift than does X-ray 
emission.  However, nearly all clusters are unresolved for $10'$ resolution 
so higher-redshift clusters occupy less of the beam and therefore their SZ
effect is in fact dimmer.  In the 4.5$'$ channels of Planck this will 
no longer be true, and SZ detection and subtraction becomes  
more challenging and potentially more fruitful as a probe 
of cluster abundance at high redshift.  An additional 
secondary anisotropy is that caused by gravitational lensing (see e.g. 
\citealp{cayonms93, cayonms94, metcalfs97, mgsc97}).  
Gravitational lensing imprints 
slight non-Gaussianity in the CMB which can be cross-correlated 
with large-scale structure templates \citep{suginoharass98} or 
perhaps analyzed directly to determine the matter power spectrum 
\citep{seljakz98, zaldarriagas98b}.

\subsection{Polarization Anisotropies}	

Polarization of the Cosmic Microwave Background radiation 
\citep{kosowsky94, kamionkowskiks97, zaldarriagas97}
arises 
due to local quadrupole anisotropies at each point on the surface 
of last scattering (see \citealp{huw97b} for a review).
  Scalar (density) perturbations generate linear 
(electric mode) polarization only, but tensor (gravitational wave) 
perturbations can generate circular (magnetic mode) polarization.  
Hence the polarization of the CMB is a potentially useful probe of 
the level of gravitational waves in the early universe
\citep{seljakz97, kamionkowskik98}, especially 
since current indications are that the large-scale primary 
anisotropies seen by COBE do not contain a measureable fraction 
of tensor contributions (see Chapter \ref{chap:results}).

\subsection{Gaussianity of the CMB anisotropies}

The processes turning density inhomogeneities into CMB anisotropies 
are linear, so cosmological models that predict gaussian primordial 
density inhomogeneities also predict a gaussian distribution of 
CMB temperature fluctuations.  Several techniques have been developed 
to test COBE and future datasets for deviations from gaussianity
\citep[e.g.][]{kogutetal96b, ferreiram97, ferreirams97}.  Most 
tests have proven negative, but a few claims of non-gaussianity have 
been made.  \citet{gaztanagafe98} found a very marginal indication 
of non-gaussianity in the spread of results for degree-scale 
CMB anisotropy observations being greater than the expected sample 
variances.  \citet{ferreiramg98} have claimed a detection of non-gaussianity 
at multipole $\ell=16$ using a bispectrum statistic,  
and \citet{pandovf98} find a non-gaussian wavelet coefficient correlation 
on roughly $15^\circ$ scales in the North Galactic hemisphere.  Both 
of these methods produce results consistent with gaussianity, however, if 
a particular beam-size area of several pixels is eliminated from the dataset 
\citep{bromleyt99}.  A different area appears to cause each detection, giving 
evidence that the COBE dataset had non-gaussian instrument noise in at 
least two areas of the sky (a true sky signal should be larger than several 
pixels so instrument noise is the most likely source of the non-gaussianity).

\subsection{Foreground contamination}

Of particular concern in measuring CMB anisotropies is the issue of foreground
contamination (see Part II for a full discussion).  
Foregrounds which can affect CMB observations include
galactic radio emission (synchrotron and free-free), galactic infrared
emission (dust), extragalactic radio sources (primarily elliptical galaxies, 
active galactic nuclei, and quasars), extragalactic infrared sources (mostly
dusty spirals and high-redshift 
starburst galaxies), and the Sunyaev-Zeldovich (SZ) effect 
which results from the Compton scattering of CMB photons by hot gas in 
galaxy clusters.  The COBE team has gone to great lengths to analyze their
data for possible foreground contamination and routinely eliminates everything
within about $30^{\circ}$ of the galactic plane.  

An instrument with large 
resolution such as COBE is most sensitive to the large foreground structure
of our Galaxy, but small-scale anisotropy experiments need to worry  
about extragalactic sources as well.  
Because 
they are now becoming critical, these extragalactic foregrounds 
are studied in detail in Part II.  Because foreground 
and CMB anisotropies are assumed to be uncorrelated, they should add in 
quadrature, leading to an increase in the measurement of CMB anisotropy.  
Most CMB instruments, however,
 can identify foregrounds by their spectral signature
across multiple 
frequencies or their display of the beam response characteristic
of a point source.  This leads to an attempt at foreground subtraction, 
which can cause an underestimate of CMB anisotropy if some true
signal is subtracted along with the foreground.

\chapter{Current Status of Observations}
\label{chap:obs}

\section{Cosmic Microwave Background Anisotropy Observations}
\label{sect:obs_cmb}

\begin{figure}
\centerline{\psfig{file=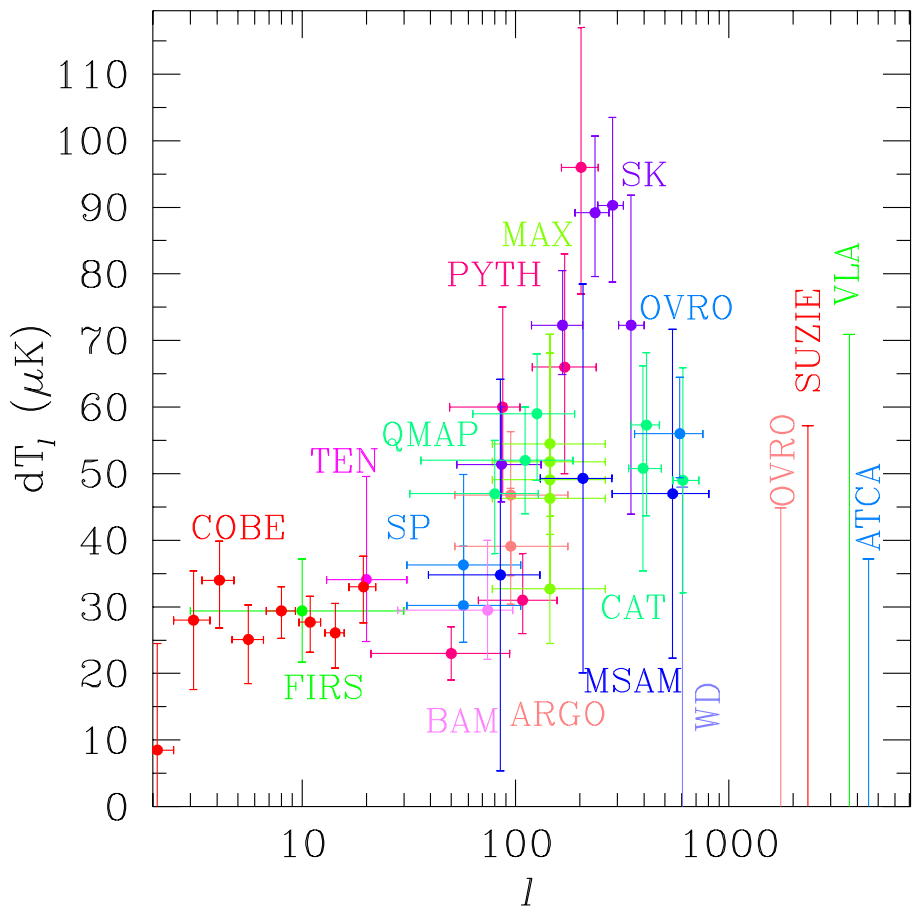,width=5in}}
\caption{Compilation of CMB anisotropy observations.}   
\mycaption{
Vertical error bars represent
$1\sigma$ uncertainties ($2\sigma$ for upper limits) 
and horizontal error bars show the full width 
at half maximum of 
each instrument's window function.
}
\label{fig:obs_cmb}
\end{figure}

Since the COBE DMR detection of CMB anisotropy \citep{smootetal92}, there have
been over thirty additional measurements of anisotropy on angular scales
ranging from $7^{\circ}$ to $0\fdg3$, and upper limits have been set
on smaller scales.  
Shown in Figure \ref{fig:obs_cmb} are COBE \citep{tegmarkh97},
FIRS \citep{gangaetal94},
Tenerife \citep{gutierrezetal97},
South Pole \citep{gundersenetal95},
BAM \citep{tuckeretal97},
ARGO \citep{masietal96},
Python \citep{cobleetal99, plattetal97}, 
MAX \citep{limetal96,tanakaetal96},
MSAM \citep{wilsonetal99},
SK \citep{netterfieldetal97}, 
CAT \citep{scottetal96,bakeretal99},
OVRO/RING \citep{leitchetal98},
WD \citep{tuckeretal93},
OVRO \citep{readheadetal89},
SUZIE \citep{churchetal97},
ATCA \citep{subrahmanyanetal93},
and VLA \citep{partridgeetal97}\footnote{CMB 
observations have also been compiled by \citet{smoots98} and
at\\ http://www.sns.ias.edu/\~{}max/cmb/experiments.html and\\
http://www.cita.utoronto.ca/\~{}knox/radical.html.}.  
This figure shows our compilation of CMB anisotropy observations without 
adding any theoretical curves to bias the eye.  It is clear that a straight 
line is a poor but not implausible fit to the data. 
There is a clear rise around $\ell=100$ and then a 
drop by $\ell=1000$.  This is not yet good enough to give a clear 
determination of 
 the 
curvature of the universe, let alone fit several cosmological parameters.  
However, the data are good enough to prefer certain models of structure 
formation when a quantitative comparison is made, as 
will be discussed more in the next few chapters.

The COBE DMR observations 
were pixelized into a skymap, from which it is possible to analyze any 
particular multipole within the resolution of the DMR.  
Current small angular scale 
CMB anisotropy observations are insensitive to both high $\ell$ and 
low $\ell$
multipoles because they cannot measure features smaller than their
resolution and are insensitive to features larger than the 
size of the patch of sky observed or the angle
covered as they ``chop'' from one direction to another on the sky to 
eliminate instrumental and atmospheric systematics.  
The next satellite mission, NASA's 
Microwave Anisotropy Probe 
(MAP), is scheduled for launch in fall of 2000
and will map angular scales down to $0\fdg2$ with high precision over most of 
the sky.  An even more precise satellite, ESA's Planck, is scheduled 
for launch in 2007.  
  Because COBE observed such large angles, the DMR data can only  
constrain the amplitude $A$ and index $n$ of 
the primordial power spectrum in wave number $k$ ($P_p(k) = A k^{n}$), and  
these constraints are not tight enough 
to rule out very many classes of cosmological models.  
Throughout this
thesis, $k$ is given in its observed units of $h$/Mpc.

Until the next satellite is flown, the promise
of microwave background anisotropy measurements to measure important 
cosmological parameters rests with a series of ground-based and 
balloon-borne
anisotropy instruments which have already published results (shown 
in Figure \ref{fig:obs_cmb})  
or will report results in 
the next few years (MAXIMA, BOOMERANG, TOPHAT, ACE, 
VIPER, MAT, VSA, CBI, DASI; see 
\citealp{leeetal99} and \citealp{halperns99}).
Because they are not satellites, these instruments face the problems of 
shorter observing times and less sky coverage, although significant 
progress has been made in those areas.  They fall into 
three main categories:  high-altitude balloons, ground-based
instruments, and 
interferometers.  Past, present, and future balloon-based instruments are 
FIRS, MAX, MSAM, ARGO, BAM, MAXIMA, BOOMERANG, TOPHAT, and ACE.  Ground-based 
interferometers include CAT, VSA, CBI, and DASI, and other ground-based 
instruments are TENERIFE, SP, PYTHON, SK, OVRO/RING, VIPER, and MAT.  
Taken as a whole, they have the potential to yield very useful 
measurements of the radiation power spectrum of the CMB on degree and 
subdegree scales.  Ground-based non-interferometers have to discard a large
fraction of data and undergo careful further data reduction to eliminate 
atmospheric contamination.  Balloon-based instruments need to keep a careful 
record of their pointing to reconstruct it during data analysis.  
Interferometers may be the most promising technique at present but they 
are the least developed, and most instruments are at radio frequencies 
and have very narrow frequency 
coverage, making foreground contamination a major concern.  
In order to use small-scale CMB anisotropy measurements to constrain 
cosmological models we need to be confident of their 
validity and to trust the error bars.  This will allow us to discard badly
contaminated data and to give greater weight to the more precise measurements 
in fitting models.  Correlated noise is a great concern for instruments 
which lack a rapid chopping because the $1/f$ noise causes correlations 
on scales larger than the beam 
in a way that can easily mimic CMB anisotropies if not 
analyzed carefully.  
Additional issues are sample variance caused by the combination of 
 cosmic variance and limited sky coverage and foreground contamination.

\subsection{Window Functions}

The sensitivity of these instruments to various multipoles is called
their window function.  
%Figure 3 shows window functions for the 
%relevant CMB anisotropy observations.  
These window functions 
are important in analyzing anisotropy measurements because
the small-scale experiments do not measure enough of the sky to produce
skymaps like COBE.  Rather they yield a few 
``band-power'' measurements of rms temperature anisotropy which reflect 
a convolution over the range of multipoles contained in the window 
function of each band.  Some instruments can produce limited 
skymaps \citep{whiteb95}.  The window function $W_\ell$ shows
how the total power observed is sensitive to the anisotropy on 
the sky as a function of angular scale:

\begin{equation}
Power = \frac{1}{4 \pi} \sum_\ell (2 \ell + 1)C_\ell W_\ell = \frac{1}{2}
(\Delta T/T_{CMB})^2 \sum_\ell \frac {2 \ell + 1}{\ell(\ell+1)} W_\ell 
\end{equation}

\noindent where the COBE normalization is $\Delta T = 27.9 \mu K$ and
$T_{CMB}=2.73 K$ \citep{bennettetal96}.  
This allows the observations of broad-band 
power to be reported as observations of $\Delta T$, and knowing the window
function of an instrument one can turn the predicted $C_\ell$ spectrum
of a model into the corresponding prediction for $\Delta T$.  
This ``band-power'' measurement 
is based on the standard definition that for a ``flat'' power spectrum,
$\Delta T = (\ell (\ell + 1) C_\ell )^{1/2}T_{CMB}/2\pi$ (flat
actually means that $\ell(\ell+1)C_l$ is constant).

The autocorrelation function for measured temperature anisotropies
is a convolution of the true expectation values for the anisotropies 
and the window function.  Thus we have \citep{whites95}

\begin{equation}
 \left \langle 
\frac{\Delta T}{T} (\hat{n}_{1} )
\frac{\Delta T}{T} (\hat{n}_{2} ) \right \rangle = 
\frac{1}{4 \pi} \sum_{\ell=1}^{\infty} (2 \ell + 1) C_\ell 
W_\ell ( \hat{n}_{1}, \hat{n}_{2}), 
\end{equation}

\noindent
where the Gaussian forms above give $W_{\ell}$ as a 
function of separation angle only.  In general, the window function 
results from a combination of the directional response of the antenna,
the beam position as a function of time, and the weighting of each 
part of the beam trajectory in producing a temperature measurement 
\citep{whites95}.  Strictly speaking, $W_\ell$ is the diagonal part of 
a filter function $W_{\ell \ell'}$ that reflects the coupling of 
various multipoles due to the non-orthogonality of the spherical 
harmonics on a cut sky and the observing strategy of the 
instrument 
\citep{knox99}.      
It is standard to assume a Gaussian beam response of width $\sigma$, 
leading to a window function 
\begin{equation}
 W_{\ell}  = \exp [ - \ell ( \ell + 1 )\sigma^{2}]. 
\end{equation} 
The low-$\ell$
cutoff introduced by a 2-beam differencing setup comes from the window
function \citep{whitess94}  
\begin{equation}
 W_{\ell}  = 2 [ 1 - P_{\ell}(\cos \theta) ] 
\exp [ - \ell ( \ell + 1 )\sigma^{2}]. 
\end{equation} 

\subsection{Sample and Cosmic Variance}

The multipoles $C_\ell$ can be related to the expected
value of multipole moments by 
\begin{equation}
 \langle \sum_m{a_{\ell m}^2}\rangle = (2 \ell + 1) C_\ell 
\end{equation}
since there are $(2 \ell + 1)$  $a_{\ell m}$ for each $\ell$ and each has
an expected autocorrelation of $C_{\ell}$.  In a theory such as inflation,
the temperature fluctuations follow a Gaussian distribution about 
these expected ensemble averages.  This makes the $a_{\ell m}$ Gaussian 
random variables, resulting in a $\chi^{2}_{2 \ell + 1 }$ distribution 
for $\sum_m{a_{\ell m}^2}$.  The width of this distribution leads to a 
cosmic variance in the estimated $C_\ell$ of 
$\sigma^{cv}_\ell$ proportional to 
$(\ell + \frac{1}{2})^{-\frac{1}{2}}C_\ell$, 
which 
is much greater for small $\ell$ than for large $\ell$ (unless $C_\ell$ is rising in a 
manner highly inconsistent with theoretical expectations).  So, although
 cosmic variance is an unavoidable source
of error for small-scale anisotropy measurements,
it is much less of a problem for small scales 
than for COBE.  

Despite our conclusion that cosmic variance is a greater concern on 
large angular scales, Figure \ref{fig:obs_cmb} 
shows a tremendous variation in the 
level of 
anisotropy measured by small-scale experiments.  Is this evidence
for a non-Gaussian cosmological model such as topological
defects?  Does it mean we cannot trust the data?  Neither conclusion 
is justified (although either could be correct) because we do in 
fact expect a wide variation among these measurements due to their
coverage of a very small portion of the sky.  Just as it is difficult to 
measure the $C_{\ell}$ with only a few $a_{\ell m}$, 
it is challenging to 
use a small piece of the sky to measure multipoles whose spherical
harmonics cover the sphere.  It turns out that 
limited sky coverage leads to a sample variance for a particular 
multipole related to 
the cosmic variance for any value of $\ell$ by the simple formula 
\begin{equation}
 \sigma^{2}_{sv} \simeq \left ( \frac {4 \pi}{\Omega} \right ) 
\sigma^{2}_{cv}, 
\end{equation} 
where $\Omega$ is the solid angle observed \citep{scottsw94}.

\section{Observations of Large-Scale Structure}

Theories of structure formation predict that the matter
 power spectrum is given by 
 $P(k) = T^2(k) P_p(k)$ for matter transfer function 
$T(k)$ and primordial power spectrum $P_p(k)$.
The transfer function describes
the processing of initial density perturbations from the Big Bang during
the era of radiation domination; the earlier a spatial scale came within the 
horizon, the more its power was dissipated by radiation 
(and in the CHDM model, by relativistic neutrinos as well).  
If the baryon fraction is large, the same acoustic oscillations of the
photon-baryon fluid that give rise to peaks in the radiation power spectrum
are visible in the matter power spectrum, otherwise the baryons fall into
the potential wells of the dark matter.    
Once matter domination and recombination arrive, 
$P(k)$ maintains its shape and grows
as $(1+z)^{-2}$; hence determining its shape today allows us to extract
the power spectrum of primordial density fluctuations that existed when
the universe was over a thousand times smaller.   

\begin{figure}

\centerline{\psfig{file=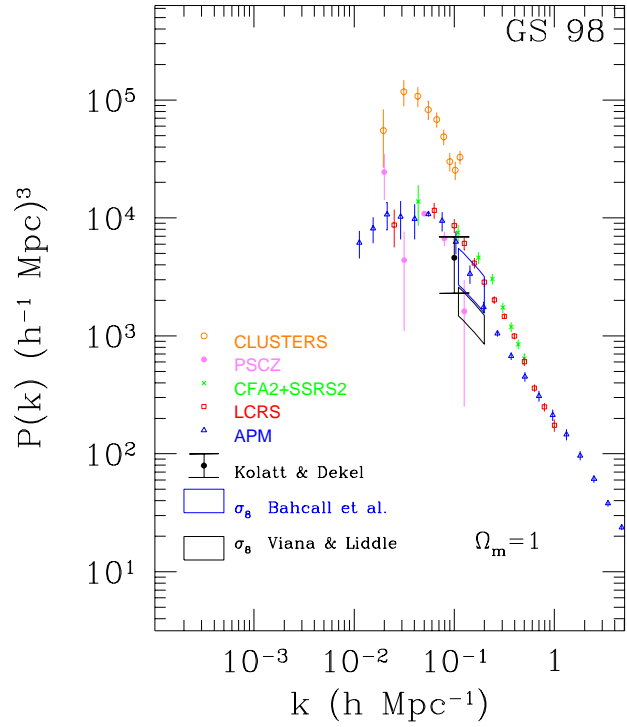,width=4in}}

\caption{Compilation of large-scale structure observations.}  
\mycaption{
No corrections for redshift distortions, non-linear evolution, or bias
have been made.  $k$ is the wave number in comoving units of $h$/Mpc. 
The black and blue boxes are measurements of $\sigma_8$ from the present-day
number abundance of rich clusters and its evolution
\citep[][respectively]{vianal96, bahcallfc97} and the large black 
datapoint is from peculiar velocities \citep{kolattd97}.   
Power
spectra shown are:  the APM galaxy survey (blue triangles), 
Las Campanas (red squares), PSCZ (filled pink circles), 
APM clusters (orange circles), 
and SSRS2+CfA2 (green crosses).   
 }
\label{fig:lss}
\end{figure}

Figure 
\ref{fig:lss} shows our compilation of observations of fluctuations
in the large-scale distribution of galaxies and galaxy clusters.   
In this figure, none of the galaxy surveys have been corrected for bias, 
redshift distortions, or nonlinear evolution. 
This figure 
demonstrates impressive agreement among the current large-scale
structure observations in terms of the shape of the 
uncorrected matter power spectrum.
Clusters are biased compared
to galaxies by about a factor of 3 (about a factor of 10 in 
$P(k)$) and appear to 
have a slightly steeper fall-off
and a small-scale feature in their power 
spectrum which is possibly an artifact of the survey 
window function \citep{tadros99}.  There is no clear evidence,
however,
for
scale-dependence in the bias of the various galaxy surveys on linear
scales.

Figure \ref{fig:lss} includes 
the determination of $\sigma_8$, the rms density variation 
in spheres of radius $8 h^{-1}$ Mpc, by \citet{vianal96} based on 
 the abundance of rich galaxy clusters
with pre-collapse radii $8 h^{-1}$ Mpc.    
The 
width of the box represents the range of spatial scales to which $\sigma_8$ is
sensitive 
(the half-max 
window for $\sigma_8$ is from $k=0.05$ to $k=0.3$ but it 
has been narrowed for clarity) 
and the height shows the $68 \% $ confidence interval.
As an observation of $P(k)$, this 
scales roughly as $\Omega_m^{-1}$ due to the relationship 
between the observed mass and the pre-collapse radius of rich clusters.  
Another measurement of $\sigma_8$ is shown;
this analysis by \citet[][see also \citealt{fanbc97}]{bahcallfc97}
 is based upon the evolution of the abundance of 
rich clusters from redshift 0.3 until now and is 
essentially independent of $\Omega_m$.
Much attention has recently been paid to using 
the abundance of galaxy clusters at $z=0.2-0.8$ as a probe of $\Omega_m$ 
\citep[e.g.][]{carlbergetal97}.  
However, some authors 
\citep{colafrancescomv97, vianal99} 
argue that the systematic uncertainties are too large at present for 
a clear determination of $\Omega_m$.  We account for this constraint 
by using both the value of $\sigma_8$ derived from the $z=0$ 
cluster abundance and that preferred by the evolution of the 
cluster abundance; models with the preferred value 
of $\Omega_m$ will have these two observations of $\sigma_8$ agree.
  Both observations of $\sigma_8$ 
are bias-independent.  However, it is important to remember that, unlike 
measurements of $P(k)$, this measurement of $\sigma_8$ depends on more 
than the second moment of the distribution of density variations.  A 
non-gaussian model of structure formation will in general predict a 
different ratio of the rare overdensities which lead to clusters to the 
root mean square density variation which is described by the power 
spectrum.  Hence the interpretation of this data is highly dependent 
on the assumption of gaussianity (see \citealp{robinsongs98, robinsongs99} 
for a full discussion).  

Our compilation also includes a measurement of the
power spectrum based on peculiar velocities  
\citep{kolattd97}.  
We plot only the
point at $k=0.1$ which is confirmed by the likelihood analysis of 
\citet{zaroubietal97a}.  
A full discussion of peculiar
velocity data is given by \citet{gramann98}.  
Peculiar velocities arise due to the gravitational potential of the underlying
density field, so this observation is bias-independent.  
It scales as $\Omega_{m}^{-1.2}$ (the square of the growing
mode) for different models.

In addition, Figure 
\ref{fig:lss} shows power spectra from four redshift surveys, 
the Las Campanas Redshift Survey (LCRS, \citealt{linetal96}),  
the real-space analysis of the IRAS PSCZ survey \citep{tadrosetal99}, 
a cluster sample selected from the APM Galaxy Survey \citep{tadrosed98},
and the combined SSRS2+CfA2 survey \citep{dacostaetal94}.    
We use the $101 h^{-1}$ Mpc
version of the SSRS2+CfA2 survey to avoid luminosity bias present in 
the deeper sample noted by \citet{parketal94}.
We use the APM Clusters $P(k)$ only for $k\leq 0.12$ to avoid possible 
artifacts of the survey window function at higher $k$.    
The APM cluster $P(k)$ has been
analyzed for several background cosmologies, and we use the version
most appropriate to each model.

We also show the 
power spectrum resulting from the Lucy inversion of the angular correlation
function of the APM catalog \citep{gaztanagab98, baughe93}.  
We drop the first 4 reported APM points because the
analysis is biased compared to numerical simulations on such large scales, 
where plate matching and galactic extinction are also a concern 
\citep{gaztanagab98}.    
The APM power spectrum is measured in 
real space, as is the version of PSCZ we use, 
whereas the others are given in redshift space. 
Each of these power spectra can be scaled by the square of 
an unknown linear bias parameter, where the
bias is the enhancement of galaxy (baryonic) density perturbations over 
the density variations in the underlying dark matter distribution.
  It is
expected that the bias is near unity for each of the galaxy 
survey power spectra,
with IRAS having the least bias because of the long correlation
length of infrared-bright galaxies and with clusters having a large bias
because they trace high density peaks of the primordial density 
distribution and such peaks are themselves highly clustered 
\citep[see][]{kaiser84}. 
We have purposely rebinned some of the galaxy survey data
to make the points independent.  
%*[describe binning procedure]

Structure formation observations at high redshift are improving 
rapidly and may soon be able to add an intermediate probe between 
low-redshift large-scale structure and the early universe density 
perturbations imprinted in the CMB.  
Observations of high-redshift damped Lyman $\alpha$ systems 
are a concern for models such as Cold+Hot Dark Matter and Tilted CDM, 
which have little small-scale power \citep{gardneretal97, maetal97}.  
 The 
improved hydrodynamic simulations of  
\citet{haehneltsr98}
indicate, however, that all 
typical structure formation models have enough small-scale power to 
produce the observations of high-redshift damped Lyman $\alpha$ absorption 
systems.  Improved methods and observations of the Lyman $\alpha$ forest 
\citep{croftetal98a, croftetal98b, hui98} have made measurements of 
the linear power spectrum on sub-Mpc scales possible, although there 
are significant uncertainties at present, particularly in the normalization 
of the measured power spectrum.  The 
clustering of Lyman break galaxies 
observed at redshifts above $z=3$ 
\citep{steideletal98, giavaliscoetal98, adelbergeretal98} indicates 
that these galaxies are highly biased, perhaps the precursors 
of present-day giant ellipticals in clusters.  A wide range of 
cosmological models, however, are consistent with 
these observations, as the first galaxies that form should be highly biased 
in any hierarchical model of structure formation.  

\section{Direct Observations of Cosmological Parameters}
\label{sect:obs_direct}

Observations not based on structure formation can 
probe cosmological parameters without being dependent 
on the primordial power spectrum.  These ``direct'' observations 
include classical cosmological tests of the 
expansion rate of the universe and its acceleration, the 
luminosity distance-redshift 
and angular size distance-redshift relations, and  geometrical 
tests sensitive to the volume-redshift relationship.  As is the nature 
of astronomy, almost all of these observations are really indirect, 
requiring the assumption of some set of objects as standard rods or 
candles and being dependent on various systematic uncertainties in 
those assumptions.  This is the challenge of being an observational science 
where performing controlled experiments is far beyond human capabilities!

  A Hubble constant of $65 \pm 15$km/s/Mpc
 encompasses the range of 
systematic variations between different observational approaches 
\citep{branch98}.  
The former 
age ``crisis'' has disappeared with recent recalibration of the 
distance to the oldest Galactic globular clusters leading to a new estimate 
of their age of $11.5 \pm 1.3$ Gyr \citep{chaboyeretal98}.  
Recent observations of primordial deuterium abundance 
place tight 
constraints on the baryon density from Big Bang nucleosynthesis, 
$0.012 < \Omega_b h^2 < 0.026$ \citep{tytlerfb96}.  
\citet{bartelmannetal98} use numerical simulations to 
compare the observed abundance of arcs from
strong lensing by galaxy clusters with the predictions of various models
and conclude that only Open CDM works, and they find that critical density
($\Omega_m=1$) 
models underpredict the number of arcs by orders of magnitude.   
Further support for low-$\Omega_m$ models comes from the 
cluster baryon fraction found by \citet{evrard97}, 
$\Omega_m/\Omega_b \leq 23 h^{3/2}$ with a best fit value of 
$11.8 h^{4/3}$.    
This favors the ratio of total matter to baryons in models with 
low matter density  and appears to disallow
models with $\Omega_m=1$ and $\Omega_b \leq 0.05$ such as Standard CDM. 
Models with a high Hubble constant and high baryon fraction 
are also in trouble.

Observations of 
Type Ia supernovae at high redshift 
are progressing rapidly, and results
argue in favor of a positive cosmological constant and appear to 
rule out $\Omega_m = 1$ \citep{garnavichetal98a, perlmutteretal98, 
riessetal98}.  
The amount of cosmological constant, however,
 is constrained to be $\Omega_\Lambda \leq
0.7$ by QSO lensing surveys \citep{kochanek96}, although other 
analyses of QSO lensing \citep[e.g.][]{chibay97, chibay99} support the 
supernova results.   
There are also a host of unresolved systematic concerns with the 
Type Ia supernova analysis, including possible evidence for  
evolution of the supernovae, the presence of gray dust 
\citep{aguirre99a, aguirre99b}, and the need to account 
for mass inhomogeneities in using the magnitude-redshift relation 
\citep{kantowski98, moffatt95}.   
While further consideration of systematic issues is needed, there is 
not yet any evidence that the results are biased, and the Type Ia 
supernovae observations are the strongest direct probe of 
cosmology available at present.

\section{Sample Variance of Cosmological Surveys}

Beyond the expected 
variations in Hubble's constant that have been calculated in 
the literature using analytic approximations and numerical simulations
\citep{turnerco92, shiwd96, wuqf96, shit98, wangst98}, 
there are significant effects caused by both sample variance and cosmic 
variance on a number of measured cosmological parameters.
\citet{zehavietal98} find evidence from the velocity flow 
of local Type Ia supernovae that there is a large-scale void, a 
20\% underdensity, surrounding us.  This underdensity causes the local 
value of $H_0$ to be higher than the global value by $\sim 6$\%.   
Whether this void is real or not, inhomogeneities at this level are 
expected in our universe, creating sample variance in measurements of 
the Hubble constant and the magnitude-redshift relation.

Cosmic variance is important when we have only a limited volume of 
the universe that can possibly be observed at a given epoch 
(at least without waiting a very long time to see a different region of 
the universe at that epoch.)  For example the abundance of 
galaxy clusters at a given redshift is only measured for the portion 
of the universe we currently observe near that redshift.  This turns out to be 
a small effect on the inferred number abundance for moderate to high redshifts,
but it adds a few percent cosmic variance error to determination of 
$\sigma_8$ based on the $z=0.05$ cluster 
number abundance.

Sample variance is a greater concern.  The high redshift cluster abundance 
is determined from only a small fraction of the available volume at that 
redshift, although it turns out to be a sufficient volume for sample variance  
to be much less important than the Poissonian variance which is already 
taken into account in these analyses.  In fact, the biggest uncertainty 
here is what actual volume has been surveyed, which is under 
some debate in the literature.  None of these effects seems enough to 
reconcile $\Omega_m=1$ with the very massive clusters seen at high 
redshift, as long as gaussianity is assumed (see 
\citealp{robinsongs99}
for the additional consideration of non-gaussianity).

Sample variance has been overlooked far too often in analyses of 
small regions of high-redshift sky, most egregiously the Hubble Deep Fields.  
Judging the cosmic star formation history from those regions is quite 
dangerous.  The angular size is that of a cluster of galaxies, at basically 
all redshifts, so we expect significant density fluctuations at any 
coordinate distance into the field.  The effect can be calculated for 
an assumed model of density inhomogeneities just as we calculate $\sigma_8$ 
but by using the Fourier transform of the survey volume in place of a 
spherical top-hat.  
%
%[give equation for this]  
%
For biased tracers of 
the dark matter such as galaxies, the expected density variation is just 
given by the bias times this $\sigma$ so it is easy to make a prediction 
to within a factor of two.  Such error bars should be added to any 
cosmological conclusions made using the Hubble Deep Field and similar surveys.

\cleardoublepage

\cleardoublepage
\chapter{Models Tested and Methodology}
\label{chap:method}

%\documentstyle[12pt,aasms4]{article}

%\textwidth 6.00in
%\oddsidemargin 0.25in
%\evensidemargin 0.25in
%\textheight 8.75in
%\topmargin 0.0in

%\begin{document}

%\title{\large
%\bf\sc Extracting Primordial Density Fluctuations}

\section{Structure Formation Models}

We examine ten models of structure formation, which represent the range of 
cosmological parameters currently considered viable.  
Due to theorists' productivity, there are an endless number of 
exotic models we could have used, but we include the most popular variations 
upon Standard Cold Dark Matter (SCDM) 
and a sampling of some promising alternatives.  
Starting from a primordial power spectrum of infinitesimal
 density perturbations in the early universe, each
model predicts how those density perturbations
create anisotropies in the Cosmic Microwave Background radiation and 
inhomogeneities in the distribution of galaxies.
By comparing each model with both kinds of data, we investigate whether it 
provides a consistent picture of structure formation on scales ranging from 
galaxy clusters to the present horizon size.

\begin{table}[h] 
\caption{Values of cosmological parameters for our models.}
\mycaption{  $\tau$ is optical depth to the last-scattering surface due to 
reionization, and x=0.001 is the residual ionization fraction in the 
ICDM model after recombination.  
Parameters marked with a $^*$ were optimized.  
}
\label{tab:models}
\begin{center}
\begin{tabular}{|r|c|c|c|c|c|c|c|c|c|}
\hline
Model &  $\Omega$& $\Omega_{\Lambda}$ & 
$\Omega_{m}$ & $\Omega_c$ & $\Omega_{\nu}$ & 
   $\Omega_{b}$ & h & n & Age (Gyr) \\
\hline
SCDM         & 1.0 & 0   & 1.0   & 0.95   & 0 & 0.05 & 0.5     & 1.0     & 13\\
 
TCDM         & 1.0 & 0   & 1.0   & 0.90  & 0 & $0.10^*$ & 0.5  & $0.8^*$ & 13\\

CHDM	     & 1.0 & 0 & 1.0 & 0.70 & $0.2^*$ & $0.10^*$ & 0.5 & $1.0^*$ & 13\\
 
OCDM       & 0.5 & 0 & $0.5^*$ & 0.45 & 0 & 0.05$^*$ & $0.6^*$ & $1.0^*$ & 12\\

$\Lambda$CDM & 1.0 & 0.5 & $0.5^*$ & 0.45 & 0 & 0.05$^*$ & $0.6^*$ & $1.0^*$ 
& 14\\

$\phi$CDM    & 1.0 & 0   & 0.92  & 0.87   & 0 & 0.05 & 0.5     & 1.0     & 13\\

BCDM         & 1.0 & 0.88 & 0.12 & 0.08   & 0 & 0.04 & 0.8     & 1.6     & 15\\

ICDM         & 1.0 & 0.8 & 0.2   & 0.17   & 0 & 0.03 & 0.7     & 2.2    & 15\\

PBH BDM      & 1.0 & 0.6 & 0.4   & 0    & 0 & 0.10 & 0.7     & 2.0    & 13\\

Strings+$\Lambda$ 
             & 1.0 & 0.7 & 0.3   & 0.25   & 0 & 0.05 & 0.5    & $\sim 1$ & 19\\
\hline
\end{tabular}
\end{center}
\end{table}

Table \ref{tab:models} shows the cosmological parameters of our models, where 
$\Omega = \Omega_m + \Omega_{\Lambda}$ is the ratio
of the energy density of the universe to the critical density necessary to 
stop its expansion, given by $\rho_c = 3 H_0^2 / 8 \pi G$ for a Hubble 
constant of $H_0 = 100h$ km/s/Mpc.  The portion of this critical 
energy density contained in matter is $\Omega_m = \Omega_{CDM} + \Omega_\nu
+ \Omega_b$, the sum of the contributions from CDM, Hot
Dark Matter (HDM) in the form of massive neutrinos, and baryonic matter.  
$\Omega_\Lambda = \Lambda /3 H_0^2$ 
is the fraction of the critical energy density contained
in a smoothly distributed 
vacuum energy referred to as a cosmological constant, $\Lambda$.
  The age of the universe in each model
is a direct consequence of the values of $h$, $\Omega_m$, 
and $\Omega_\Lambda$;
for a critical matter density universe, it is $2H_0^{-1}/3$ 
\citep{weinberg72,carrollpt92}.  
All of our models have an age of at least 13 Gyr except OCDM (12 Gyr).

%*[add other formulas]

Each model has a primordial power spectrum of density perturbations 
given by $P_p(k) = A k^n$ where $A$ is a free normalization parameter and
$n$ is the scalar spectral index.  
 The primordial power spectrum does not have to be an exact power 
law; inflationary models predict slight variation of the power-law 
index with scale.  Allowing this spectral index to vary freely makes it 
more difficult to constrain cosmological models, as shown by 
\citet{gawiser98} and discussed in detail in Chapter \ref{chap:ppk}.    
We normalize to the dataset as a whole,  
except for the galaxy surveys whose free bias parameter gives them no impact
on normalization.  
Scale-invariance corresponds to 
$n=1$.
Topological defect models have scaling solutions, so $n=1$ is approximately
correct in an $\Omega_m=1$ model, but defects will deviate from this 
scaling in low-matter-density models such as the one considered here.   

%*[change all to n=1, add footnote about isocurvature definitions]

\subsection{Standard Cold Dark Matter And Its Variants}

The first seven models are based on the Standard Cold Dark 
Matter (SCDM) 
model \citep{davisetal85} and assume that the initial density perturbations
in the universe were adiabatic (constant entropy), 
as is generally predicted by the 
inflationary universe paradigm.  
Tilted CDM (TCDM) and Cold + Hot Dark Matter (CHDM)  
each correspond to changing the shape of the SCDM matter power spectrum to 
eliminate its well-known problem of 
excess power on small scales relative to large scales 
\citep{whiteetal95,davisss92,klypinetal93,pogosyans93,choir98}. 
The CHDM model has one family of massive neutrinos which contributes
20\% of the critical density.  
The neutrino mass is given by $m_\nu = 94 h^2 \Omega_\nu$ eV $= 4.7$ eV.
Strictly speaking, it is possible for particles other than neutrinos 
to comprise the hot dark matter; \citet{brusteinh98} offer a mechanism by which a single species of particles with a bimodal energy distribution can 
simultaneously comprise cold and hot dark matter.  
 Setting 
$\Omega_b=0.1$
helps TCDM and CHDM agree with the high level of 
CMB anisotropy observed by Saskatoon (SK, \citealp{netterfieldetal97}; 
we use the recalibration of the SK data from \citealp{leitch97}).
The cosmological constant ($\Lambda$CDM) 
and open universe (OCDM) models are similarly motivated; 
$\Omega_m=0.5, h=0.6$ guarantees roughly 
the right shape of the matter power
spectrum \citep{liddlelrv96,liddlelvw96,klypinph96}.

We have optimized some 
parameters of these models:  $n$ for TCDM, $\Omega_\nu, n$, and the number of 
massive neutrino families for CHDM, and $\Omega_m, \Omega_b, h$ and 
$n$ for OCDM and $\Lambda$CDM.   
\citet{primacketal95} 
suggest using two equally massive 
neutrinos to meet the observed cluster abundance, but including 
the higher value of 
$\sigma_8$ implied by the observed cluster abundance at $z=0.3$ 
 makes a single massive neutrino slightly favored. 
 For a single massive neutrino, 
we find $\Omega_\nu = 0.2$ preferable to $\Omega_\nu = 0.3$ by a small
margin.  The parameters for TCDM 
and CHDM are within the range found acceptable by \citet{liddlelssv96}.

The $\phi$CDM model of \citet{ferreiraj98} contains a significant
energy density contribution from the vacuum energy of a late-time
scalar field, $\Omega_\phi = 0.08$.  This energy behaves like matter today,
but during matter-radiation equality and recombination 
it alters the shape of the 
matter and radiation power spectra from the otherwise
similar SCDM model.  Several other types of scalar fields have been proposed
recently 
\citep{caldwellds98, vianal98, hu98, turnerw97, cobledf97}, 
and this model is an example that appears to 
agree well with the observations.

The BCDM model is from \citet{eisensteinetal98} and contains 
nearly equal amounts of baryonic and non-baryonic 
matter.  Its parameters have been tuned to produce a 
peak due to baryonic acoustic oscillations 
in the matter power spectrum at $k=0.05h$/Mpc, 
where a similar peak is seen 
in the 3-dimensional power spectrum of rich Abell clusters 
\citep{einastoetal97} 
and the 2-dimensional power spectrum of the Las Campanas Redshift Survey 
\citep{landyetal96}. 
  BCDM has a scalar spectral index of $n=1.6$, which is 
barely consistent with constraints from
the COBE FIRAS spectral distortion limits \citep{huss94}, although those 
limits come from small spatial scales so varying $n$ could relax that 
constraint.  
%*[mention PBH constraints, point out same workaround]
To reduce the level of CMB anisotropy that results from this tilt, BCDM 
has reionization with optical depth to the last scattering surface of 
$\tau = 0.75$.  The Tilted CDM model has scalar spectral index $n=0.8$.  
The other SCDM variants have scale-invariant 
primordial power spectra with $n=1$ 
\citep{harrison70, zeldovich72, peeblesy70}.   
For the Open CDM
model, we have used the primordial power spectrum from \citet{liddlelrv96},
where the scale-invariance of gravitational potential perturbations 
causes a rise in the matter power spectrum beyond the  
curvature scale.   

\subsection{Isocurvature Models}

The Isocurvature Cold Dark Matter (ICDM) 
model we consider has been proposed by \citeauthor{peebles97} 
(\citeyear{peebles97}, see also \citealp{peebles99a,peebles99b}).
  Its primordial power spectrum is tilted so 
that $\sigma_8=1$ when it is normalized to 
COBE.  Early structure formation, 
in agreement with observations of galaxies at high 
redshift and the Lyman $\alpha$ forest, is caused by 
a non-Gaussian ($\chi^2$) distribution of 
isocurvature (constant potential) 
density perturbations of the CDM produced by a massive scalar field
frozen during inflation.  This model has a residual ionization 
fraction $x=0.001$ after decoupling. 

The Primordial Black Hole Baryonic Dark Matter
(PBH BDM) model of \citet{sugiyamas99}  
has isocurvature perturbations but no Cold Dark Matter.  The 
primordial black holes form from baryons 
at high density regions in the very early 
universe during the quark-hadron phase transition 
and thereafter act like CDM.  Only a tenth
of the critical energy density remains outside the black holes to participate
in nucleosynthesis.  These black holes have the appropriate mass 
($\sim 1 M_\odot$) to be
the Massive Compact Halo Objects (MACHOS) which have been detected in 
our Galaxy \citep{alcocketal97}.

\subsection{Topological Defect Models}

Rough agreement has now been reached 
on the radiation and matter power spectra predicted for models in 
which the primordial density perturbations are seeded by topological
defects originating in a symmetry-breaking in the early universe.  
Topological defects are active sources and exist at low redshift, 
but they play 
a significant role in structure formation only at relatively early times.  
Thus their  
relative predictions for matter and radiation fluctuations are similar
to inflationary models, although non-Gaussianity is expected 
in the matter distribution.   
\citet{albrechtbr97}
take a range of 
defect models in critical-density cosmologies and claim 
that standard topological defect models are ruled out.  
The Strings + $\Lambda$CDM model we use is from 
\citet[][see also \citealt{avelinocm97}]{battyera98},
% was \citet{battyera98} followed by avelino  
where the change in cosmology causes a deviation 
from scaling and makes cosmic strings a viable model.

\section{Comparison with Observations}

In selecting these models, we have not used the current set of 
direct observations of cosmological parameters to eliminate 
critical density models, although the Type Ia supernovae results 
are strong enough to discriminate against even low-density non-cosmological 
constant models with good statistics.  We are more interested at this 
point in seeing what the structure formation data say directly, and in 
seeing if there is agreement between the structure formation data and 
the direct observations of cosmological parameters.
\citet{steigmanhf99} 
combined a number of observational constraints 
and found that the strongest constraint comes from the shape parameter
of the matter power spectrum, which we treat in much greater detail here.  
When they relaxed that constraint, all of our models were 
allowed at the $2 \sigma$ level. 
We find that the current discriminatory power of observations of 
structure formation outweighs that of direct parameter observations other 
than the Type Ia supernovae observations and is roughly equal 
to that of the supernovae results.

\citet{scottsw95} 
illustrated the comparison between CMB anisotropy and 
fluctuations in the galaxy distribution.  Several 
similar analyses 
\citep{taylorr92, whites96, dodelsongt96, 
liddlelrv96, liddlelvw96, liddlelssv96, whiteetal96, klypinph96,
hancocketal97, bondj97, lineweaverb98} 
have been performed  
but have used only a portion
of the compilation of observations that we present.
\citet{websteretal98}  includes a more detailed analysis 
of the galaxy distribution of the IRAS redshift survey
 than is performed here.  When 
we restrict our analysis to the datasets they consider we find that several 
models are in good agreement with the data.

  We have developed an expanded technique for testing models 
of structure formation.  Our
procedure is as follows:  
we start with a primordial power spectrum of density 
fluctuations imprinted during the Big Bang.
Each cosmological
model gives transfer functions that develop this primordial
power spectrum into predictions for temperature fluctuations in the microwave
background and fluctuations in the distribution of galaxies.  
We use the CMBfast code developed by 
\citet[see also \citealt{zaldarriagasb98}]{seljakz96} 
to calculate the predicted 
CMB angular power spectrum and matter power spectrum
for the SCDM, TCDM, CHDM, OCDM, $\Lambda$CDM, and BCDM models and have 
obtained the power spectra for the other models from their authors.  
We are able to compare the two types of observations visually by plotting 
35 detections of CMB anisotropy, ranging from COBE
at $7^{\circ}$ to CAT at $0\fdg3$, as model-dependent estimates of
the matter fluctuation power spectrum on the same spatial scales.    
We also derive the variance in
density fluctuation amplitudes from the various galaxy redshift
surveys.  Several complications
 arise here, which we treat in detail, 
following the method developed by \citet{peacockd94}.
  The fluctuations are often in redshift
space, which must therefore be corrected to physical space.  On smaller
scales, the effects of non-linearity are important.  Finally there is selection
bias:  IRAS-selected galaxies are more uniformly distributed than galaxies
selected optically, and each morphological type 
is expected to have a scale-independent bias which reflects the enhancement 
in clustering
of density peaks of that mass versus the clustering of the underlying dark
matter.   
Our assumption that bias is constant for a given morphology of 
galaxies except on non-linear scales is supported by 
\citet{kauffmannns97}, \citet{mannph98}, and \citet{scherrerw98}.  
We address
these issues by correcting the data for redshift distortions and 
non-linear clustering and finding the best-fit bias of each type of galaxies.
These procedures are now well understood on scales above 10 Mpc 
($k=0.2$), so we retain
information only on that scale and larger for our statistical analysis.  
We obtain information on the amplitude of matter density fluctuations
from measurements of peculiar velocities and galaxy cluster abundances, which
are expected to reflect the underlying dark matter distribution
and are therefore bias-independent.  We compare the predictions
of each model with the corrected observations and compute the probability
of obtaining the data given the model.

%*[which models are consistent with observations and how]

\subsection{Cosmic Microwave Background Anisotropy Detections}

In order to compare the
predictions and observations for each structure formation model on one plot,
we have translated these observations of the angular power spectrum 
 into  estimates of the matter power spectrum
on the same range of spatial scales.  
We could instead compare both data sets as reconstructions of 
the primordial power spectrum which existed in the very early universe, but 
here we choose to plot the data as measurements of the matter power spectrum,
whose shape has been constant since the era of matter domination began.  
Due to projection effects, 
fluctuations at the last-scattering surface on a given angular scale, 
represented by a spherical harmonic multipole 
$\ell \simeq 180^{\circ} / \theta$, 
are generated by density perturbations from a range of spatial scales centered
on a wavenumber $k$, where $\ell \simeq k \eta_0$ and the 
distance to the last scattering surface is given by 
$\eta_0 = 2 c H_0^{-1} \Omega_m^{-\alpha}$; $\alpha \simeq 0.4$ 
in a flat universe
and $\alpha =1$ in an open universe \citep{vittorios92}.  

\begin{figure}
\centerline{\psfig{file=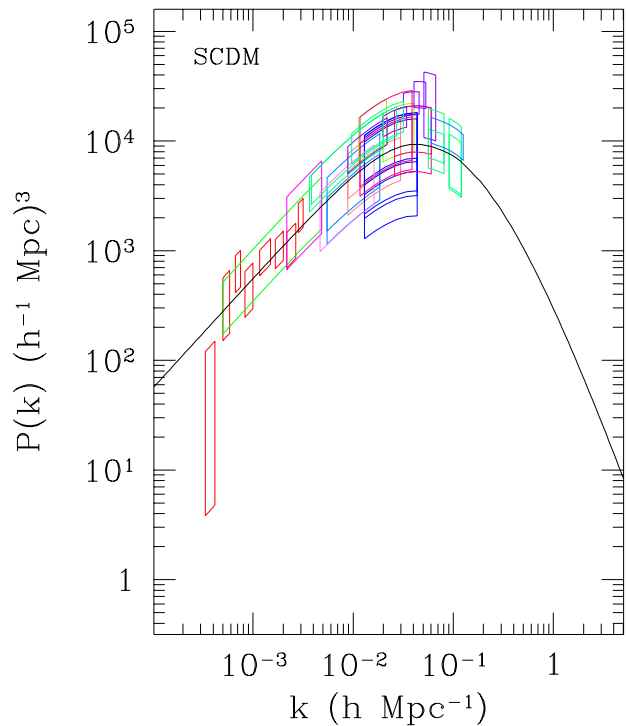,width=4in}}
\caption{
CMB anisotropy observations shown with 
SCDM power spectrum.  
}
\mycaption{
Each box has a width 
indicating the range of $k$ to which the corresponding observation 
is sensitive and a height indicating the $1 \sigma$ error bars of the 
band-power observed.  The amount by which the box is above or below the 
curve indicates how much more or less power that observation prefers 
versus the normalization of the theory shown here.
}
\label{fig:scdm_cmb}
\end{figure}

We use this correspondence to plot each CMB anisotropy detection 
as a box in Figure \ref{fig:scdm_cmb}  
where the width of the box represents the range of $k$ to which that
experiment is most sensitive, and the height of the box shows the $68\%$ 
confidence interval.  
The boxes 
follow the local shape of each model's prediction for $P(k)$ to 
indicate that they 
are a model-dependent averaging of the power over a range of $k$.
The range of uncertainty includes 
the statistical uncertainty
in the detection, the calibration uncertainty of the instrument, the sample
variance from observing only part of the sky, and the cosmic variance
from observing at only one location within the universe.  
The calibration errors, although systematic, have been treated as 
statistical and added in quadrature.  Although calibration errors 
are correlated for multiple observations
by the same instrument, they have been treated as independent, which after 
the recalibration of SK by \cite{leitch97} is a good approximation.

This translation from angular to spatial power spectra is 
model-dependent in two ways.  
Each model has a particular value of $\eta_0$ 
which moves the entire set of boxes horizontally.
Additionally, each model gives
a prediction for the angular power spectrum of CMB anisotropy whose comparison
with observations gives the vertical placement of the box, showing
the inferred amplitude of matter density fluctuations at that
range of spatial scales.

\subsection{Observations of Large-Scale Structure}

\begin{figure}
\centerline{\psfig{file=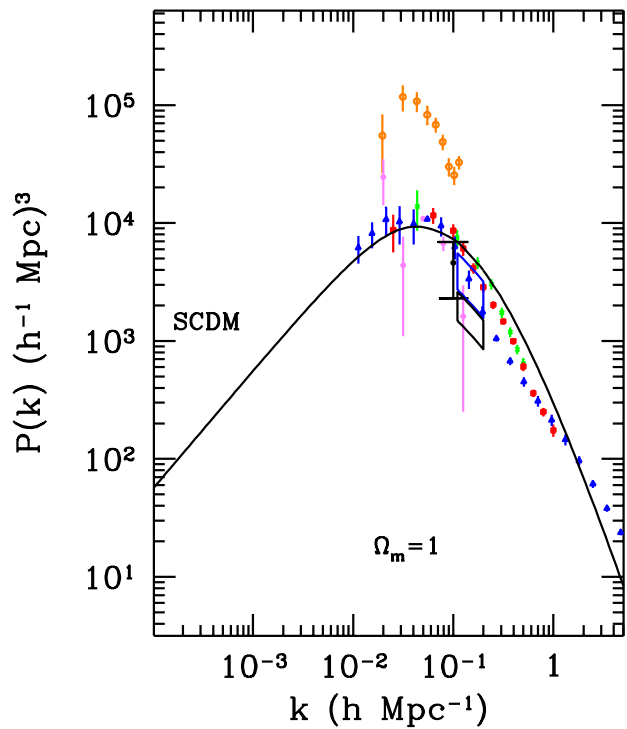,width=4in}}
\caption{
  Compilation of large-scale structure observations 
with SCDM $P(k)$.}
\mycaption{
No corrections for redshift distortions, non-linear evolution, or bias
have been made.
Black and blue boxes are measurements of $\sigma_8$ and the large 
black point with error bars is from peculiar velocities.
Power
spectra shown are:  the APM galaxy survey (blue triangles), 
Las Campanas (red squares), IRAS (pink circles), 
APM clusters (orange circles), 
and SSRS2+CfA2 (green crosses).  
}
\label{fig:lss_scdm}
\end{figure}

We plot our large-scale structure data compilation in 
Figure \ref{fig:lss_scdm}, along with 
the theoretical matter power spectrum $P(k)$ of the
SCDM model, 
with its best-fit normalization.   
The prediction of $\sigma_8$
is an integral over the matter power spectrum using a spherical 
top-hat window function given by \citet{peacockd94}
\begin{equation}
\sigma_R^2 = \frac{1}{2 \pi^2} \int dk k^2 P(k) \frac{9}{(kR)^6}
(\sin kR - kR \cos kR)^2. 
\end {equation}
The boxes for $\sigma_8$ thus 
follow the local shape of each model's prediction for $P(k)$ to 
indicate that they 
are a model-dependent averaging of the power over a range of $k$.
We use the method proposed by \citet{chiuos98} and generalized by 
\citet{robinsongs98} 
to interpret the observed 
cluster abundances and their evolution as measurements of $\sigma_8$ 
for the non-Gaussian 
Strings+$\Lambda$CDM model.  For ICDM, we use a single
determination of $\sigma_8=0.9 \pm 0.1$ \citep{peebles99b}.
The error bars on the measurement of $P(k)$ from 
peculiar velocities include cosmic variance.  To 
be precise, the cosmic variance should be calculated from the model, but in 
this case the cosmic variance of a model with the observed level of power has 
been added in quadrature into the error bars of the observation.    

For each galaxy redshift survey, we perform the model-dependent 
redshift distortion and 
non-linear evolution corrections for a range of biases and find the 
best fit bias parameters for each model. 
The power spectrum observed
in redshift space is related to that in real space by 
\begin{equation}
\frac{P_z(k)}{P_{real}(k)} = (1 + \beta \mu^2)^2 D(k \mu \sigma_p), 
\end{equation}
where the first term gives the Kaiser distortion \citep{kaiser87}
from coherent infall
of galaxies with bias $b$ as a function of $\beta = \Omega_m^{0.6} / b$ 
and the second term is the 
damping of such distortions by the rms pairwise galaxy velocity dispersion 
$\sigma_p$ measured in units of $H_0$.  This velocity dispersion leads to
the so-called fingers-of-God effect in redshift surveys.
For an exponential velocity distribution,
\begin{equation}
D(k \mu \sigma_p) = (1 + (k \mu \sigma_p)^2/2)^{-1}.
\end{equation}  
We couple these terms and average over $\mu$, the cosine of the 
angle between the line of sight and a given wave vector $\mathbf{ k}$, 
to produce
an estimate of the unbiased real-space spectrum  
$P_{real}(k) = P_z(k) / f(k,b)$.  Defining $K = k \sigma_p / \sqrt{2}$, 
we have \citep{ballinger97}
\begin{equation}
f(k,b) = \frac{b^2}{K}\left [ \tan^{-1}(K) \left (1 - \frac{2 \beta}{K^2} + 
\frac{\beta^2}{K^4} \right ) + \frac{2 \beta}{K} + \frac{\beta^2}{3 K} - 
\frac{\beta^2}{K^3} \right ].
\end{equation}

For the pairwise velocity dispersion, we use the observation of 
\citet{landysb98} 
of $\sigma_p = 3.63 h^{-1}$Mpc; their determination that 
the velocity distribution is exponential motivates using that form for  
the damping term.  We have tried this analysis
using the higher value of $\sigma_p = 5.70 h^{-1}$Mpc claimed 
by \citet{jingmb98} 
and it makes only a small difference on quasi-linear
scales; at $k=0.2$  the scale-dependence of the redshift
distortions is a 15\% effect for the Landy et al. value and twice that
for the higher one. This systematic uncertainty in the pairwise 
velocity dispersion makes the corrected real-space power spectrum somewhat 
unreliable on smaller scales.
\citet{smithetal98}
give a full discussion of the effects of varying 
$\sigma_p$ and propagating this systematic uncertainty through the 
linearization procedure; their results confirm that the systematic 
uncertainty is small up to $k=0.2$.  They also find good agreement 
between the linearized observations and linear theory under the CHDM 
model when optically-selected galaxies have bias $b=1.1$.

\begin{figure}
\centerline{\psfig{file=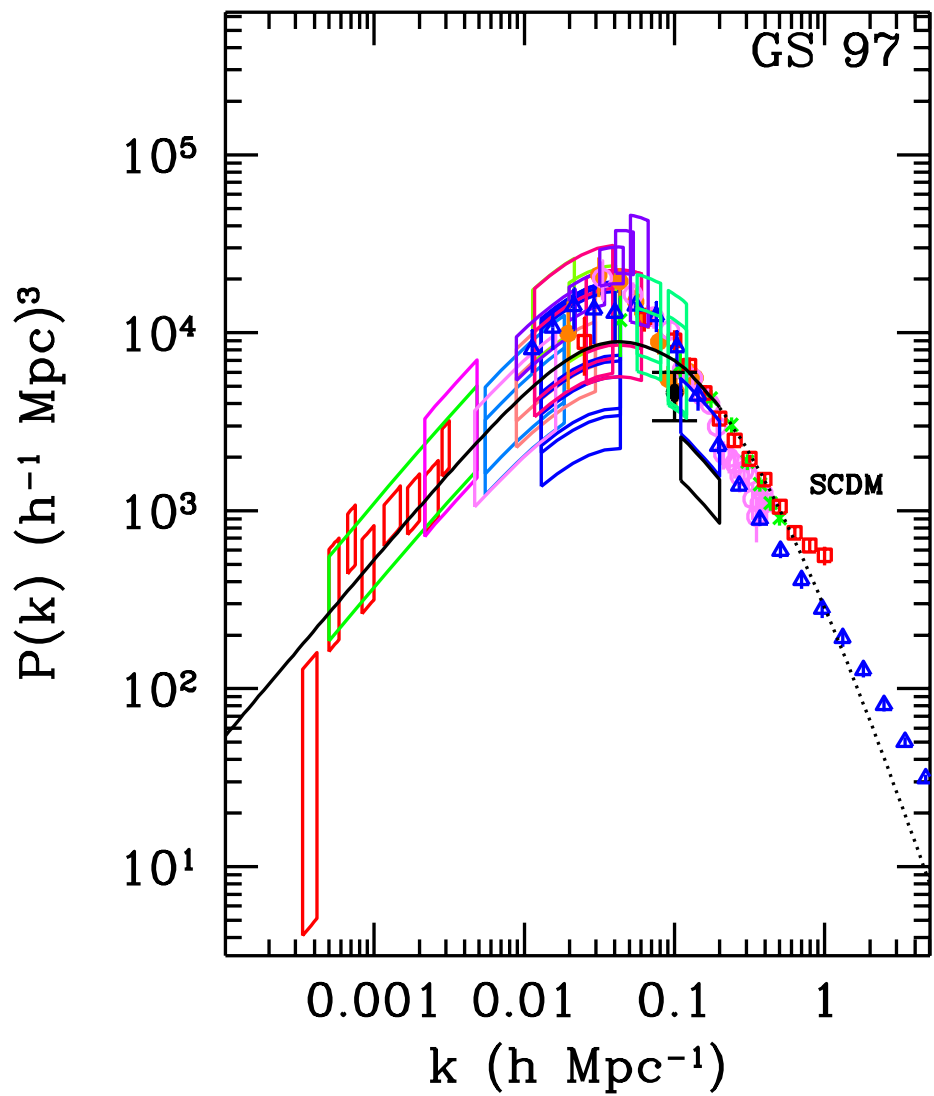,width=6in}}
\caption{
SCDM compared to CMB and LSS data, after the LSS data have been 
corrected for redshift distortions, but not yet for non-linear evolution.  
This represents our reconstruction of the 
real-space non-linear power spectra, but it is compared with the linear theory
prediction for the SCDM power spectrum.
}
\label{fig:scdm_nl}
\end{figure}

After correcting for redshift distortions and bias, 
we have the unbiased real-space non-linear $P(k)$ represented
by a given galaxy survey (shown in Figure \ref{fig:scdm_nl}).  
Next, 
we correct for non-linear evolution
to produce estimates of the linear power spectrum
from these galaxy surveys.  Because collapsing structure leads to 
a change of physical scale, observed $k_{nl}$ can be corrected to their
linear values, given by

\begin{equation}
k_l = (1 + \Delta^2_{nl})^{-\frac{1}{3}}k_{nl},  
\end{equation}

\noindent using the observed power per logarithmic $k$-interval, 
$\Delta^2=k^3P(k)/2\pi^2$.  
The non-linear evolution from $k_l$ to $k_{nl}$ is given by 
$\Delta^2_{nl} = f(\Delta^2_l)$.  A semi-analytic fit for this 
function with 10\% accuracy compared to numerical simulations is given 
by \citet{peacockd96}.  
By inverting their formula numerically,
we linearize the unbiased real-space non-linear $P(k)$ to extract
the primordial density fluctuations.  
The accuracy of this formula is confirmed by  
\citet{smithetal98}.

This correction for non-linear
evolution is model-dependent, as it assumes a local 
slope for the original
linear power spectrum based on the model being tested.  On scales 
$k \leq 0.2$ this linearization preserves the shape of the observed 
non-linear $P(k)$ while sliding the data points to smaller $k$ values;
the main effect of linearization is to shrink the error bars.
  The APM and PSCZ real-space galaxy power spectra are corrected
for bias  and then linearized.  For the clusters, we use only the 
scale-independent Kaiser distortion term to correct for redshift distortions
as the clusters are engaged in coherent infall onto superclusters.     
The magnitude of the variation due to 
systematic uncertainty in the pairwise velocity distribution at $k=0.2$ is 
not changed by this linearization procedure, as $k=0.2$ is roughly the 
scale where $\Delta^2=1$ and non-linear evolution is therefore significant
only at smaller scales.    

\begin{figure}
\centerline{\psfig{file=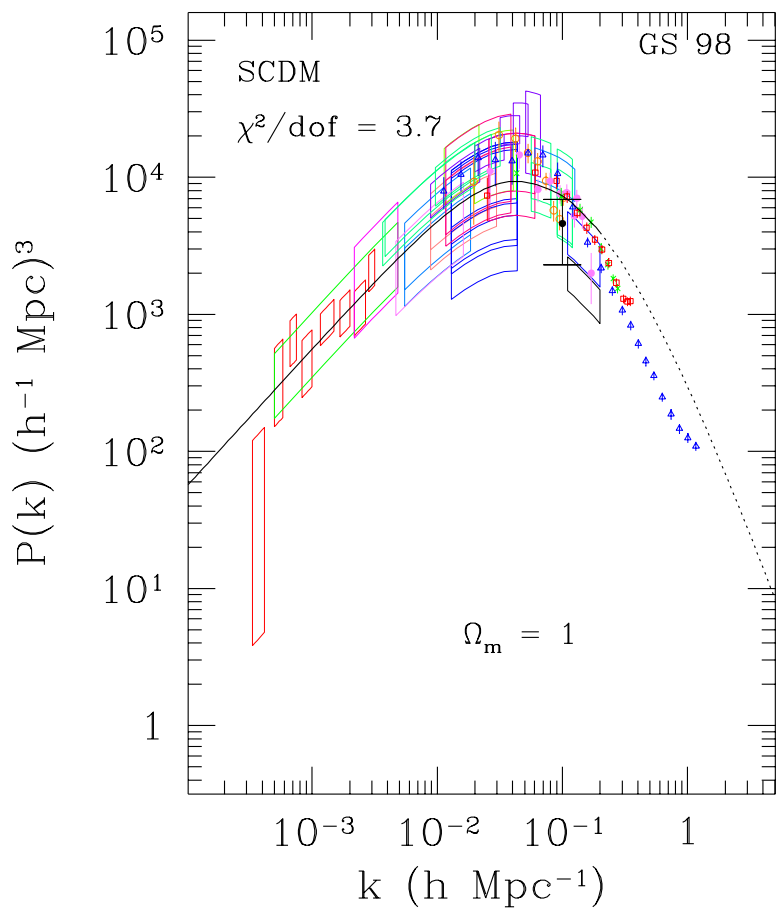,width=5in}}
\caption{
The SCDM model compared to the data, which has undergone 
model-dependent corrections for 
redshift distortions, bias, and non-linear evolution.  
  Beyond $k=0.2$, the predicted matter power spectrum
 curve is dotted to indicate uncertainty in the data corrections.
}
\label{fig:scdm}
\end{figure}

Figure \ref{fig:scdm} shows the reconstructed linear real-space power spectrum
data for the SCDM model with CMB anisotropy constraints shown as well.  
The non-linear evolution causes
a kink in the observed APM $P(k)$ at $k=0.2$.
We find that the lack of a kink in the observed LCRS and SSRS2+CfA2
data results from the rough cancellation of non-linear evolution and 
velocity dispersion damping on scales between $k=0.2$ and $k=1$.  
The high-k 
end of the LCRS data shows that the combination of deconvolving the 
fingers-of-God and linearizing the data has kept the shape
relatively the same but moved the points along that curve and 
reduced the error bars.  The linearization
of the APM dataset has removed the kink at $k=0.2$ but gives a slight curve
around $k=1$, an indication of a problem in either the model or the 
procedure at those scales.  
The error bars are generally smaller between $k=0.1$ and $k=0.2$ than
at larger scales, giving this region the most weight in selecting the 
best-fit biases.  
The dataset as a whole is rather smooth, showing no
strong indications of peaks or troughs; this contradicts the 
finding of \citet{einastoetal97} based on the power spectrum of rich
Abell clusters.     
As noted by 
\citet{gaztanagab98}, 
a clear peak is 
taking shape in the matter power spectrum 
around $k=0.03$, which constrains
$\Omega_m h$ by identifying the epoch of matter-radiation equality.  
The large-scale structure observations 
contain too much information 
to be summarized by a single shape parameter; no value of 
the traditional CDM shape parameter, $\Gamma$ \citep{efstathioubw92}, 
can simultaneously 
match the location of this peak and its width.

%\begin{references}
%\begin{thebibliography}{}
%\item \ldots

%\end{references}
%\end{thebibliography}

%\newpage

\cleardoublepage

\cleardoublepage

\cleardoublepage

\cleardoublepage

%\end{document}

\cleardoublepage
\chapter{Results}
\label{chap:results}

\section{Goodness of Fit}

Because we have predictions, observations, and error bars for each 
data point, it is straightforward to calculate the
$\chi^2$ value and the resulting probability of obtaining the data for 
each model.  
Only points observed at $k \leq 0.2$
are used in calculating $\chi^2$; 
our conclusions are unaffected by varying this non-linear 
cutoff between
$k=0.15$ and $k=0.25$.  On smaller scales, 
the linearization process allows us to gain
qualitative information despite the systematic uncertainties on smaller 
scales, and we show the model power spectrum as a dotted line on 
these scales to indicate that only qualitative comparison is 
possible.    
CMB predictions are calculated and compared to observations in $\ell$-space 
but using the $k$-space boxes shown would yield an identical value for 
$\chi^2$.  
Instead of normalizing to 
the COBE result alone \citep{bunnw97}, 
we give each model a free normalization parameter.  Our
rationale is that COBE is just one subset of the available 
data, albeit with small error bars, and is in fact
the data most likely to be affected by a possible contribution of gravitational
waves to microwave background anisotropies.  These gravitational
waves from inflation would have a significant
 impact only on large angular scales
and are not traced by the large-scale structure observations.  By 
normalizing to the data set as a whole, we make our results relatively 
insensitive to the possible contribution of gravitational waves.  
Because of the range
of spatial scales and amplitudes covered, our figures are log-log plots,
so the models which seem by eye to be the best fits may not in fact 
have the best linear chi-squared value.  We have tested the 
effect of asymmetry in the error bars
 by calculating $\chi^2$ on the log of the data (which greatly 
overestimates that asymmetry), and we find no significant change in 
our conclusions.  The precision of the next generation of
observations will 
require full knowledge of the covariance matrices of the observations and
the asymmetry of the error bars 
(see \citealp{bondjk98b, bartlettetal99}).

\begin{table}[htb] 
\caption{
Best-fit normalizations and biases.  The 
normalization of each model is given by $\sigma_8$ or 
the value of $dT$ at $\ell = 10$, which can be compared to 
the COBE normalization of $dT = 27.9 \mu$K.}  
\label{tab:bias}
\begin{center}
\begin{tabular}{|r|c|c|c|c|c|c|c|}
\hline
Model &  $dT_{10}$ ($\mu$K) & $\sigma_8$ & b$_{clus}$ & b$_{cfa}$ & 
   b$_{lcrs}$ & b$_{apm}$ & b$_{iras}$ \\
\hline
SCDM              & 25.4 & 1.08 & 2.12 & 0.83 & 0.72 & 0.89 & 0.57 \\

TCDM              & 31.2 & 0.79 & 2.73 & 1.13 & 1.01 & 1.18 & 0.83 \\

CHDM	          & 27.1 & 0.75 & 2.52 & 1.11 & 1.01 & 1.13 & 0.78 \\

OCDM              & 29.0 & 0.77 & 2.67 & 1.25 & 1.11 & 1.10 & 0.93 \\

$\Lambda$CDM      & 26.8 & 1.00 & 2.14 & 0.91 & 0.82 & 0.87 & 0.68 \\

$\phi$CDM         & 27.6 & 0.74 & 3.12 & 1.35 & 1.20 & 1.31 & 0.98 \\ 

BCDM              & 24.8 & 1.76 & 1.30 & 0.48 & 0.40 & 0.41 & 0.37 \\

ICDM              & 28.2 & 0.83 & 2.95 & 1.25 & 1.12 & 1.02 & 0.97 \\

PBH BDM           & 29.9 & 0.78 & 2.74 & 1.21 & 1.09 & 1.10 & 0.92 \\

Strings+$\Lambda$ & 21.2 & 0.32 & 6.95 & 3.10 & 2.86 & 2.62 & 2.48 \\
\hline
\end{tabular}
\end{center}
\end{table}

\section{Evaluating the Models}

\begin{figure}

\centerline{\psfig{file=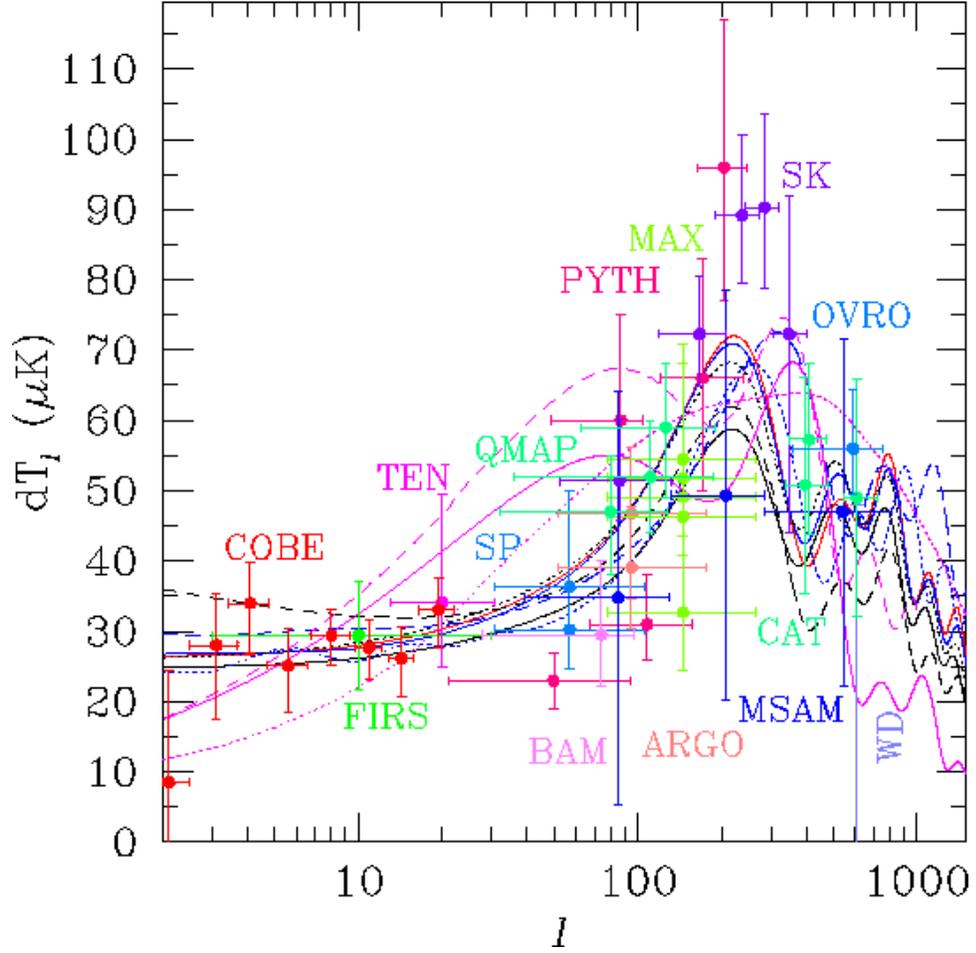,width=5in}}
%was 6in wide but too tall then - did this fix it?  
\caption{Compilation 
of CMB anisotropy results compared to model $C_\ell$.}
\mycaption{
Vertical error bars represent 
$1\sigma$ uncertainties and horizontal error bars show the width 
at half maximum of 
each instrument's window function.  
Model predictions 
are plotted as $\Delta T_\ell 
= (\ell(\ell+1)C_\ell/2 \pi)^{1/2} T_{CMB}$.  
Models and curves are SCDM (solid black), TCDM (dashed black), CHDM (solid
red), OCDM (dashed blue), $\Lambda$CDM (solid blue), $\phi$CDM (dotted 
black), BCDM (dotted blue), ICDM (dashed magenta), PBH BDM (solid 
magenta), and Strings + $\Lambda$ (dotted magenta).  }
\label{fig:models_cmb}
\end{figure}

Figure \ref{fig:models_cmb} contains nearly 
all of these detections and
the theoretical curves of our models, with their best-fit 
normalizations (listed in Table \ref{tab:bias}).  
All of the theory curves are consistent 
with the upper limits, which we disregard hereafter.  
This figure 
shows that current CMB anisotropy detections by themselves
cannot constrain models well, although the shape of the adiabatic 
radiation power spectra is preferred.  Each of our models has a distinct
curve; the first three acoustic peaks will be sufficient to distinguish
them at the precision of forthcoming satellite observations.

\begin{figure}
\centerline{\psfig{file=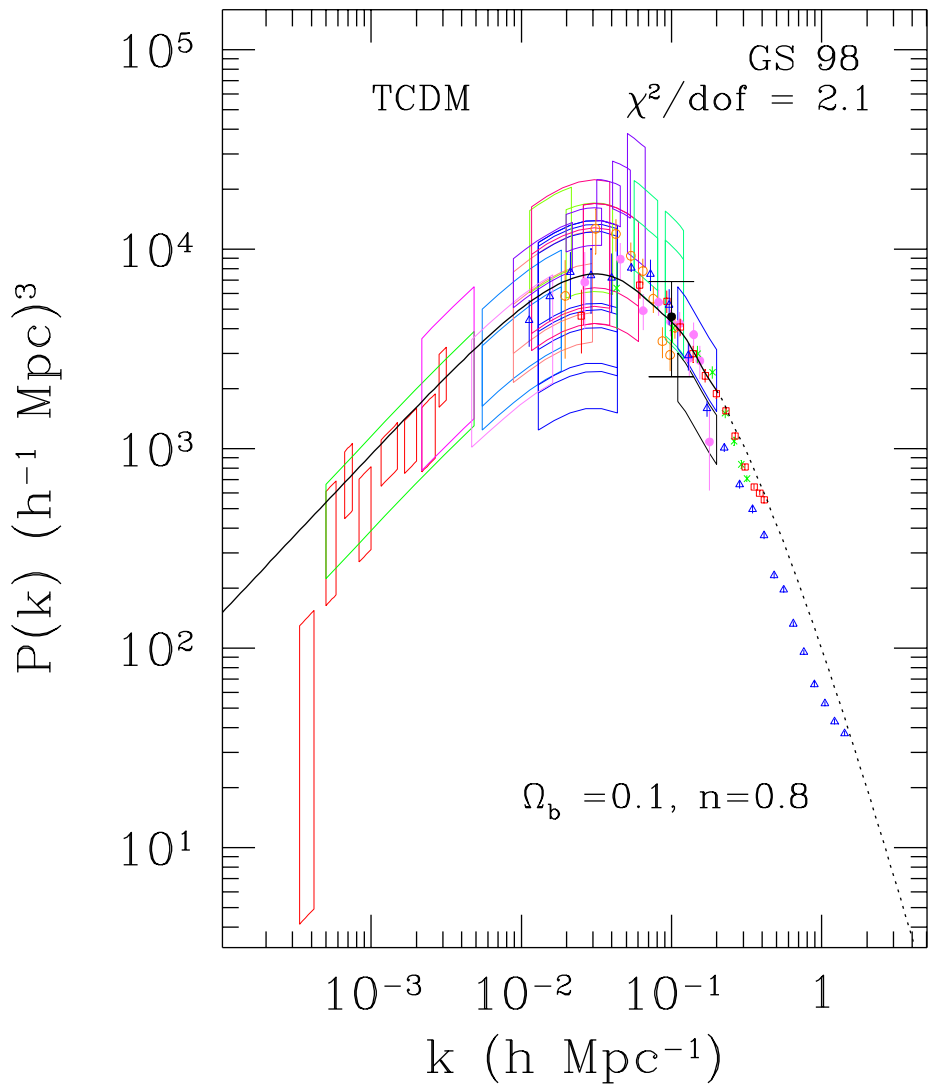,width=6in}}
\caption{The Tilted CDM model compared with CMB and LSS data.}
\label{fig:tcdm}
\end{figure}

\begin{figure}[htb]
\centerline{\psfig{file=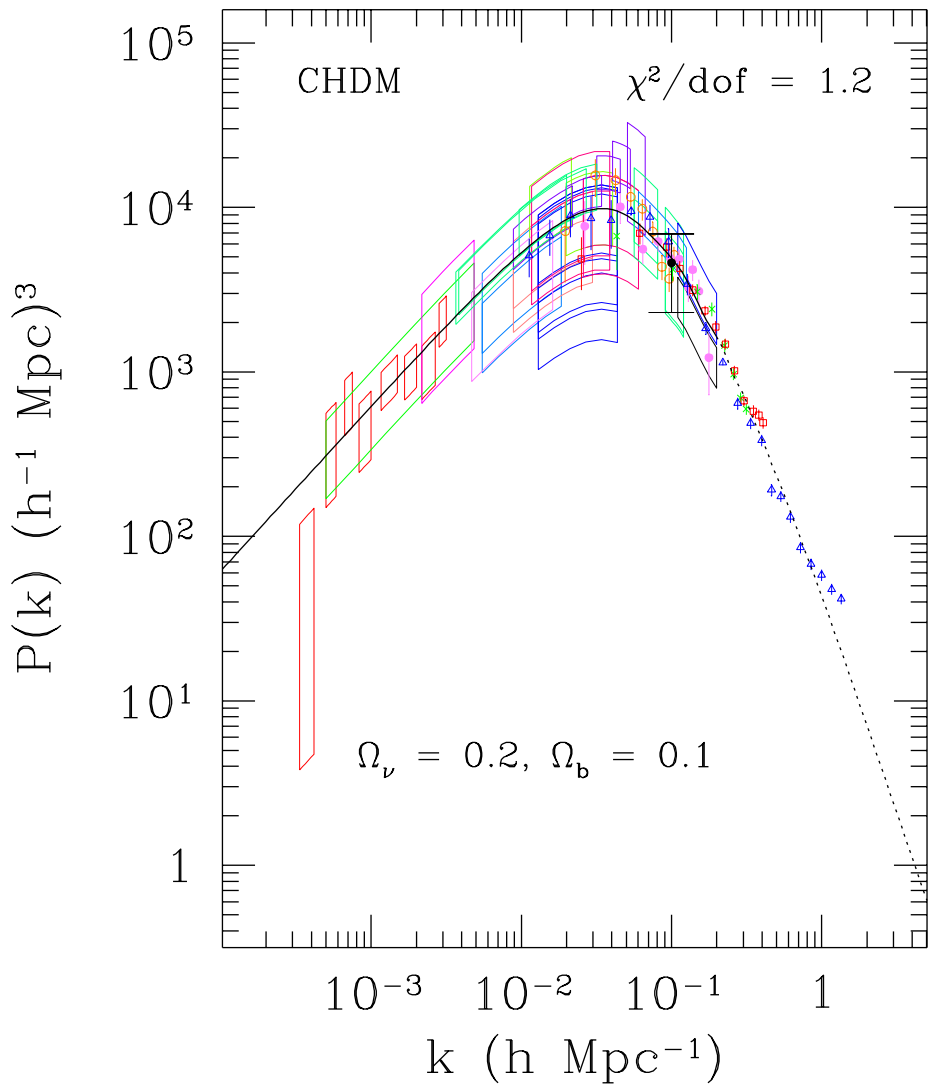,width=6in}}
\caption{
CHDM, our best-fit model.  Note agreement
even on non-linear scales. }
\label{fig:chdm}
\end{figure}

\begin{figure}[htb]
\centerline{\psfig{file=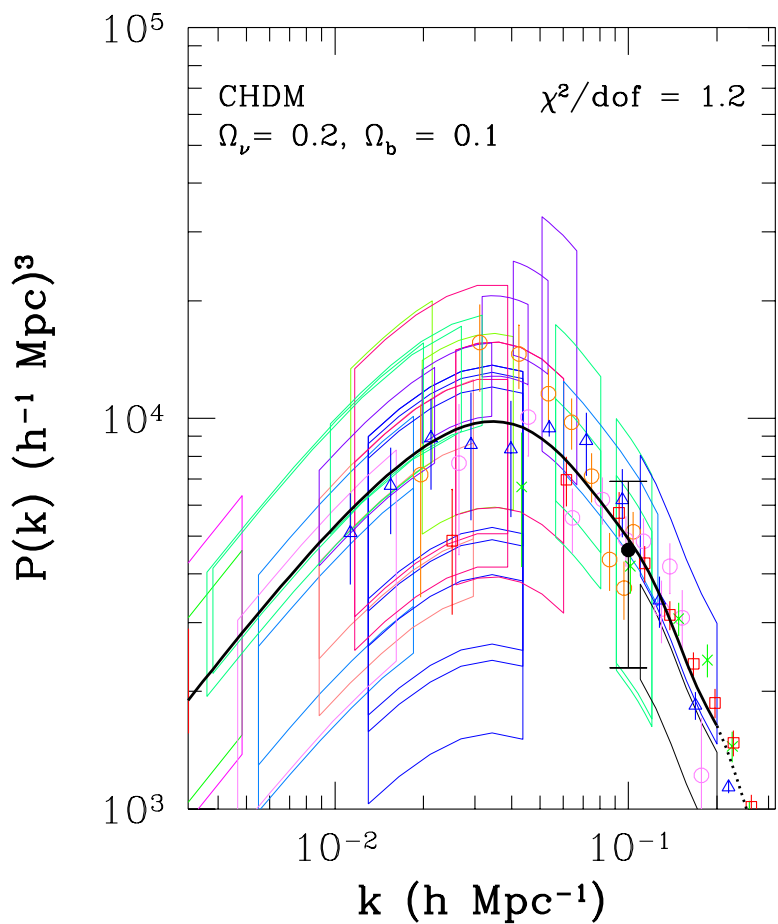,width=5in}}
\caption{Constraints from LSS and CMB on CHDM model expanded in 
region of highest precision LSS data near $k=0.1$.  The location 
and shape of the peak in the matter power spectrum and the steepness 
of its falloff towards higher $k$ are the features that make this 
model by far the best fit to the data of those we have tested.} 
\label{fig:chdm_zoom}
\end{figure}

\begin{figure}[htb]
\centerline{\psfig{file=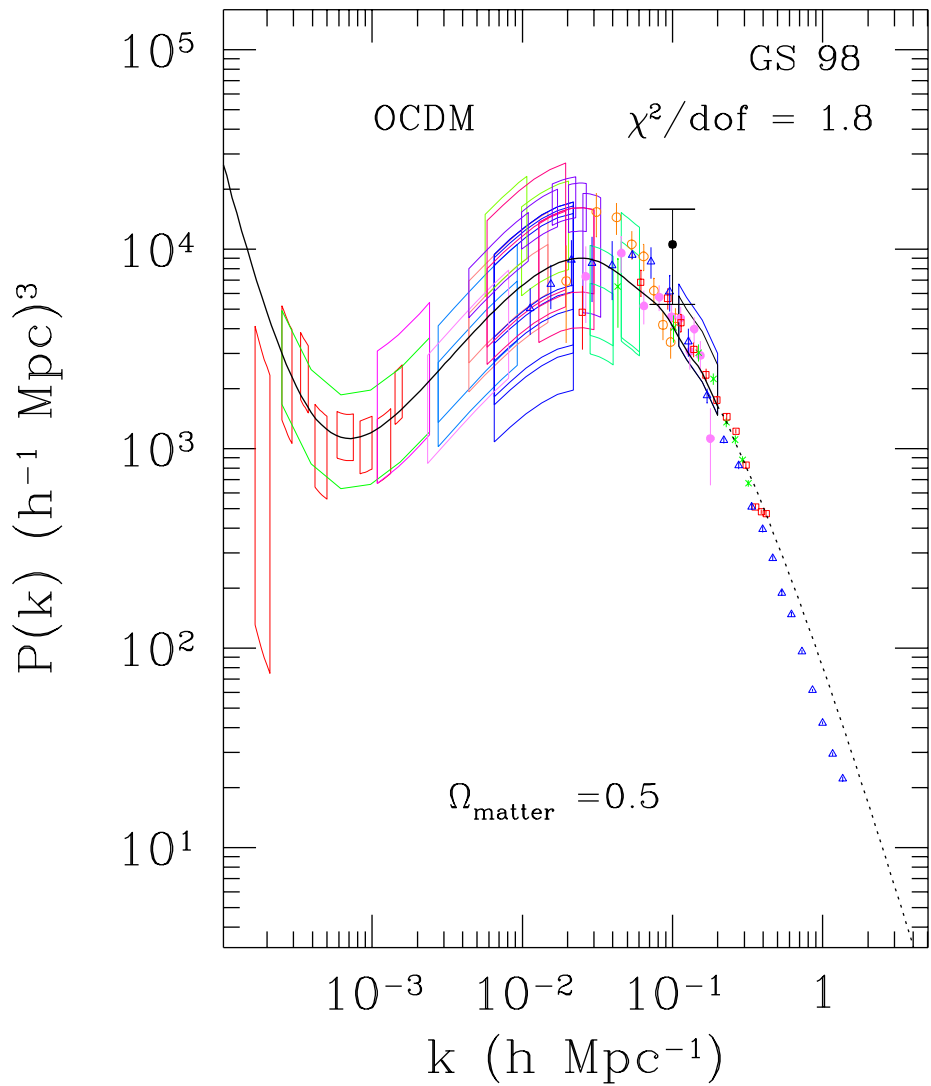,width=6in}}
\caption{Constraints from LSS and CMB on the Open CDM model, 
with scale-invariance of potential perturbations causing
an increase in the matter power spectrum beyond the curvature scale.  
} 
\label{fig:ocdm}
\end{figure}

\begin{figure}[htb]
\centerline{\psfig{file=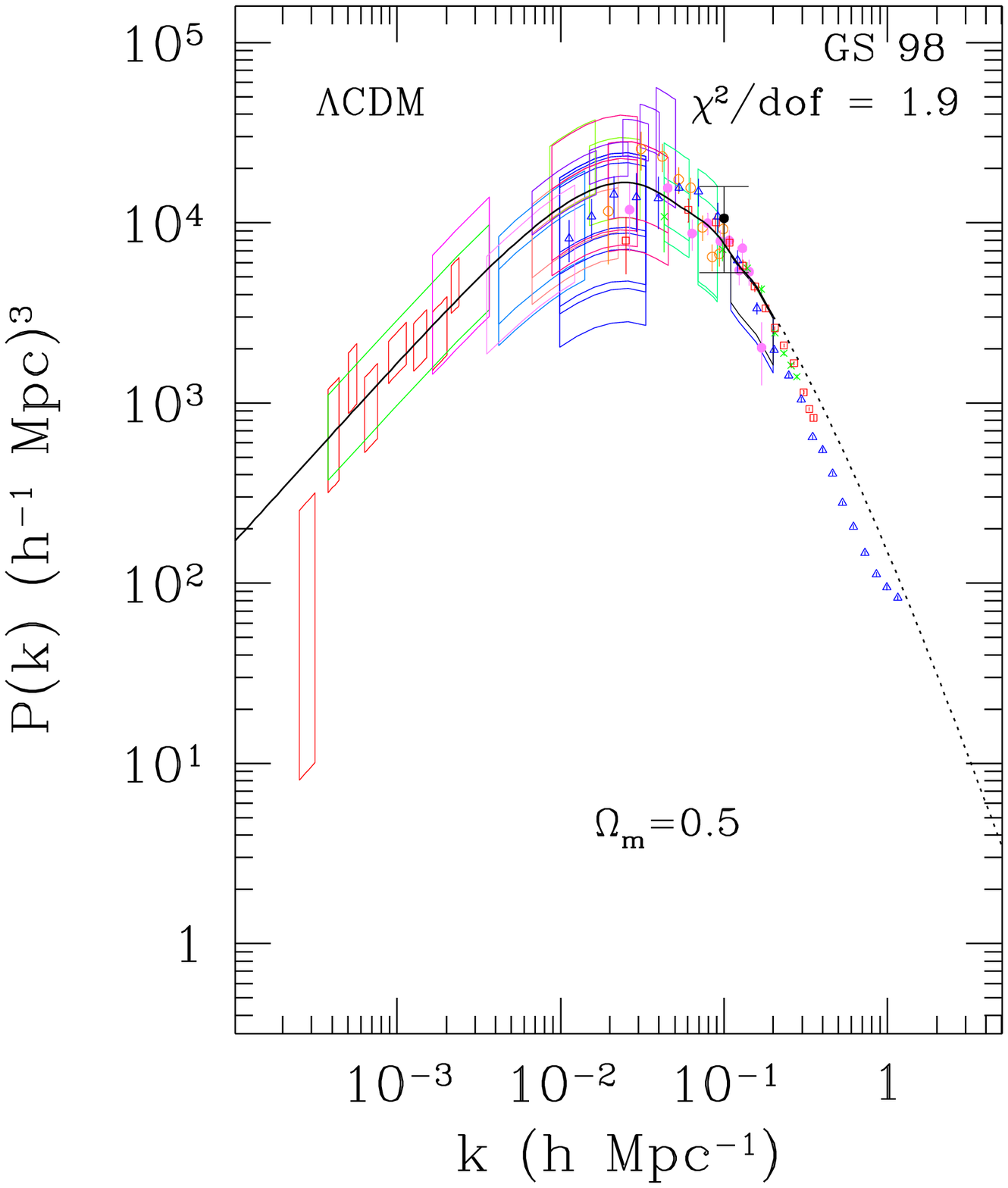,width=6in}}
\caption{
Constraints from LSS and CMB on $\Lambda$CDM model.} 
\label{fig:lcdm}
\end{figure}

\begin{figure}[htb]
\centerline{\psfig{file=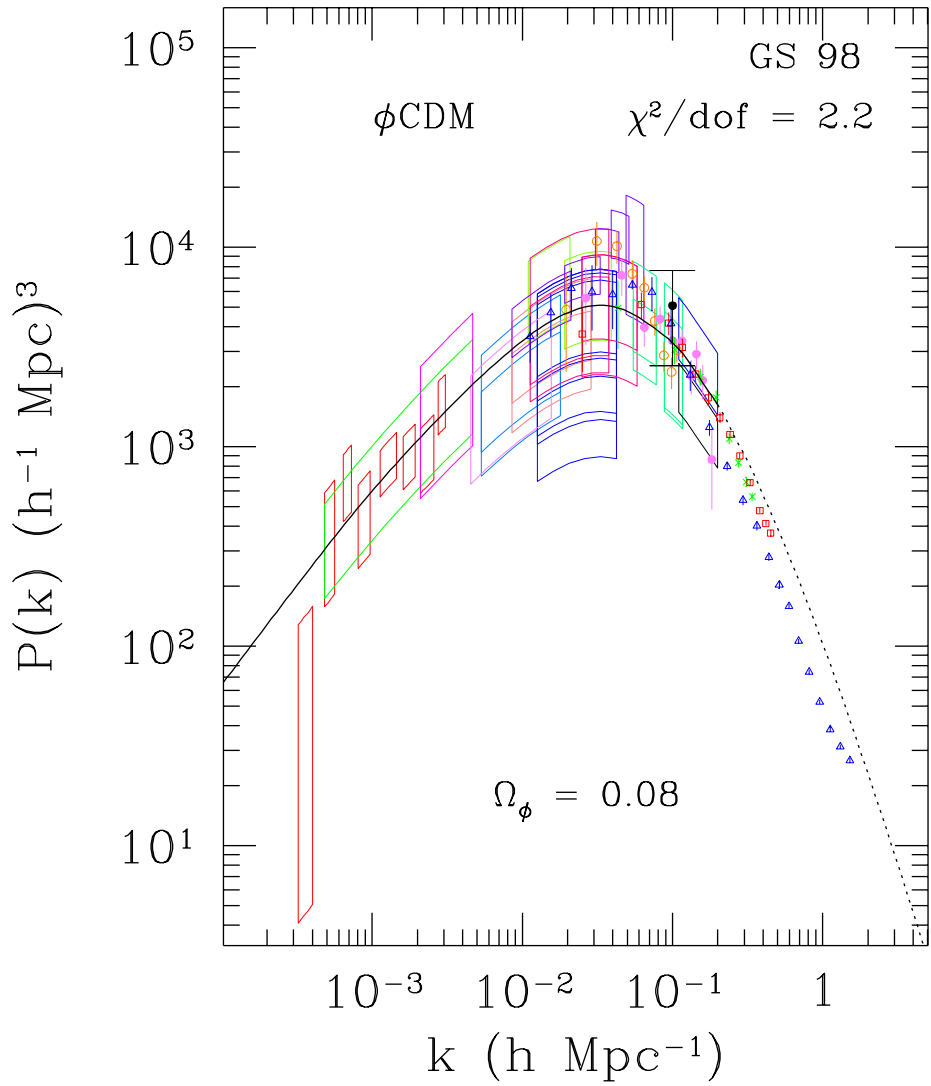,width=6in}}
\caption{Constraints from LSS and CMB on $\phi$CDM model.} 
\label{fig:pcdm}
\end{figure}

\begin{figure}[htb]
\centerline{\psfig{file=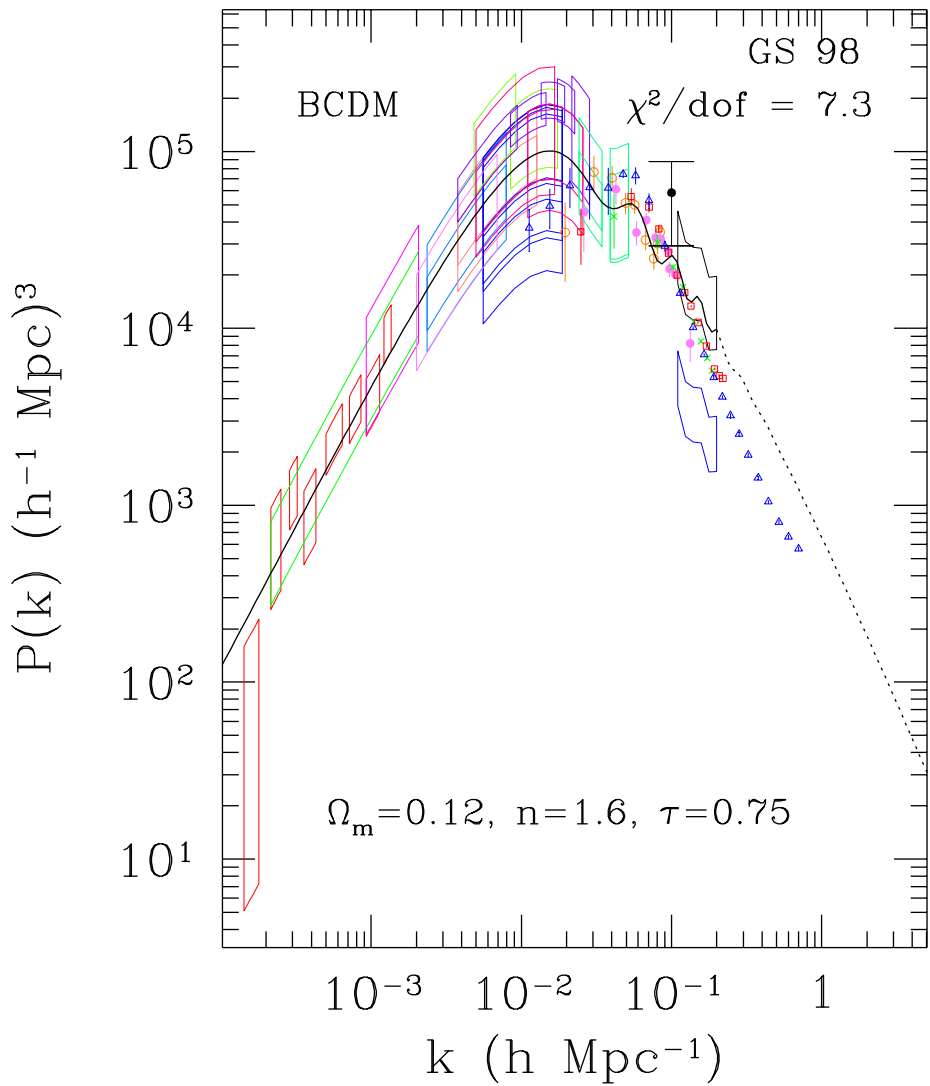,width=6in}}
\caption{The BCDM model.  Note the 
poor agreement at the main peak of the power spectrum.}
\label{fig:bcdm}
\end{figure}

\begin{figure}[htb]
\centerline{\psfig{file=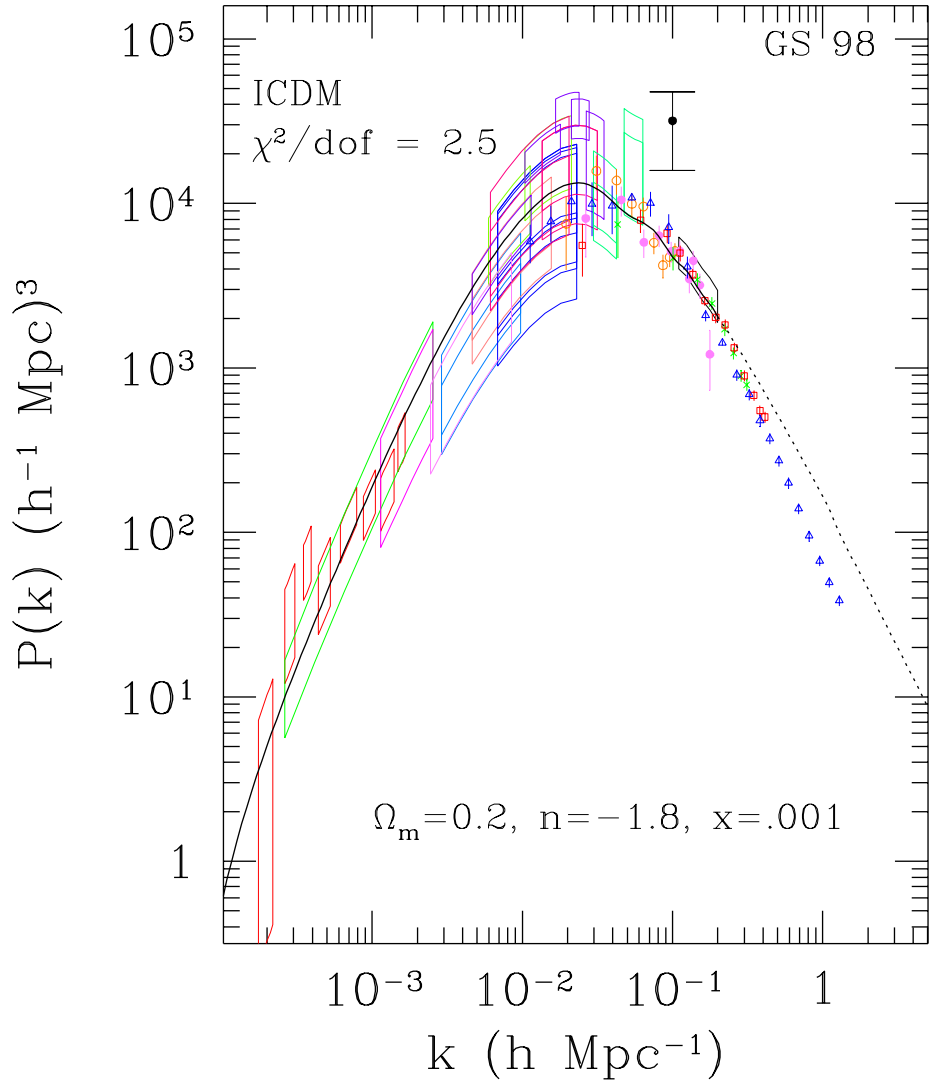,width=6in}}
\caption{Constraints from LSS and CMB on Isocurvature CDM model.} 
\label{fig:icdm}
\end{figure}

\begin{figure}[htb]
\centerline{\psfig{file=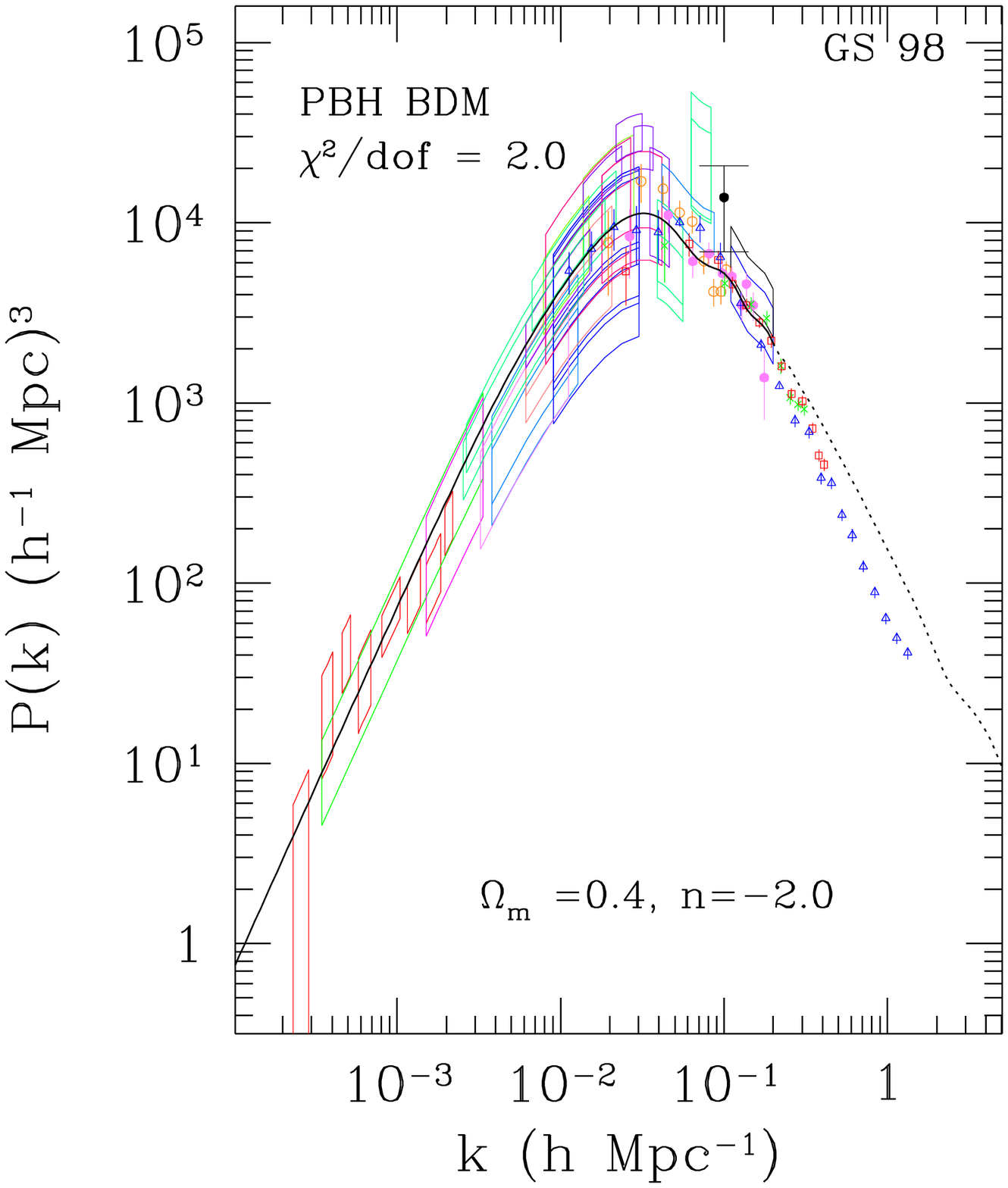,width=6in}}
\caption{Constraints from LSS and CMB on PBH BDM model.} 
\label{fig:pbhbdm}
\end{figure}

\begin{figure}[htb]
\centerline{\psfig{file=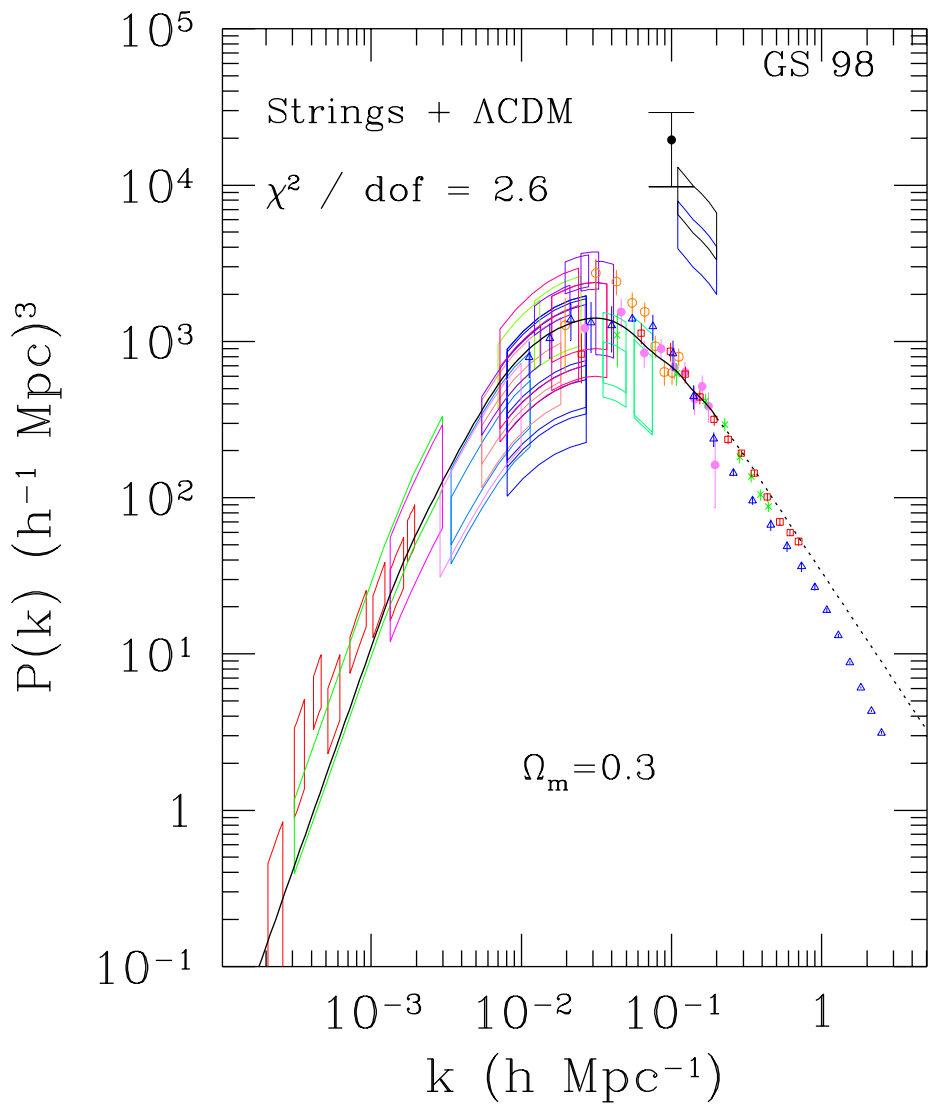,width=6in}}
\caption{
Strings + $\Lambda$CDM.  Notice the deficit of power versus the 
bias-independent observations at $k=0.1$.}
\label{fig:lstrings}
\end{figure}

Figures \ref{fig:tcdm} through \ref{fig:lstrings} 
show the large-scale structure 
data after model-dependent corrections
 compared with the CMB 
anisotropy observations and the theoretical matter power spectra $P(k)$ of 
each model.\footnote{
Full-size color figures are also available at
http://cfpa.berkeley.edu/cmbserve/fluctuations/figures.html}  
  The $\chi^2$ values for each model versus subsets of 
the data compilation are given in Table \ref{tab:chisq}.  
The present large-scale structure data have more power to discrimate 
among models of structure formation than 
do the present CMB anisotropy detections.

\begin{table}[htb] 
\caption{Chi-squared values for our models.}  
\mycaption{Results are based on 
data at $k \leq 0.2h$Mpc$^{-1}$.    
The $\chi^2_{\sigma_8}$ category includes the contribution from peculiar
velocity measurements.  The degree of freedom used by normalizing is
counted under $\chi^2_{CMB}$, and each galaxy survey loses one 
degree of freedom in choosing a best-fit bias.  The ICDM model has one less
degree of freedom in the $\chi^2_{\sigma_8}$ column and a total of 69.
P is the probability of getting $\chi^2$ greater than or equal to the 
observed value given that a model is correct.}
\label{tab:chisq}
\begin{center}
\begin{tabular}{|r|r|r|r|r|r|r|r|r|r|r|}
\hline
Model & $\chi^2_{CMB}$  & $\chi^2_{\sigma_8}$ & $\chi^2_{clus}$ & 
$\chi^2_{cfa}$& $\chi^2_{lcrs}$ & $\chi^2_{apm}$ & $\chi^2_{iras}$ & 
$\chi^2_{total}$ & $\chi^2~~~~~~~~~~$ & P~~~~~\\

 d.o.f.  & 34 & 3 & 8 & 2 & 5 & 9 & 9 & 70 &/d.o.f.& \\
\hline
SCDM             & 46 & 36 & 37 & 0.2 & 8 & 121 & 18 & 266 &3.8 & $<10^{-7}$ \\

TCDM             & 51 & 5 & 27 & 0.4 & 6 & 49 & 11  &148  & 2.1 
&$1.8\times 10^{-7}$ \\

CHDM	         & 30 & 4 & 20 & 3 & 9 & 10 & 11 &  86 & 1.2 & 0.09 \\

OCDM             & 36 & 2 & 24 & 2 & 11 & 42 & 12 & 128  & 1.8 
&$2.9\times 10^{-5}$ \\

$\Lambda$CDM    & 30 & 3 & 26 & 2 & 12 & 46 & 13 & 132  & 1.9 
&$1.1 \times 10^{-5}$ \\

$\phi$CDM        & 32 & 4 & 30 & 0.1 & 5 & 71 & 12 & 155  & 2.2 &$<10^{-7}$ \\ 

BCDM             & 32 & 38 & 33 & 1 & 125 & 225 & 56 & 511 & 7.3 &$<10^{-7}$ \\

ICDM             & 61 & 3 & 17 & 2 & 21 & 50 & 16 & 170 & 2.5 & $<10^{-7}$ \\

PBH BDM          & 65 & 4 & 22 & 2 & 9 & 30 & 11 & 142  & 2.0 
&$8.3 \times 10^{-7}$ \\

Strings+$\Lambda$ &64 & 37 & 20 & 0.3 & 8 & 43 & 10 &182 & 2.6 &$<10^{-7}$ \\
\hline
\end{tabular}
\end{center}
\end{table}

We find a very poor fit for SCDM (shown in Figure \ref{fig:scdm}), 
which is unsurprising
given the difference in shape between the theory curve and 
the data and the disagreement with the bias-independent measurements at 
$k=0.1$.  The fit to the CMB is poor, because SK would prefer
more power and the best-fit normalization is only 0.82 that of COBE. 
Figure 
\ref{fig:tcdm} shows the TCDM model. The fit is drastically improved versus
SCDM, although the peak of the matter power spectrum 
is still broader than that found in the data.  
The fit with the CMB is harmed by the high normalization versus COBE 
and by the tilt on medium scales,
but reaching $\sigma_8$ without missing the peak of the matter
power spectrum
makes this the optimal amount of tilt for an $\Omega_m=1$ CDM model.   

The best-fit model, Cold + Hot Dark Matter, is shown in 
Figure \ref{fig:chdm} and in expanded detail in Figure \ref{fig:chdm_zoom}.
  The agreement
with the location and shape of the peak of the matter power spectrum
is remarkable, with the exception of the APM cluster power spectrum, 
which seems to have a narrower peak than the other surveys.  The agreement
with CMB is excellent as well, 
although several other models do equally well.
  As noted by \citet{peacock97} 
and \citet{smithetal98}, 
the theoretical curve
for CHDM matches the APM galaxy power spectrum down to scales well into
the non-linear regime, making this model a good explanation of 
structure formation far beyond the scales used for our statistical
analysis.  Small variations of the parameters of the model do 
not improve the fit.  Observations of Damped Lyman $\alpha$ absorption 
systems at high redshift, which probe linear power on presently 
non-linear scales and are sensitive to the growth function of a particular 
cosmology, however, are much less favorable for CHDM.  

The OCDM model is shown in Figure 
\ref{fig:ocdm}.  
The high value of $\Omega_m$ is
favored by the location of the peak in the matter power spectrum and
the constraints of the SK and CAT CMB anisotropy detections.
However, the shape
is still not right for the peak.  
The shape of the APM data is in good agreement with the model 
on non-linear scales although the data falls consistently below
the prediction.  $\Omega_m=0.5$ generates excellent
agreement between the two observations of $\sigma_8$.  
This model is our second best fit, but it is 
statistically much worse than CHDM.

The $\Lambda$CDM model of 
Figure
\ref{fig:lcdm} is nearly as successful as OCDM.  
It is a slightly better fit to the CMB but is 
worse in comparison to large-scale structure. 
The observations of
$\sigma_8$ are again in good agreement, but the location of the peak
in the matter power spectrum appears wrong, and the shape does not 
compare well with that of the APM galaxy survey.  

Figure 
\ref{fig:pcdm} shows the $\phi$CDM model, which is too broad at the peak 
and misses a number of APM datapoints.  Its agreement with the other
datasets is rather good; excluding the APM galaxy survey would make
 $\phi$CDM a much better fit.  It remains to be seen whether other 
variations of scalar field models 
can match the observations better (see 
\citealp{caldwellds98, chibasn98, vianal98, huetal99}).

By far the worst disagreement with the data is seen in 
Figure \ref{fig:bcdm} for 
BCDM.  Choosing parameters to place an acoustic oscillation peak 
near $k=0.05$ has clearly generated the wrong shape, even
though the APM galaxies and clusters seem to fit the first and second
oscillations, respectively.  We average the 
predictions of the matter power spectrum over the rough 
window function of the 
observations to take into account the possible smoothing of 
these oscillations during observation.  
In order to make the linearization procedure work smoothly, we fixed
the local slope of the linear power spectrum; otherwise the oscillating
slope produces a mess.  
It is not clear whether this is a 
flaw in the procedure, which was only tested on convex power spectra,
or whether indeed the non-linear evolution predictions of such a 
model are in clear disagreement with the data.  This 
issue has been investigated by \citet{meiksinwp98}.  
The critical problem with the model, however, is that its main 
peak is in the wrong
place; no model with similar oscillations and a baryon content consistent
with Big Bang Nucleosynthesis can fix that problem 
\citep{eisensteinetal98}. 
In general, the data are smooth enough to 
set a limit on the baryon fraction $\Omega_b/\Omega_m$; 
when that fraction gets higher
than about 0.1 the fit worsens.  \citet{goldbergs98} discuss 
the future prospects of this constraint.

In the Isocurvature CDM model of Figure 
\ref{fig:icdm}, the 
strong
rise of the matter power spectrum is caused by the sharp tilt 
($n=2.2$) of the model
away from scale-invariance.  The fit to the CMB is poor, due 
to the rise of $C_\ell$ on COBE scales and too little power compared to 
Saskatoon.  The fit to large-scale structure is mediocre as well, although
the narrow peak agrees with the APM Clusters $P(k)$ as well as any model.  
The linearization procedure seems to work fine; although it was calibrated
for Gaussian models, the linearization is expected to be rather
similar under the $\chi^2$ distribution \citep{stirling98}.

Figure 
\ref{fig:pbhbdm} shows that the PBH BDM model has similar problems to ICDM,
although the placement of the peak is a bit improved.  The more prominent
acoustic oscillation in the matter power spectrum of this model is placed
well compared to the data but fails to produce a good fit or break
the linearization procedure.  With its best-fit normalization, this model 
seriously underpredicts $\sigma_8$.  Overall, this model agrees with 
the data almost as well as OCDM and $\Lambda$CDM despite the 
difficulty of reconciling isocurvature perturbations with current CMB 
data.  This success of isocurvature versus the large-scale structure data, 
however, implies that it is premature to restrict our consideration to 
adiabatic models at present.  

The Strings + $\Lambda$CDM model (\ref{fig:lstrings}) 
requires a very large bias
for all types of galaxies; it is difficult to explain this bias.  
Correspondingly, the amplitude of bias-independent
measurements at $k=0.1$ is sorely underestimated, and the fit is not 
very good to the CMB or the APM galaxy survey.  However, the model
 has roughly the right
shape for the matter power spectrum.  It is possible that the linearization
procedure needs to be adjusted to account for the level of non-Gaussianity
in the matter distribution, but that only affects the smallest scales 
considered in our statistical analysis, and the reduced power of this
model weakens the effects of non-linear evolution.  Overall, this 
model agrees with the data almost as well as OCDM and $\Lambda$CDM; 
if an explanation can be found for the large galaxy bias needed, topological
defects may still be a viable paradigm for structure formation.

\section{Discussion}

Perhaps the most impressive result 
is the rough agreement of the two datasets over
a wide range of models, giving strong 
evidence that the gravitational instability paradigm of cosmological 
structure 
formation is 
alive and well.  Models in which the density perturbations arise
from inflationary adiabatic, isocurvature, and topological defect initial
conditions all succeed in meeting the qualitative standard of rough
agreement with the data.  The current set of CMB anisotropy detections may be 
a poor discriminator among adiabatic models, but it prefers
them to non-adiabatic models.  

Several models (SCDM, TCDM, BCDM, ICDM, PBH BDM, and Strings+$\Lambda$) 
have a best-fit 
normalization significantly different from normalizing to COBE and would
have been unfairly penalized if forced to that normalization.  The Strings
model already includes a tensor contribution, but SCDM, BCDM, and PBH 
would benefit from adding a gravitational wave component that brought
them into better agreement with COBE without changing the amplitude of
their smaller-scale 
scalar perturbations.  Adding gravity waves is not, however, a panacea
for any of those models.  In general, the models which are the best fits
to the shape of the matter power spectrum have chosen to be close to their
COBE normalization, which argues against there being a significant
tensor contribution to large-angle CMB anisotropies.  

The average ratio of best-fit biases in Table 4.1 is 
$b_{clus}:b_{cfa}:b_{lcrs}:b_{apm}:b_{iras} = 
3.4:1.4:1.2:1.1:1$; this is roughly 
consistent with the bias ratios found by \citet{peacockd94}.  
The CfA galaxies are more strongly clustered than 
other optically-selected galaxies, which   
could result from residual 
luminosity bias in the 101$h^{-1}$Mpc sample.  
Most models 
allow optical galaxies to be almost unbiased tracers of the dark 
matter distribution.

By restricting our analysis to the linear regime 
and carefully correcting for the minor effects of scale-dependent 
redshift distortions and non-linear evolution on those scales, we have
made it possible to test models quantitatively.  Because of the history
of systematic errors in observational cosmology, our conclusions must 
in the end offer some qualitative interpretation.  If all these
models are a priori equally likely, the most likely cosmology by a 
tremendous factor is 
Cold + Hot Dark Matter, which is the only model consistent with 
the observations at the 95\% 
confidence level.    
The disagreement between the data and the other models is 
sufficient to rule out all of them 
unless there
are severe unanticipated systematic problems in the data.  Undoubtedly,
there are some small systematic errors; in trusting Gaussian 
statistics we are essentially relying on a central limit theorem
for these systematics.  
It is difficult 
to argue that the error bars have been severely underestimated; expanding
them enough to bring OCDM into concordance at 5\% confidence 
would make the data a better fit to CHDM than is expected 80\% of the time.
All models except CHDM are ruled out at above 99\% confidence by our 
comparison, and the agreement of the SCDM variants cannot be improved 
by simple variations of their parameters.  

As it stands, CHDM itself
is not statistically very likely, 
but this is only because of the APM cluster
survey, which no model fits much better, and which seems to disagree
somewhat with the shape of the galaxy surveys.  Dropping the APM 
cluster $P(k)$ would give CHDM a $\chi^2$/d.o.f. 
of 1.06, which is within the 68\% 
confidence level for the resulting 62 degrees of freedom.  We cannot
justify doing that, but it is worth investigating whether
the APM Cluster survey contains a scale-dependent bias not accounted 
for by the non-linear evolution formula or if its errors have 
somehow been underestimated.  
The two-dimensional APM galaxy survey
is by far the strongest discriminator between models; removing it would
allow $\phi$CDM, OCDM, $\Lambda$CDM, and TCDM to be much better fits,
although of those models 
only $\phi$CDM would be allowed at 99\% confidence.

Future observations may well 
show that a variation upon CHDM or a completely different model
that imitates its predictions for structure formation is correct, 
but we find that the amplitude and shape of its spectrum of primordial
density fluctuations 
agree well with the data.  
This does
not provide direct evidence for the existence of 
hot dark matter, which requires experimental confirmation of neutrino
mass (this issue will be discussed in more detail in 
the next chapter).  
The CHDM model has other observational hurdles to 
overcome, including indications of early galaxy formation for 
which this model has too little power on correspondingly small scales, 
although it is impressive that CHDM agrees with the linearized APM 
data so well out to $k=1$. 

If the current 
results from Type Ia supernovae observations hold up there will 
be enough statistical power in the direct observations of 
cosmological parameters to make OCDM and $\Lambda$CDM 
preferred to CHDM, although in that case 
no model would be a satisfactory fit to both the supernovae and 
structure formation observations.  The next two 
chapters explore possible resolutions of 
this conflict.

%\newpage

\cleardoublepage

\cleardoublepage

\cleardoublepage

\cleardoublepage

\cleardoublepage

\cleardoublepage

\cleardoublepage

\cleardoublepage

\cleardoublepage

\cleardoublepage

\cleardoublepage

\cleardoublepage
\chapter{Cosmological Limits on the Neutrino Mass}
\label{chap:nu}

% add refs to Tritium and beta decay for direct mass limits
% explain that neutrinos heavier than cosmologically allowed range 
% are too heavy to be HDM
%Also add delta m^2 refs 
% mention lyman alpha forest, DLAS, sigma_8 constraints
% ~30eV is cosmological bound for pure HDM closing universe

\section{Is Cold+Hot Dark Matter Compatible with a Cosmological Constant?}

\begin{figure}[htb]
%\figurenum{1}
%\epsffile{lchdm00_ebb.ps}
\centerline{\psfig{file=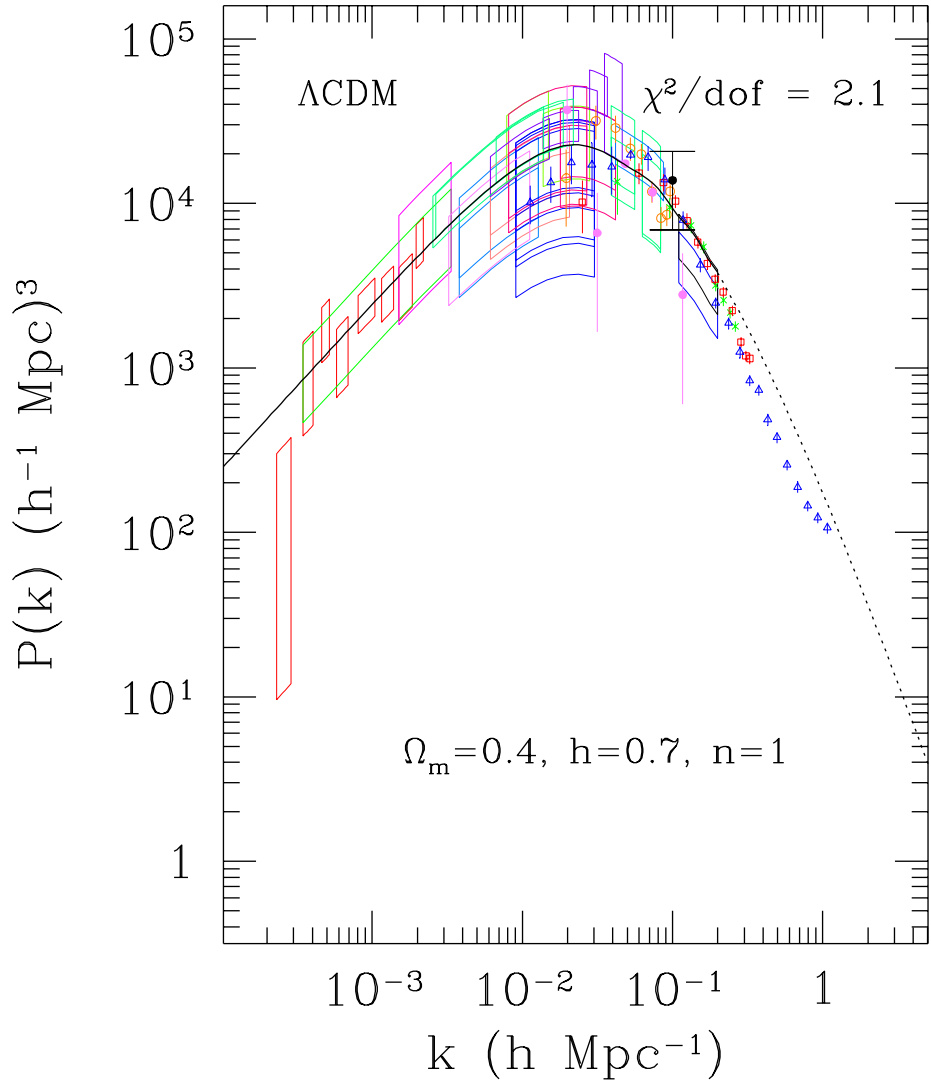,width=6in}}
\caption{Constraints from LSS and CMB on $\Lambda$CDM model.} 
\label{fig:lchdm00}
\end{figure}

The conflict between the $\Lambda$CDM model favored by direct 
observations of cosmological parameters (especially 
the recent Type Ia supernovae results) and the current set of 
structure formation observations that we have found motivates us to 
explore whether there is a simple way to reconcile $\Lambda$CDM with 
those observations.  
We start with a version of $\Lambda$CDM which is in 
good agreement with all direct parameter observations, with $\Omega_m=0.4$, 
$\Omega_b=0.04$, and $h=0.7$.   
Figure \ref{fig:lchdm00} 
shows the $\Lambda$CDM power spectrum compared 
with observations of Large-Scale Structure and CMB anisotropy.
The problem of the peak location and shape of the matter power spectrum 
versus the data is the same as that identified in the preceding chapter;
the era of matter-radiation equality occurs later in this $\Lambda$CDM 
model than the data appears to prefer, and improving this requires increasing
$\Omega_m$ which is the opposite of what direct parameter observations 
recommend.

\begin{figure}[htb]
%\figurenum{2}
\centerline{\psfig{file=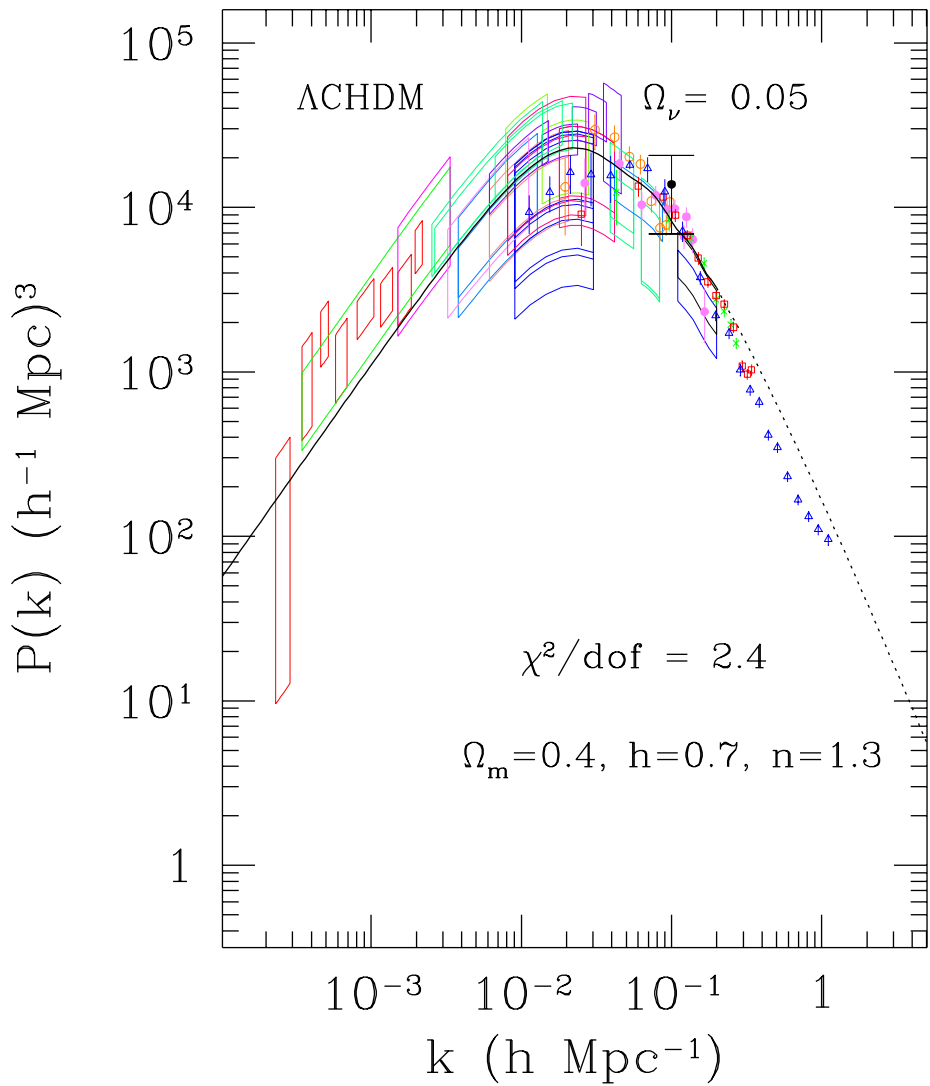,width=6in}}
\caption{$\Lambda$CDM model with $\Omega_\nu=0.05$.}
\label{fig:lchdm05}
\end{figure}

\begin{figure}[htb]
%\figurenum{3}
\centerline{\psfig{file=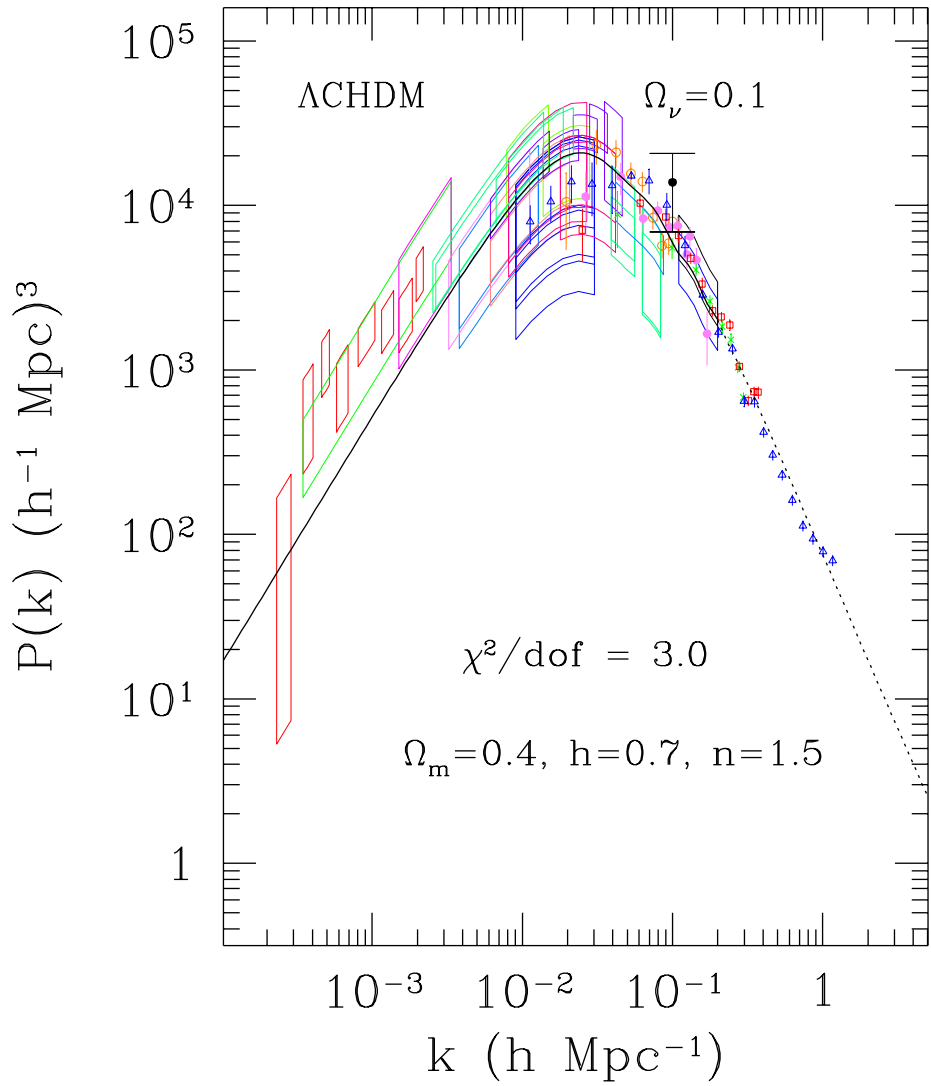,width=6in}}
\caption{$\Lambda$CDM model with $\Omega_\nu=0.10$ and $n=1.5$.} 
\label{fig:lchdm10}
\end{figure}

One can pose the following question: does adding a hot component improve the  
marginally acceptable LSS fit?  
We find that as 
HDM is added, the combined fit to CMB and LSS deteriorates
(see Figures \ref{fig:lchdm05} and 
\ref{fig:lchdm10}).  
This occurs because adding HDM reduces the power on physical 
scales shorter than the neutrino free-streaming length, 
which 
exacerbates
the mismatch in peak locations that occurs for pure CDM low-density models.  
A blue tilt of the 
primordial power spectrum ($n=1.3$) 
is necessary to counteract the damping 
of small-scale perturbations by free-streaming of the 
massive neutrinos, which makes the peak of the model fall 
even farther below that of the data unless $n>1$.
Even with this best-fit value of $n$, the 
fit to the data is worse than with no HDM, because 
CMB observations disfavor such a high value of $n$.  
For a higher HDM fraction, an even higher value of $n$ is required 
($n=1.5$ for the $\Omega_\nu=0.10$ model of Figure \ref{fig:lchdm10}), 
leading to an even worse fit to the data.  
$\Lambda$CHDM has also been explored by 
\citet{valdarninikn98} and \citet{primackg98} with different analysis 
methods and significantly smaller data compilations.

\section{Limits on the Neutrino Mass}

Neutrino masses imprint a distinct signature on $P(k)$, 
the reduction in power on scales larger 
than the free-streaming length of 
41 Mpc ($m_\nu$/30eV)$^{-1}$ \citep{bondes80}.  
Present cosmological bounds on the mass of a light neutrino are 
stricter than those from laboratory experiments; a 30eV neutrino 
would lead to $\Omega_\nu=1$, so for a universe at less than 
critical density the neutrinos must all be lighter than this.  
The exception to this is if the neutrino is so massive 
that it was non-relativistic during matter-radiation equality, i.e. 
Cold Dark Matter.   
For massive CDM neutrinos, the abundance drops enough to no longer overclose 
the universe; however, 
laboratory limits rule out the possibility of an electron 
neutrino more massive than 15eV \citep{primackg98}.   
The shape of the radiation power spectrum of CMB anisotropies 
\citep{dodelsongs96} 
and of 
the matter power spectrum from large-scale structure are both sensitive 
to the mass of neutrinos, and 
the LSS probe may potentially
be more powerful \citep{huet98}.
 There is more dynamical
range available in probing $P(k)$ with 
LSS on the neutrino free streaming scale, 
where the primary signature should be present.  
Even current data is sensitive to a neutrino 
mass of around an eV; 
we find that the fit changes significantly between 0.1 and 1 eV.  
While there are 
considerable systematic uncertainties in this approach, it is promising as a 
complement to the direct evidence for mass difference between neutrino 
species from SuperKamiokande \citep{fukudaetal98} and the solar neutrino 
problem \citep{bahcallks98}, 
and is already beginning to conflict with results from 
LSND 
\citep{athanassopoulosetal96} that require a large mass difference.

We have assumed here that  
$\Lambda$CDM is the correct model of structure formation and 
that the primordial power spectrum is well-described by a power-law. 
Our limits on the neutrino mass are based upon an attempt to search 
the reasonable parameter space around this fiducial model to 
produce the best fit possible to the data for a given neutrino 
mass.  Since disagreement with CMB data is the main problem once 
a blue tilt is considered, we have tried to alleviate this by 
adding a significant tensor component or early reionization.  Each of these 
effects reduces the small-scale CMB power relative to COBE scales, which 
helps to reconcile $n>1$ with the CMB data.  However, no parameter 
combination helps enough to make $\Lambda$CHDM a better fit than 
the fiducial $\Lambda$CDM model, and this allows us to set 
upper limits on the neutrino mass.  
For $n=1$ we find that $\Omega_\nu<0.05$ i.e. 
the mass of the most massive neutrino must be 2 eV or less (the limit is 
tighter, of course, if there are at least 2 massive neutrinos with nearly 
equal masses).  For a scale-free primordial power spectrum, there is more 
freedom to increase $n$ to counteract the effect of HDM but this leads to 
conflict with CMB anisotropy observations, limiting $\Omega_\nu<0.1$ i.e. 
the mass of the most massive neutrino must be 4 eV or less.  This is 
compatible with the recent claim by 
\citet{crofthd99} 
that the Lyman $\alpha$ 
forest power spectrum limits the neutrino mass to 3 eV or less, but our method 
appears more robust as the normalization of the Lyman $\alpha$ 
forest power spectrum is quite uncertain and it covers a more narrow 
range of scales than our large-scale structure compilation.  Future constraints
from combining CMB and large-scale structure are discussed by \citet{huet98} 
and a more speculative method using weak gravitational lensing surveys 
is presented by 
\citet{cooray99}
.  

\cleardoublepage

\cleardoublepage

\cleardoublepage

%\cleardoublepage
%\begin{figure}[htb]
%\figurenum{1}
%\epsffile{lchdm00_ebb.ps}
%\centerline{\psfig{file=tlchdm_cmb.ps,width=6in}}
%\caption{Constraints from CMB on $\Lambda$CHDM model with tensor 
%perturbations.} 
%\label{fig:tlchdm_cmb}
%\end{figure}

%\cleardoublepage
%\begin{figure}[htb]
%%\figurenum{1}
%%\epsffile{lchdm00_ebb.ps}
%\centerline{\psfig{file=n15tlchdm10.ps,width=6in}}
%\caption{Constraints from LSS and CMB on $\Lambda$CHDM model with 
%tensor perturbations.} 
%\label{fig:tlchdm10}
%\end{figure}

%%\cleardoublepage
%\begin{figure}[htb]
%%\figurenum{1}
%%\epsffile{lchdm00_ebb.ps}
%\centerline{\psfig{file=tau05n15lchdm10.ps,width=6in}}
%\caption{Constraints from LSS and CMB on reionized $\Lambda$CHDM model} 
%\label{fig:taulchdm10}
%\end{figure}

\cleardoublepage
\chapter{Reconstructing the Primordial Power Spectrum}
\label{chap:ppk}

\section{Motivation}

The initial density perturbations in the universe are believed to have
originated from quantum fluctuations during  
inflation or from active sources such as topological defects.  
  Inflationary models predict rough
scale-invariance; the shape of the inflaton potential
leads to tilting as well as  
variation of the degree of tilt with
spatial scale.
The assumption of scale-invariance \citep{harrison70, zeldovich72, peeblesy70}
 is no longer acceptable because 
the data are now accurate
enough to reveal the predicted deviations from $n=1$.

The combination
of Cosmic Microwave Background (CMB) anisotropy measurements and 
Large-Scale Structure observations
has for several years caused dissatisfaction with the standard Cold
Dark Matter (sCDM) cosmogony, leading some to advocate a 
 ``tilt'' of the primordial power
spectrum away from scale-invariant ($n=1$) 
to $n=0.8-0.9$ \citep{whiteetal95}.  
Other CDM cosmogonies have not commonly
been tilted but their agreement with the data might also improve.  
Because the 
primordial power spectrum is an inherent set 
of degrees of freedom in all CDM cosmogonies, we can 
find the best-fit primordial power spectrum for 
any cosmological model.  
We have seen that adding Hot Dark Matter 
does not resolve the disagreement of $\Lambda$CDM with 
observations of structure formation, 
but utilizing a non-scale-free primordial power 
spectrum may be a solution.

\section{Methodology}

We can adopt a parameterization of the primordial
power spectrum as a polynomial in log-log space versus wave number $k$
\citep{kosowskyt95, gawiser98, copelandgl98}:  
\begin{equation} \log P_p(k)\; =\; \log A\; +\; n\log k\; +\; \alpha (\log k)^2\; +\; \ldots ~ ~ ~ . \label{eq:pp} \end{equation}
Inflation predicts that successive terms will have rapidly diminishing 
coefficients, i.e. the deviation from a power-law should be even less 
than the deviation from scale-invariance.  This is generally required 
by the slow-roll condition, however, hybrid inflationary models and 
phase transitions during inflation can introduce greater curvature and 
even rather sharp features into the resulting power spectrum of 
density perturbations.  Hence, 
one can also consider a free-form primordial power spectrum, to be fitted 
by eye to the data, and schemes that bin the primordial power into 
``band-powers'' \citep{wangss99, souradeepetal98}.

Each cosmogony has transfer functions, $T(k)$ and $C_{\ell k}$. 
CMB anisotropies are
given by
\begin{equation} C_\ell=
%\frac{1}{8 \pi}
 \sum_k d \log k \, C_{\ell k} \,P_p(k),\label{eq:cl} \end{equation}
where C$_{lk}$ is the radiation transfer function after Bessel
transformation into $\ell$-space.  The matter power spectrum is  
\begin{equation} 
P(k) = \frac{2 \pi^2 c^3}{H_0^3}\, T^2(k)\, P_p(k). \label{eq:pk} 
\end{equation}

\noindent We use the compilations of CMB anisotropy 
and large-scale structure observations
introduced in Chapter \ref{chap:obs}.
For a given  
$P_p$(k), we compare predictions with observations using the $\chi^2$ 
statistic using the analysis method described in Chapter \ref{chap:method}.  
We can vary the coefficients of $P_p$(k) given in Equation~\ref{eq:pp} to 
find the best fit for each cosmogony.  

\begin{figure}[htb]
\centerline{\psfig{file=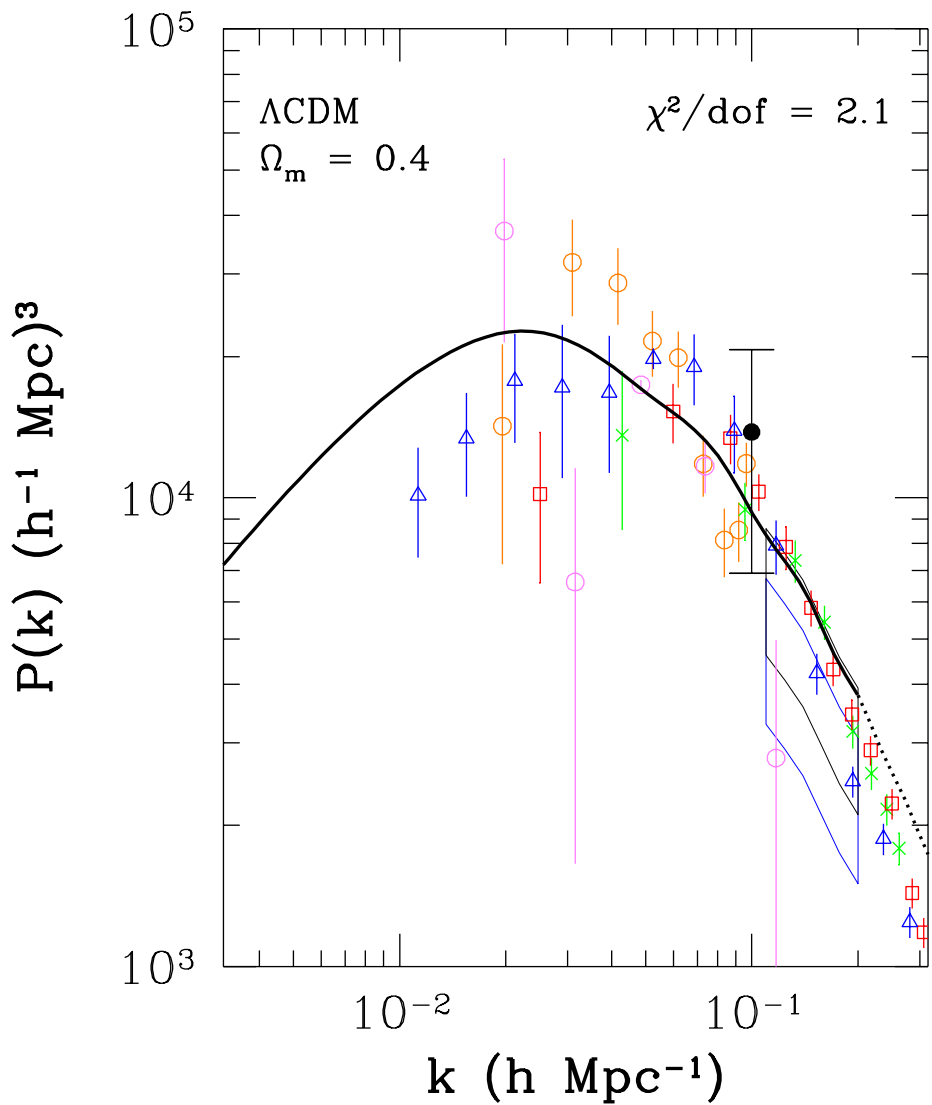,width=6.0in}}
\caption{
Detailed constraints from LSS and CMB on $\Lambda$CDM model 
showing apparent misplacement of peak in the matter power spectrum.}
\label{fig:lchdm00_zoom}
\end{figure}

\begin{figure}[htb]
%\figurenum{4}
\centerline{\psfig{file=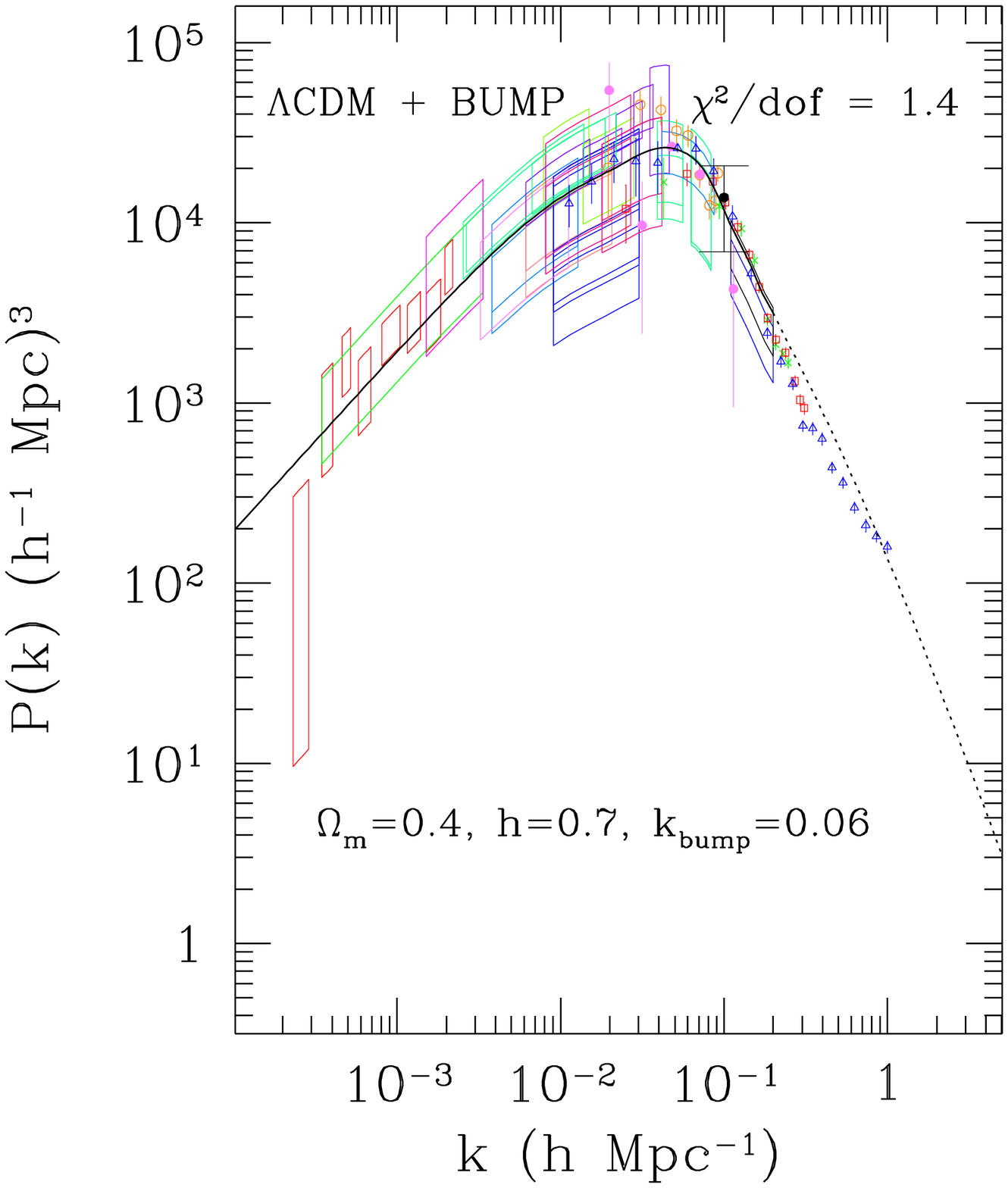,width=6.0in}}
\caption{Constraints from LSS and CMB on $\Lambda$CDM model with 
%* feature
a broad enhancement centered at $k=0.06h^{-1}$Mpc 
added to the primordial power spectrum.}
\label{fig:lchdm00b}
\end{figure}

\begin{figure}[htb]
%\figurenum{4}
\centerline{\psfig{file=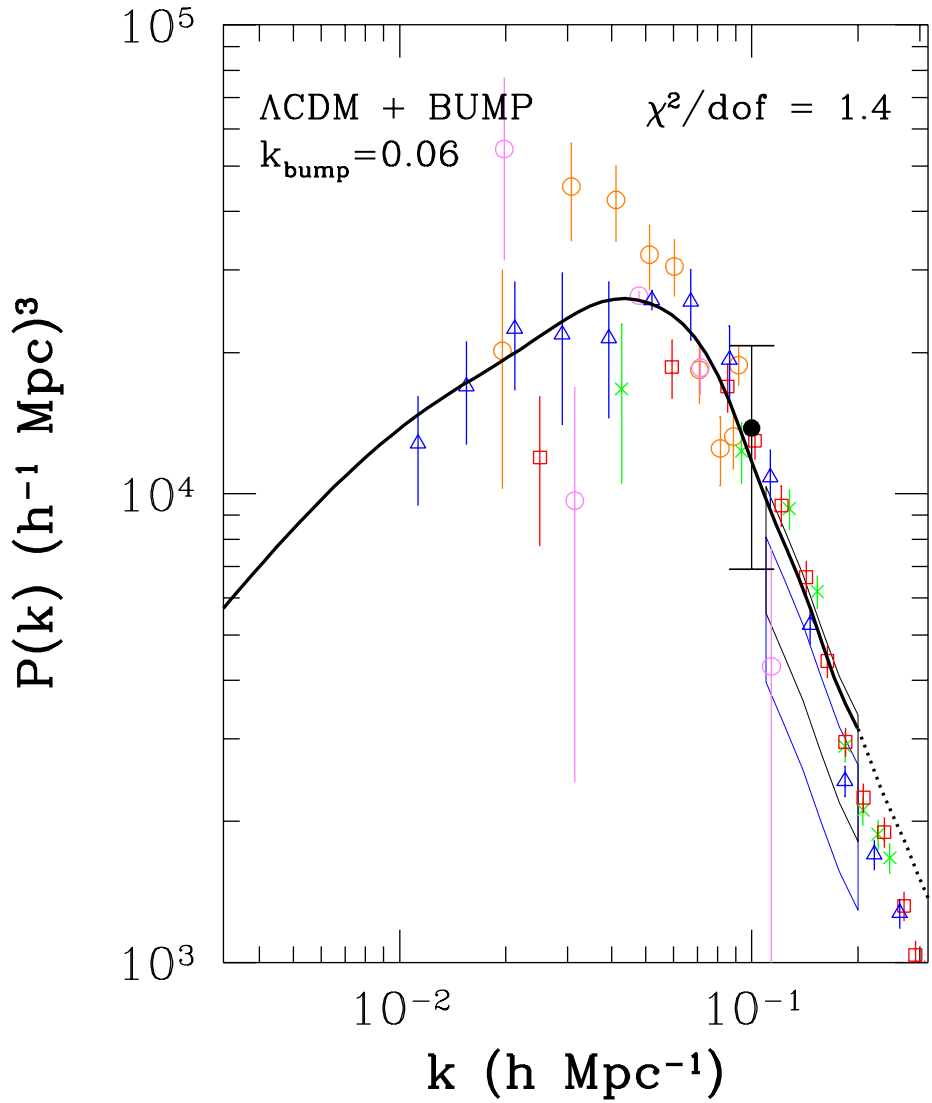,width=6.0in}}
\caption{Detailed constraints from the peak of the matter power spectrum 
on $\Lambda$CDM model with 
%* feature
a broad enhancement centered at $k=0.06h^{-1}$Mpc 
added to the primordial power spectrum.}
\label{fig:lchdm00b_zoom}
\end{figure}

Figure \ref{fig:lchdm00_zoom} shows the detailed constraints on the 
$\Lambda$CDM model we adopted in the preceding chapter in order to fit 
direct observational constraints on cosmological parameters.  It 
appears to have 
the wrong peak location and shape, but with our newfound freedom 
in the primordial power spectrum we can imagine changing the effective 
peak location and shape by adding a ``bump'' of primordial power near 
$k=0.05$.  We adopt the parameterization, 
\begin{equation}
P_p(k) = A k \left( 1 + A_{bump} 
\exp \left( - \frac{(\log k - \log k_0)^2}{ 2 (\sigma_k / k_0)^2}\right) 
\right),
\end{equation}
and find the best-fit parameters $A_{bump}=1$, 
$k_0=0.06$, and $\sigma_k=0.03$, 
i.e. a Gaussian amplification of the primordial power spectrum of width 
0.03 $h^{-1}$Mpc centered at 0.06 $h^{-1}$Mpc with peak amplification of 
a factor of 2 versus the best-fit overall normalization $A$.  This is 
shown in Figures \ref{fig:lchdm00b} and \ref{fig:lchdm00b_zoom}.  The 
improvement is tremendous, from a $\chi^2$ per degree of freedom of 2.1 
to a reduced $\chi^2$ of 1.4 using just three parameters (compared to 70 
degrees of freedom in the data).  While this is not quite as good of a 
fit to the structure formation data as CHDM, it is in good agreement with 
nearly all direct observations of cosmological parameters.  Perhaps the 
primordial power spectrum is radically non-scale-invariant; if so, we 
appear to have reconciled the previous disagreement between direct 
parameter observations and the structure formation data.  It is, however, 
a rather undesirable coincidence to be adding primordial power at 
scales so close to the horizon size at matter-radiation equality, so 
we will need to confirm this feature with improved future data before 
arguing that the model is really the answer.  

\begin{figure}[htb]
%\figurenum{4}
\centerline{\psfig{file=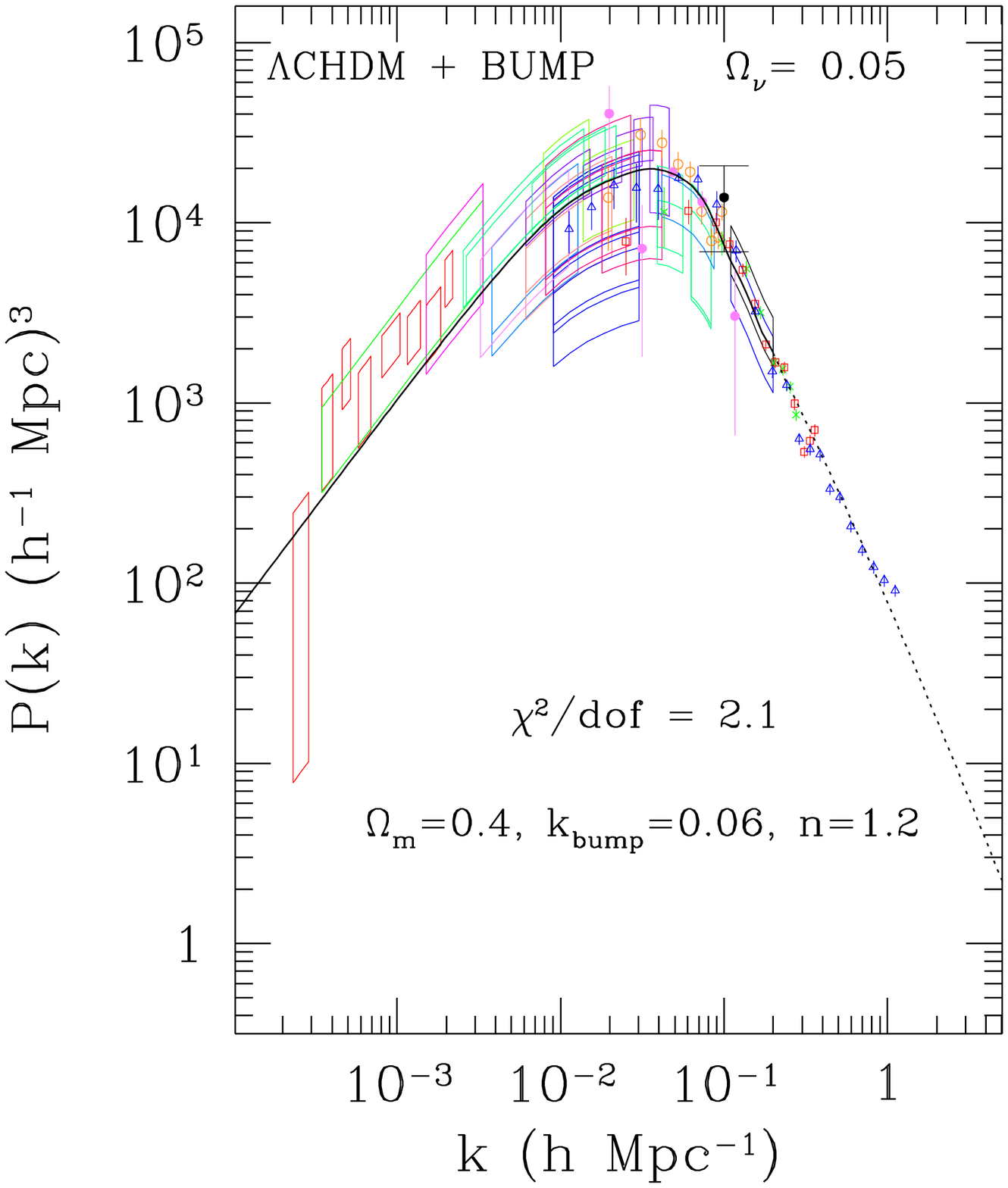,width=6.0in}}
\caption{Constraints from LSS and CMB on Tilted $\Lambda$CHDM model with 
%* feature
a broad enhancement centered at $k=0.06h^{-1}$Mpc 
added to the primordial power spectrum.}
\label{fig:tlchdm05b}
\end{figure}

\begin{figure}[htb]
%\figurenum{1}
%\epsffile{lchdm00_ebb.ps}
\centerline{\psfig{file=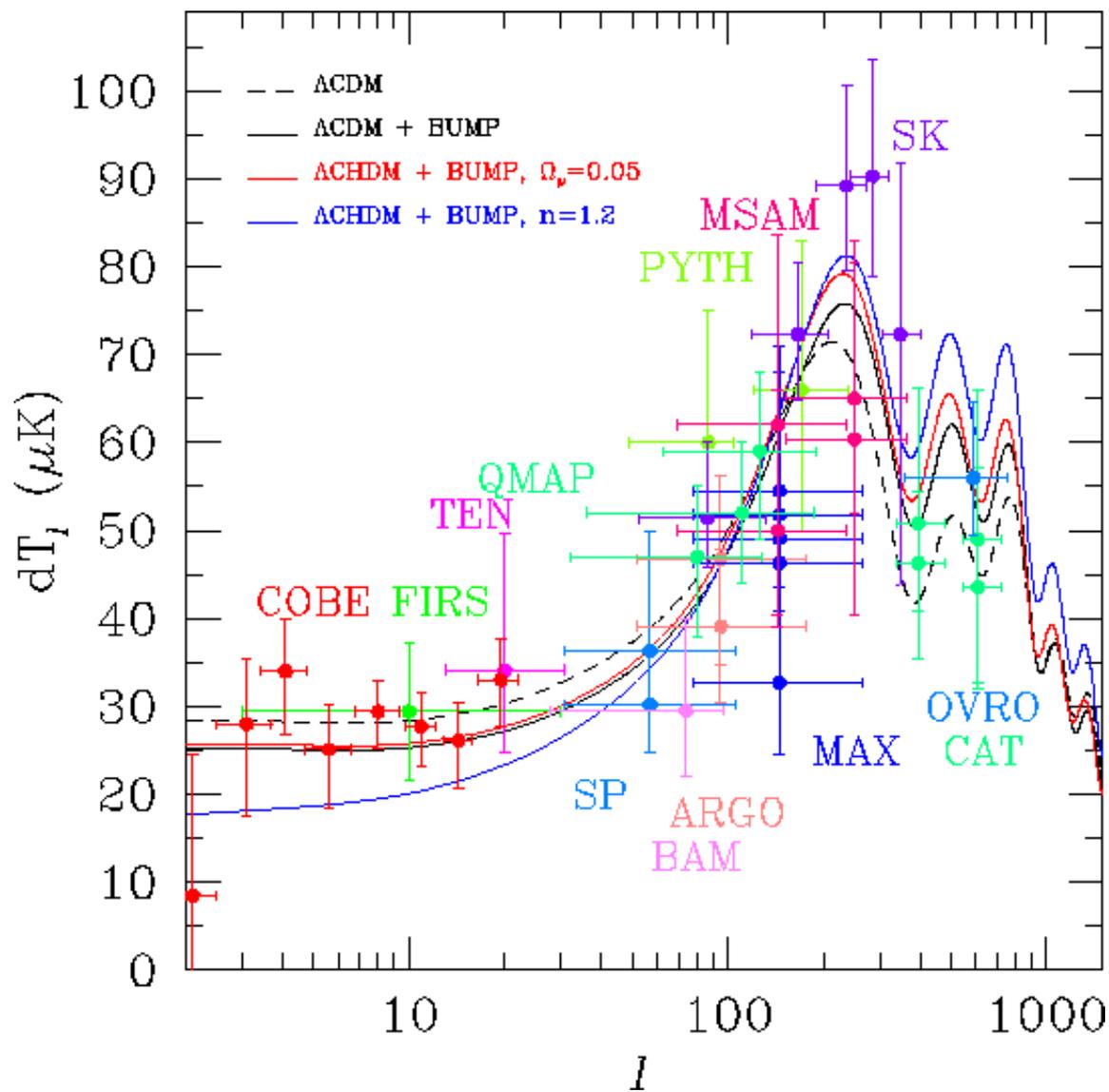,width=6in}}
\caption{Constraints from CMB on tilted $\Lambda$CHDM + BUMP models.} 
\label{fig:lchdm_cmb}
\end{figure}

An important consideration at this point is whether our newfound 
freedom in the primordial power spectrum has radically altered our 
limits on the neutrino mass of the previous chapter.  At least for 
the $\Lambda$CDM+BUMP model, this is not the case, as shown in Figure 
\ref{fig:tlchdm05b}, where the addition of HDM rapidly degrades the 
fit just like before.  The impact of a bump in the primordial power 
spectrum on the CMB predictions of these models is shown in Figure 
\ref{fig:lchdm_cmb}, which illustrates that the enhancement in power 
actually helps reach towards the high SK data in the first acoustic 
peak and is not much higher than the CAT and OVRO measurements 
at smaller angular scales.  Once HDM is added, the shape of the matter
power spectrum must be restored with a blue tilt of $n=1.2$ and 
this starts to lead to serious disagreement with the COBE results 
versus the amplitude of smaller-scale observations as discussed in 
Chapter \ref{chap:nu}.  Once the precision of CMB anisotropy observations 
improves, it should be quite possible to constrain or detect this 
sort of enhancement in power on $\sim 100$ Mpc scales.  

\section{Discussion}

This ad-hoc model, although not aesthetic, 
is physically possible.  
It could be generated, for example, by 
incomplete coagulation of bubbles of new phase 
in a universe that already has been homogenized by a previous episode of
inflation \citep{amendolab94, baccigalupiao97}. 
One can tune the bubble size distribution to be sharply peaked at any 
preferred
 scale.  In this case, voids of the observed sizes of tens of Mpc are 
sufficient to create roughly the right shape of the matter power spectrum.
This results in nongaussian features and excess power where needed. 
The non-gaussianity provides a distinguishing characteristic.

 Other suggestions that might create such a ``bump'' 
appeal to an inflationary  relic of excess power from 
broken scale invariance, 
arising from  double inflation  in a $\Lambda$CDM model,
which results in a spike due to a sharp change in 
amplitude of the inflaton potential at the desired 
physical scale
\citep{lesgourguesps98}. 
This 
 improves the fit in much the same way as adding a hot component to CDM 
 improves the empirical fit.
Such ad hoc fits may seem  unattractive, 
but at present they are needed 
in order to fit the data.  
Moreover, there are interesting, testable predictions 
that arise from the tuned void
approach.
The bubble-driven shells provide a source of overdensities on large scales. Rare shell interactions  could produce nongaussian  massive galaxies or  clusters at low or even high redshift: above a critical surface density threshold
gas cooling would help concentrate gas and aid collapse.
 If massive
galaxies were discovered at say $z>5$ or a massive galaxy cluster at $z>2$ this
would be another indication that the current library of cosmological models is
inadequate. 
If the data are accepted as mostly being free of systematics
and ad hoc additions to the primordial power spectrum are avoided,
 there is no acceptable model for
large-scale structure.  No one LSS data set can be blamed.  
New data sets such as SDSS and 2DF are 
urgently needed to verify whether the shape discrepancies in $P(k)$ will
persist; the good news is that 
the Sloan and 2DF surveys are already acquiring
galaxy redshifts.

\section{The Primordial Power Ratio}

\begin{figure}[htb]
\centerline{\psfig{file=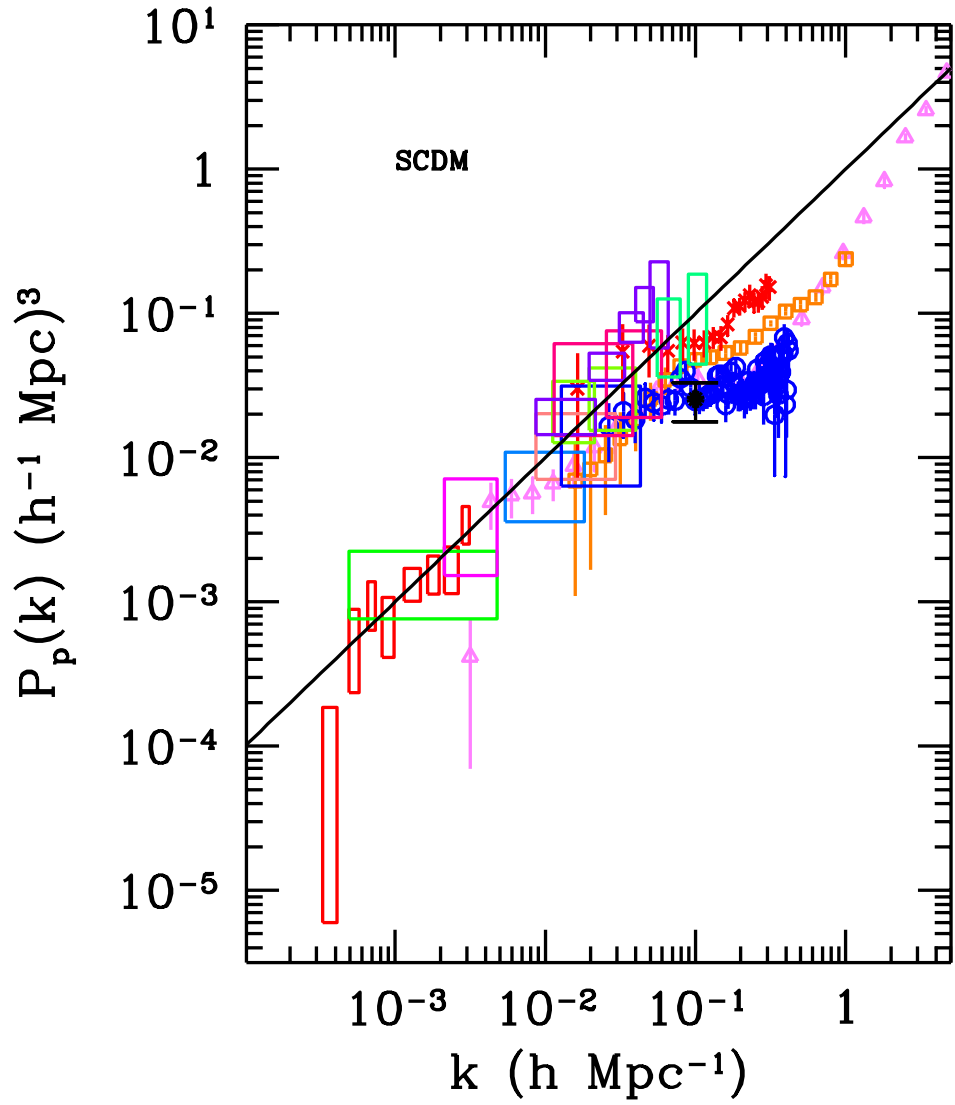,width=6.0in}}
\caption{Primordial power 
constraints from LSS and CMB for SCDM model.  The solid line 
is a scale-invariant primordial power spectrum, $n=1$.}
\label{fig:scdm_ppk}
\end{figure}

\begin{figure}[htb]
\centerline{\psfig{file=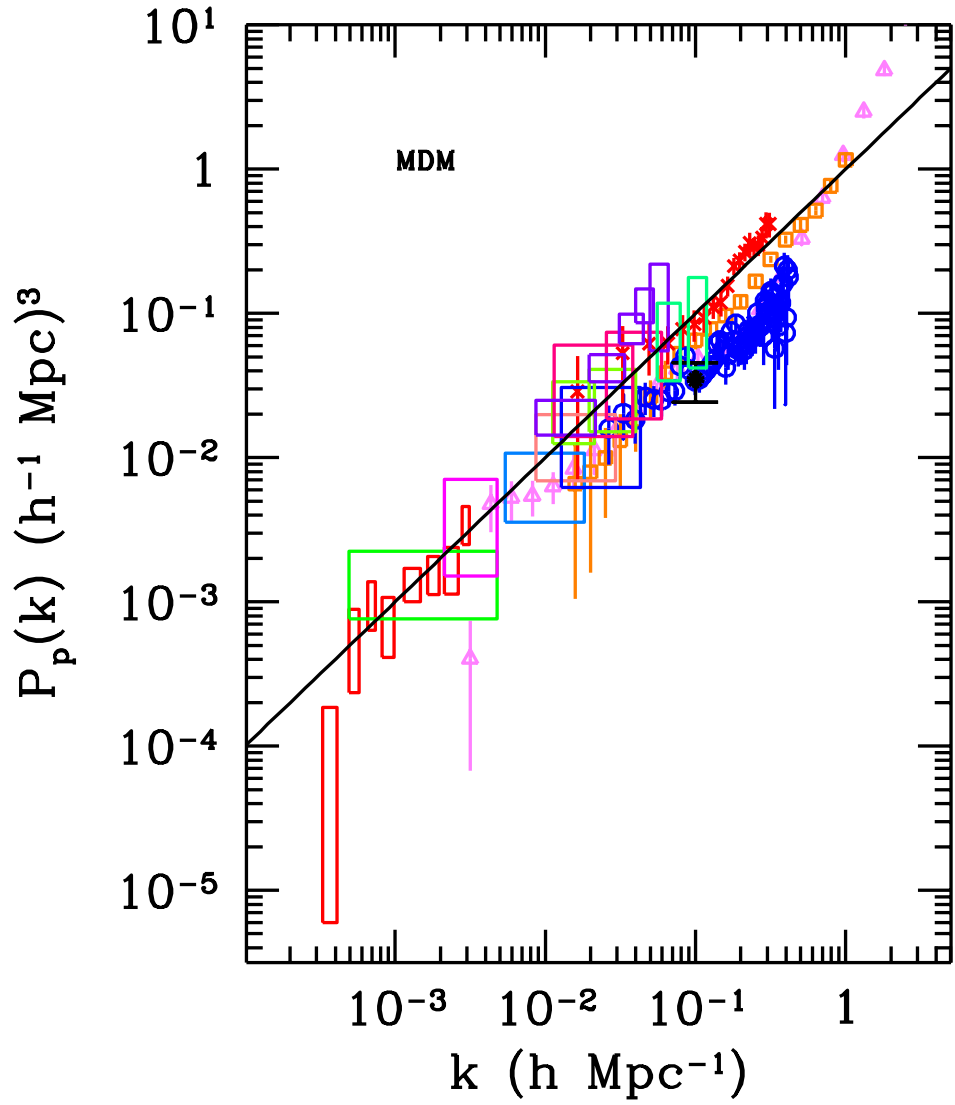,width=6.0in}}
\caption{Primordial power constraints from LSS and CMB for CHDM model.}
\label{fig:chdm_ppk}
\end{figure}

\begin{figure}[htb]
\centerline{\psfig{file=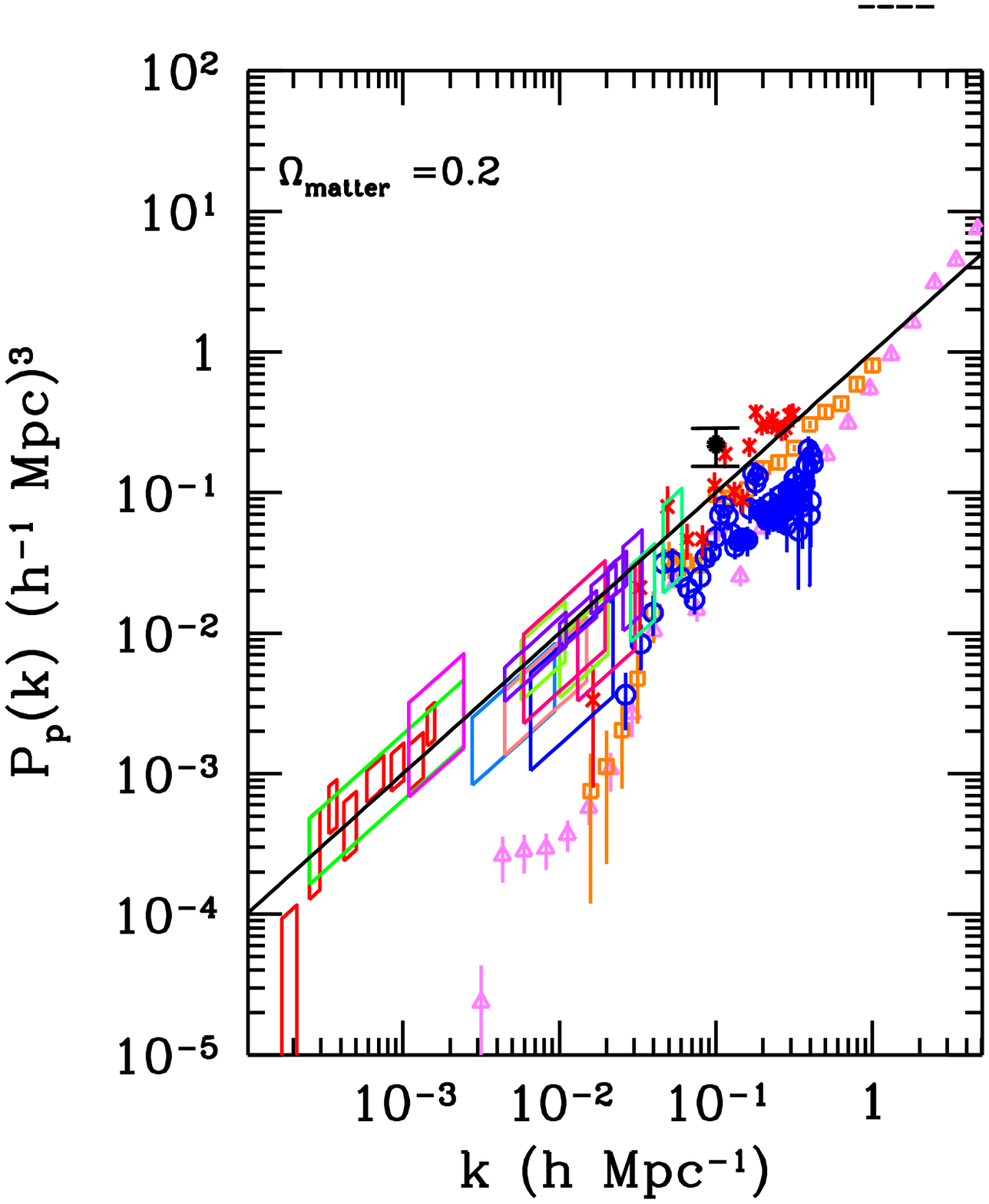,width=6.0in}}
\caption{Primordial power constraints from LSS and CMB for 
a high baryon-fraction model similar to BCDM.}
\label{fig:silk_ppk}
\end{figure}

Figure \ref{fig:scdm_ppk} illustrates the difficulties of the SCDM 
model in a new form, that of the primordial power spectrum reconstructed 
from both CMB and large-scale structure data.  The biases of 
redshift surveys in 
these plots have not been optimized, so it is their shape that is 
a strong constraint.  For SCDM, the CMB is compatible with the 
scale-invariant primordial power spectrum shown, but the redshift 
surveys prefer a tilt around $n=0.8$.  Hence the Tilted CDM model is 
better, but this plot shows that a better fit still can be achieved 
by letting the scalar spectral index vary from about $n=1.2$ on 
COBE scales to $n=0.8$ around $k=0.1h$/Mpc, as 
found by \citet{gawiser98}.  This added freedom in the primordial power 
spectrum makes SCDM a much better fit to the data, just as we have 
been able to make $\Lambda$CDM improve by adding a bump of 
primordial power.  Figure \ref{fig:chdm_ppk} shows that a model like 
CHDM which is already a good fit to the data assuming a scale-invariant 
primordial power spectrum is not helped by this added freedom, as 
a Harrison-Zeldovich form is a good fit to the data on all scales.  
Figure \ref{fig:silk_ppk} 
shows the opposite case, however, a model with a very 
high baryon fraction where the predictions for the peak of the matter 
power spectrum differ from the data so strongly that there is no 
shape of the primordial power spectrum that can simultaneously agree 
with CMB and LSS data on large scales.  This model can be falsified without assuming any particular shape for $P_p(k)$.  This occurs because CMB anisotropy 
and large-scale structure data offer us two independent probes of the 
primordial density fluctuations.  When they are sensitive to 
overlapping spatial scales, their reconstructed primordial power 
spectra can be compared to form the primordial power ratio, which should equal 
one on all scales if these density variations came from the same source.  
Thus with improved high-resolution data we can hope to expand the overlap 
range of CMB and LSS observations and discriminate among structure formation 
models using the primordial power ratio, which represents a model-independent 
method of testing the primordial power spectrum and structure formation 
models simultaneously.

More formally, the primordial power ratio $R(k)$ is formed from 
the ratio of these two independent reconstructions of the 
primordial power spectrum, 
\begin{equation}
P^M_p(k) = \frac{H_0^3}{2 \pi^2 c^3} \frac{P_p(k)}{T^2(k)} 
\end{equation}
and 
\begin{equation}
P^R_p(k) = C_{\ell k}^{-1} C_\ell 
\end{equation}
where the latter is invertible as long as few enough $k$ bins are 
desired and the bins have equal logarithmic width in $k$.  We 
need to force $P^R_p(k)$ to be positive definite and smooth, however, 
so a regularization method is required.  For gravitational instability 
resulting from passive sources that generate density perturbations 
long before recombination, $R(k)$ is a very simple function; it should 
equal one for the entire overlap range of CMB and LSS data where it 
can be defined.

We have seen that assumptions about the primordial power
spectrum make a tremendous difference in testing theories of structure
formation.  
Combining Large-Scale Structure observations with CMB anisotropy data gives
us a long lever arm in $k$-space with which to reconstruct the primordial
power spectrum.  With the next generation of observations, we hope that
our technique will prove powerful enough to either discredit inflation or
reconstruct the inflaton potential.

\cleardoublepage

\cleardoublepage

\cleardoublepage

%\cleardoublepage
%\begin{figure}[htb]
%\centerline{\psfig{file=lcdm_ppk.ps,width=6.0in}}
%\caption{Primordial power constraints from LSS and CMB for $\Lambda$CDM 
%model.} 
%\label{fig:lcdm_ppk}
%\end{figure}

\cleardoublepage

\cleardoublepage
\part{Making Realistic Simulations of the Microwave Sky}

\chapter{The WOMBAT Challenge}
\label{chap:wombat}
%\documentstyle[12pt,aaspp4,psfig]{article}
%\begin{document}
%\title{The WOMBAT Challenge:  A ``Hounds and Hares'' Exercise for Cosmology}

\section{Motivation}

Cosmic Microwave Background (CMB) anisotropy observations during the 
next decade will yield data of unprecedented quality and quantity.  
Determination of cosmological parameters to the 
precision that has been forecast 
\citep{jungmanetal96, bondet97, zaldarriagass97, eisensteinht98}
will require significant 
advances in analysis techniques to handle the 
large volume of data, subtract foreground contamination,  
and account for instrumental systematics.  To guarantee the accuracy 
required to improve upon the constraints on models of structure formation 
that were derived in Part I, 
we must ensure that the analysis techniques used on these 
forthcoming high-precision datasets  
do not introduce unknown biases into the estimation of cosmological 
parameters.  

The Wavelength-Oriented Microwave Background Analysis Team 
(WOMBAT, see http://astro.berkeley.edu/wombat, \citealt{gawiseretal98}, 
and \citealt{jaffeetal99})
 has produced state-of-the-art 
simulations of microwave 
foregrounds, using information about the 
frequency dependence, power spectrum, and spatial distribution of 
each component.  Using the phase information (detailed spatial morphology 
as opposed to just the power spectrum) 
of each foreground component 
 offers the possibility of improving upon foreground 
subtraction techniques that only use the predicted angular 
power spectrum of the foregrounds to account for their spatial distribution.
Most foreground separation techniques rely on assuming that the frequency 
spectra of the components is constant across the sky, but we  provide
 information on the spatial variation of each component's spectral index 
whenever possible.  
The most obvious advantage of this approach 
 is that it reflects our actual sky rather than just 
a statistical description of it.  With 
the high precision expected from future CMB maps we must test our 
foreground subtraction techniques on as realistic a skymap as 
possible.  A second advantage is the construction
of a common, comprehensive database for all known CMB foregrounds.  
The database includes  
predicted uncertainties in the estimation of the foregrounds.  
This should prove valuable for all groups involved in measuring 
the CMB and extracting cosmological information from it.  Section 
\ref{sect:wombat_models}
summarizes methods for generating foreground models which include 
phase information, 
and Section \ref{sect:wombat_techniques} gives a brief survey of existing subtraction techniques 
and their limitations. 

These microwave foreground models 
provide the starting point for 
the WOMBAT Challenge, 
a ``hounds and hares'' exercise for which we have simulated skymaps 
corresponding to 
various cosmological models and made them available to the cosmology community for 
analysis without revealing the input parameters.  
This exercise will 
test the efficacy of current foreground subtraction, power 
spectrum analysis, and parameter estimation techniques and will help 
identify 
the areas most in need of progress.  
This challenge is similar to the ``Mystery CMB Sky Map challenge'' posted 
by our sister collaboration, 
COMBAT\footnote{Cosmic Microwave Background Analysis Tools, 
http://cfpa.berkeley.edu/group/cmbanalysis}, except 
that our emphasis is on dealing with realistic foregrounds rather 
than the ability to analyze large data sets. 
Our simulations contain CMB anisotropies combined with all 
major expected foreground components and instrument noise, as 
described in Section \ref{sect:wombat_simulations}.   
Section \ref{sect:wombat_challenge} describes our 
plans to conduct this foreground removal challenge. 
The WOMBAT Challenge promises to shed light on several open questions 
in CMB data analysis:  What are the best foreground subtraction techniques?  
Will they allow instruments such as MAP and Planck to achieve the 
precision in $C_\ell$ reconstruction 
which has been advertised, or will the error bars increase
significantly due to uncertainties in foreground models?  Perhaps most 
importantly, do some CMB analysis methods produce biased estimates 
of the radiation power spectrum and/or cosmological parameters?  

\section{Microwave Foregrounds}	
\label{sect:wombat_models}

Phase information is now available for 
Galactic dust 
and synchrotron  
and for the brightest radio galaxies, 
infrared galaxies, and X-ray clusters on the sky.  
By incorporating known information 
on the spatial distribution of the foreground components and spatial 
variation in their spectral index, we have greatly improved 
upon previous highly-idealized foreground models.

There are four major expected sources of Galactic foreground emission at 
microwave frequencies:  thermal emission from dust, electric or 
magnetic dipole emission 
from spinning dust grains 
\citep{drainel98a,drainel98b,drainel99}, 
free-free 
emission from ionized hydrogen, and synchrotron radiation from electrons 
accelerated by the Galactic magnetic field.  Good spatial templates 
exist for thermal dust emission 
\citep{schlegelfd98}
and synchrotron emission 
\citep{haslametal82}, 
although the $0\fdg5$
resolution of the Haslam maps means that smaller-scale structure must 
be simulated or ignored.  
Extrapolation to microwave frequencies 
is possible using maps which account for spatial 
variation of the spectra 
\citep{finkbeinerds99, plataniaetal98}.
The COMBAT collaboration has recently posted a software 
package called 
FORECAST\footnote{Foreground and CMB Anisotropy Scan Simulation Tools,\\ 
http://cfpa.berkeley.edu/forecast}  
 that displays the expected dust foreground for a 
given frequency, location, and observing strategy.  
Our best-fit foreground maps will be added to this user-friendly site 
in the near future, and this should be a useful resource for planning and 
simulating CMB anisotropy observations.  

A spatial template for free-free emission 
based on observations of H$\alpha$ 
\citep{smoot98, marcelinetal98}
can be created in the near future 
by combining WHAM observations 
\citep{haffnerrt98}
with the southern celestial hemisphere H-Alpha Sky Survey 
\citep{mcculloughetal99}.   
While it is known that there is an anomalous 
component of Galactic emission at 15-40 GHz 
which is partially correlated 
with dust morphology 
\citep{kogutetal96a, leitchetal97, docetal97}, 
it is not yet clear whether this is spinning dust 
grain emission or free-free emission somehow  
uncorrelated with H$\alpha$ observations.  In fact, spinning dust grain 
emission has yet to be observed directly (see, however, \citealp{docetal99}), 
so the uncertainties in its amplitude are 
tremendous.  A template for free-free emission 
can be derived from DMR data but has tremendous uncertainties.
A better choice is to use the Galactic dust template multiplied by 
the appropriate correlation coefficent at a given frequency as a template, 
but this yields no information on any portion of ``anomalous'' emission 
that may be uncorrelated with the structure of Galactic dust.  

There exist three nearly separate categories of galaxies that
 also generate microwave 
foreground emission; they are radio-bright galaxies, low-redshift 
infrared-bright galaxies, and high-redshift infrared-bright galaxies.  
The level of anisotropy produced by these 
foregrounds is predicted by 
\citet{toffolattietal98}
using 
models of galaxy evolution to produce source counts.  
Updated models calibrated to recent SCUBA observations are also 
available 
\citep{blainetal98, scottw98}.  
For the high-redshift galaxies detected by SCUBA, no spatial template 
is available, so a simulation of these galaxies with realistic 
clustering will be necessary.  
\citeauthor{scottw98}
and 
\citeauthor{toffolattietal98}
have used very different estimates of clustering to produce divergent
results for its impact, so this issue will need to be looked at more 
carefully.  Upper and lower limits on the anisotropy generated by 
high-redshift galaxies and as-yet-undiscovered types of point sources 
are given in Chapter \ref{chap:sources} 
using recent observations over a wide range of microwave 
frequencies.  
We will need to look for these sources with direct 
observations and design analysis techniques that might manage 
to subtract them.  
The 5319 brightest low-redshift IR galaxies detected at 60$\mu$m are 
contained 
in the IRAS 1.2 Jy catalog 
\citep{fisheretal95}
 and can be extrapolated to 100 GHz with a systematic 
uncertainty of a factor of a few 
(see Chapter \ref{chap:ir}).  
Chapter \ref{chap:radio} 
discusses the catalog of 2207 bright radio sources
which has been compiled by \citet{sokasiangs98}.
This catalog appears to contain the few hundred brightest sources in 
the full sky, but at lower fluxes it contains very few sources from 
the Southern Celestial Hemisphere.  We have fixed this deficit by 
including roughly a thousand sources from the Parkes-MIT-NRAO 
catalog 
\citep{griffithw93} 
in our simulations (see Section \ref{sect:radio_pmn}).   

The secondary CMB anisotropies that 
occur when the photons of the Cosmic Microwave 
Background radiation are scattered after the original last-scattering 
surface can also be viewed as a type of foreground contamination (see 
discussion in Chapter 1 for details).
  The shape of the blackbody 
spectrum can be altered through 
inverse Compton scattering by the thermal Sunyaev-Zeldovich (SZ) effect
\citep{sunyaevz72}.  
The effective temperature of 
the blackbody can be shifted locally by 
a Doppler shift from the peculiar velocity of the scattering medium (the 
kinetic SZ and Ostriker-Vishniac effects) as well as by passage through
nonlinear structure (the Rees-Sciama effect).  
Simulations have been made of the impact of the 
SZ effects in large-scale structure 
\citep{persietal95}, 
clusters 
\citep{aghanimetal97},
groups 
\citep{bondm96c}
and reionized patches
\citep{aghanimetal96, knoxsd98, gruzinovh98, peeblesj98}.  
The brightest 200 X-ray clusters are known from the 
XBACS catalog and can be used to incorporate the locations of 
the strongest SZ sources 
\citep[see Chapter \ref{chap:sz}]{refregiersh98}.  
The SZ effect itself is independent of redshift, so it can yield 
information on clusters at much higher redshift than does X-ray 
emission.  However, nearly all clusters are unresolved for $10'$ resolution, 
so higher-redshift clusters occupy less of the beam and therefore their SZ
effect is in fact dimmer.  In the 4.5$'$ channels of Planck this will 
no longer be true, and SZ detection and subtraction becomes  
more challenging and potentially more fruitful as a probe 
of cluster abundance at high redshift.

\begin{figure}[htb]
%\figurenum{1}
%\epsffile{lchdm00_ebb.ps}
\centerline{\psfig{file=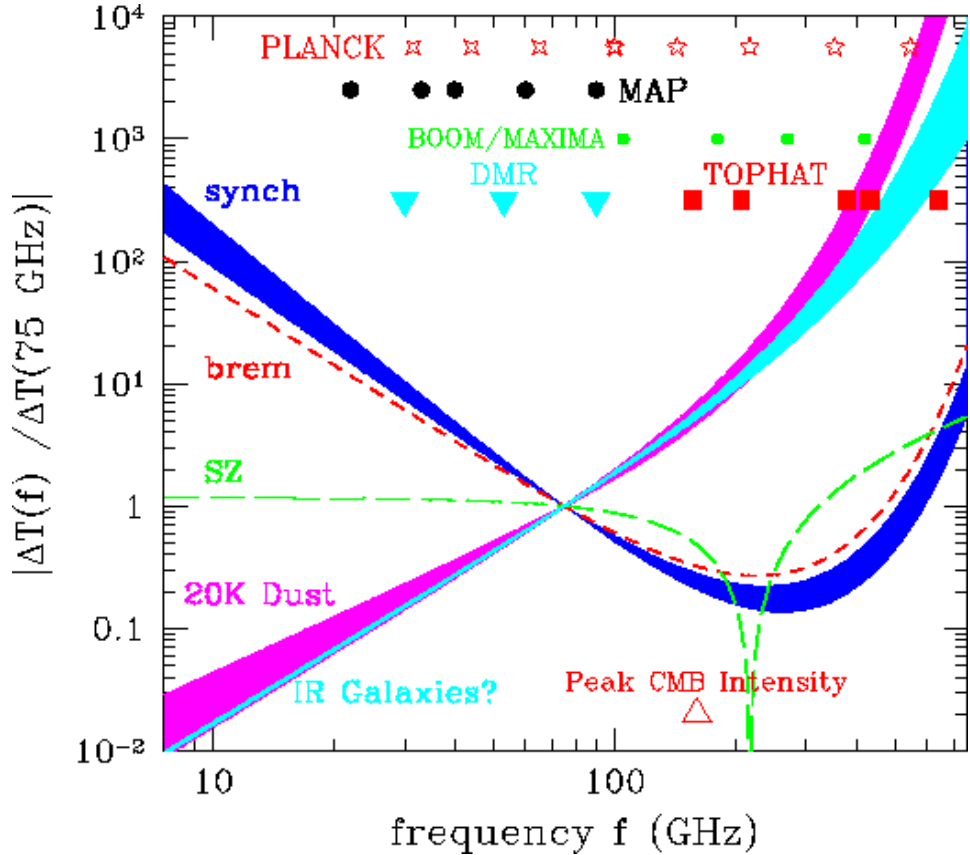,width=5in}}
\caption{Frequency spectra of microwave foregrounds.}
\label{fig:jaffe}
\end{figure}

\section{Reducing Foreground Contamination}
\label{sect:wombat_techniques}

Figure \ref{fig:jaffe} illustrates the spectra, normalized at 75 GHz, of 
the various foreground components which are expected to contribute 
significantly to microwave anisotropy.  The frequencies of several 
current and future observing instruments are shown; the amplitude of 
Galactic foregrounds will decrease at higher resolution whereas the amplitude of extragalactic point sources will increase.  Radio galaxies 
have similar spectra to those shown for synchrotron and brehmsstrahlung 
emission.  The expected 
minimum of all foreground contamination is near 100 GHz.    
Various methods have been proposed for reducing foreground contamination.  
For point sources, it is possible to mask pixels which represent 
positive $5 \sigma$ fluctuations since such fluctuations 
are highly unlikely for Gaussian-distributed
 CMB anisotropy and Gaussian-distributed instrument noise and 
can be assumed to be caused by point sources.    
This pixel masking technique can 
be improved somewhat by filtering 
(\citealt{tegmarkd98};
see 
\citealt{tenorioetal99}
for a different technique using wavelets).  
Chapter \ref{chap:radio} 
demonstrates that using prior information from good source catalogs 
may allow the masking of pixels which contain sources brighter than  
the $1 \sigma$ level of CMB fluctuations and instrument noise.  For 
 the 90 GHz MAP channel, this could reduce 
the residual radio point source 
contamination by a factor of two, which 
might significantly reduce systematic errors in cosmological 
parameter estimation.  
Galactic foregrounds with well-understood frequency spectra can be 
projected out of multi-frequency observations on a pixel-by-pixel basis 
\citep{brandtetal94, dodelsons94, dodelsonk95, dodelson97}.  
Prior information in 
the form of spatial templates can be included in this projection, but 
uncertainty in the spectral index is a cause for concern.  

Perhaps surprisingly, the methods for foreground subtraction which have 
the greatest level of mathematical sophistication and have been tested 
most thoroughly ignore the known locations on the sky of some  
foreground components.  The multi-frequency Wiener filtering approach 
uses assumptions about the spatial power spectra and frequency spectra of the 
foreground components to perform a separation in spherical 
harmonic or Fourier space 
\citep{tegmarke96, bouchetps98, bouchetg99, knox98}.  
However, it does not include 
any phase information at present.    
The Fourier-space Maximum Entropy Method 
\citep{hobsonetal98a}
can add phase information on diffuse Galactic foregrounds in 
small patches of sky but treats 
extragalactic point sources as an additional source of instrument noise, 
with good results for simulated Planck data 
\citep{hobsonetal98b}
 and worrisome systematic 
difficulties for simulated MAP data 
\citep{joneshl98}.  
Maximum Entropy has not yet been adapted to handle full-sky 
datasets.  Both methods have difficulty if pixels are 
masked due to strong point source contamination or the spectral 
indices of the foreground components are not well known 
\citep{tegmark98}.   

Since residual foreground contamination can increase 
uncertainties and bias parameter estimation, it is important to 
reduce it as much as possible.  Current analysis methods usually 
rely on cross-correlating the CMB maps with foreground templates at 
other frequencies 
(see 
\citealt{docetal97}
and 
\citealt{jaffefb99}). 
It is clearly superior to have 
region-by-region (or pixel-by-pixel) information on how to extrapolate
these templates to the observed frequencies; otherwise this cross-correlation
only identifies the emission-weighted 
average spectral index of the foreground 
from the template frequency to the observed frequency. 

Because each foreground has a non-Gaussian temperature distribution, 
the covariance matrix of its expected 
$a_{\ell m}$ coefficients is not diagonal.  
When a known foreground template 
is subtracted from a CMB map, it is inevitable that the correlation 
coefficient used for this subtraction will be slightly different than 
the true value.  This expected under- or over-subtraction of each 
foreground leads to off-diagonal structure in the ``noise'' covariance matrix 
of the remaining CMB map, as opposed to the contributions of expected  
CMB anisotropies and uncorrelated instrument noise, 
both of which give diagonal contributions to the covariance matrix of 
the $a_{\ell m}$.  Thus incomplete foreground subtraction, 
like $1/f$ noise, can introduce non-diagonal correlations into the 
covariance matrix of the $a_{\ell m}$.  
These correlations complicate the likelihood analysis necessary for 
parameter estimation 
\citep{knox98}. 
Chapter \ref{chap:pixel} presents a foreground subraction method that 
attempts to account for these correlations in pixel space.  
Having phase information on the 
brightness and spectral index of foreground emission 
 should reduce inaccuracies in foreground subtraction, and this motivates 
us to produce the best estimates we can of these quantities along with 
estimates of their uncertainties.  

The Wavelength-Oriented Microwave Background Analysis Team is dedicated 
to generating the best possible predictions of microwave emission from 
Galactic and extragalactic sources along with a clear description 
of the uncertainties.  Those predictions and uncertainties were used 
in generating the WOMBAT Challenge Simulations, and they are ready to 
be applied to real-world data as well.  
We have made our ``best-fit'' 
knowledge of the various foreground 
components available to the public, and each best-fit
foreground map (or list of predicted point source fluxes) is  accompanied 
by a map (or list) 
of its uncertainties and a discussion of possible systematic 
errors\footnote{see http://astro.berkeley.edu/wombat/foregrounds.html}.
  Each simulation of that foreground will be different 
from the best-fit map based upon a realization of those uncertainties. 
This simulates the real observing 
process in a way not achieved by previous foregrounds simulations.
Observational teams that 
wish to download our predictions for use in masking pixels, cross-correlating, 
and subtracting foreground contamination from microwave data are encouraged
to do so.

\section{The WOMBAT Challenge Simulations}
\label{sect:wombat_simulations}

We produced simulations analogous 
to high-resolution balloon observations 
(e.g. MAXIMA and BOOMERANG; 
see 
\citealt{hananyetal97}
and 
\citealt{debernardism98}) 
as well as the MAP 
satellite. 
We have used 
the publicly available HEALPIX package of pixelization and analysis 
routines\footnote{http://www.tac.dk/\~{}healpix} for CMB simulation, 
beam convolution, and basic power spectrum analysis of 
full-sky maps.  We modified 
several of the routines to input and output files of $a_{\ell m}$ 
in addition to $C_\ell$ and to create constrained realizations 
of CMB anisotropy.  We also created basic new routines which bin a HEALPIX 
map down to a lower level of resolution and perform a weighted sum of 
several maps.

Our skymaps contain simulated foreground emission from the Galaxy 
(thermal dust emission, synchrotron, free-free emission, 
spinning dust grains) and extragalactic sources 
(radio- and infrared-bright galaxies 
and the Sunyaev-Zeldovich effect from hot gas in galaxy clusters).  
Except for a few intentional minor surprises, they are well described by 
the WOMBAT foreground predictions.  
Each 
simulation has foreground contamination which is reasonable given 
our best-fit models and their uncertainties.  

All simulated skymaps have an average of zero (up to slight rounding 
errors).  This simulates the insensitivity of MAP's differential 
radiometers to the monopole and MAXIMA's subtraction of the average value 
over time to eliminate $1/f$ noise.  The practical effect of this 
is that for full-sky simulated MAP maps the Galactic plane is bright and 
positive and the rest of the sky is negative.  To see CMB fluctuations (and 
even instrument noise), 
it is necessary to use asymmetric minimum and maximum values when displaying 
the map, especially at low frequencies where the Galaxy is very bright.  
For the MAXIMA patches, which are at high Galactic latitude where the 
Galactic emission is much less, setting the average value in 
the $10^\circ \times 10^\circ$ 
degree patch to zero has fixed this, and a symmetric color 
table allows 
one to see the CMB and instrument noise rather well.  

We have not subtracted the dipole from the final maps, but the simulated 
CMB component has no dipole and the reflex dipole of the Local Group's 
motion relative to the CMB frame has not been added.  In principle, one can 
therefore use the remaining dipole in the full-sky maps 
to learn about the foregrounds alone, 
particularly the Galaxy since the Galactic center is much brighter than 
the anti-center.  This is realistic, since the best fit detection and 
subtraction of the CMB reflex dipole usually ignores low Galactic latitude 
and thus allows a Galactic-emission dipole to be analyzed.

\subsection{Constrained Realizations of CMB Anisotropy}

\begin{figure}[htb]
%\figurenum{1}
%\epsffile{lchdm00_ebb.ps}
\centerline{\psfig{file=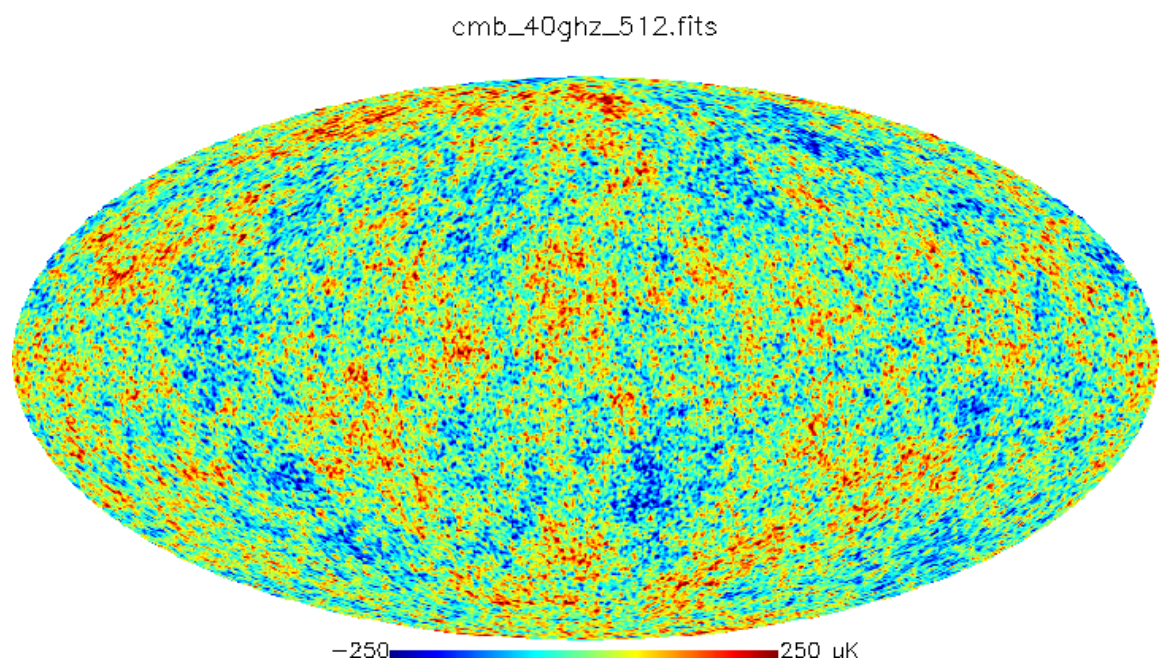,width=6in,angle=90}}
\caption{Simulated CMB component for MAP 40 GHz channel with 28' FWHM.}   
\label{fig:cmb}
\end{figure}

In choosing cosmological models to use for the WOMBAT simulations, 
we assumed 
that adiabatic CDM is probably correct but surprises are always possible.
Figure \ref{fig:cmb} shows a simulation of CMB anisotropy from 
the WOMBAT Challenge simulations for a $28^\prime$ beam for the 
40 GHz channel of the MAP satellite.  
This is a constrained realization generated 
using the method of \citet{bunnetal94}, i.e. if smoothed on $7^\circ$ scales 
it will roughly match the Wiener filtered COBE DMR map.  
This is a necessary step of making the most realistic 
microwave sky simulation possible, since we already know from 
COBE about the fluctuations in CMB intensity on $7^\circ$ scales.  
However, the Wiener-filtered version 
has a lot less information than the DMR 4-year map might appear to, and 
there is not yet any indication that having this low-$\ell$ constraint 
will prove useful in data analysis.  The Wiener filtered map consists 
of best-fit values and uncertainties in each low-$\ell$ $a_{\ell m}$, 
so a simulation consists of adding a realization of these uncertainties 
to the Wiener filtered map and then generating Gaussian random $a_{\ell m}$ 
according to the $C_\ell$ of a given model for all $\ell>30$ since COBE 
yields no information on those scales.  
The $C_\ell$ of the model being considered are used to perform the 
Wiener filtering process, as changing the assumed level of signal 
changes the resulting Wiener filtered map, and we want to be sure that 
the $C_\ell$ of the resulting simulation will be smooth around $\ell=30$ 
where the transition from large-uncertainty Wiener-filtered coefficients 
to purely Gaussian random realization occurs.  
No smaller-scale CMB observations 
have been used as constraints.

\subsection{Instrument Noise and Beam Convolution}

All skymaps are of temperature fluctuations measured in 
thermodynamic microKelvin, i.e. variations about the 
central CMB blackbody with temperature 2.73 K.  Instrument noise is 
also given in thermodynamic $\mu$K, hence there is a significant rise 
in noise in the 390 GHz MAXIMA channel 
where the blackbody of the CMB is dropping rapidly.  
The expected bandwidths of the MAP channels are 20\%, well-approximated by 
a tophat in frequency response.  The expected bandwidths of the MAXIMA 
channels are 30\%, better approximated by a Gaussian.  However, except for 
extreme spectral behaviors, these bands are well approximated by their 
central frequency, and we have done so.  The resulting error is far smaller 
than several other sources of systematic uncertainty.  
The instrument noise is a Gaussian realization, uncorrelated between pixels.
For MAXIMA, the noise is uncorrelated and uniform across the map, which is 
a gross simplification of real-world MAXIMA data.  For MAP, we have used a 
weight map generated using the planned MAP observing strategy 
to determine the noise variance\footnote{
All information used for MAP, including frequencies, beam sizes, 
observing strategy, and expected instrument noise, is as shown at  
http://map.gsfc.nasa.gov}.

We used beamsize uncertainties of $\pm1^\prime$ for the 
Full Width Half Maximum for 
the three MAXIMA frequencies and slightly 
different central values for FWHM at 
each frequency.  This is an attempt to model the real-world uncertainty 
in beam size that is typical for high-precision balloon-borne experiments.  
Beam calibration using planets and bright point sources should allow the 
FWHM to be determined to about 10\% at each frequency, and the calibrated 
value will not necessarily be the same at each frequency.  

The MAXIMA maps are a gnomic projection of a $10^{\circ} \times 10^{\circ}$ 
patch of sky centered at Galactic coordinates (90$^\circ$, 50$^\circ$) 
read from a full-sky simulation 
of CMB anisotropy and foregrounds in HEALPIX pixelization.  The instrument 
noise, however, is realized independently in each of the $4.5^\prime$ width 
pixels of the 128$\times$128 array.  
This simulation covers the 
same patch of sky as the recent MAXIMA flight.  Since we are 
including the known foregrounds on that particular patch of sky in 
our simulation, this is a very good approximation of the 
real data set taken by MAXIMA, except that we have modelled the instrument 
noise as uncorrelated whereas the noise correlations in the real data set 
are considerable.  
The gnomic projection should be flat enough to approximate
this patch of sky as 2-dimensional at the 1\% level, so two-dimensional
Fast Fourier Transforms are a 
feasible analysis method.

\subsection{Galactic foregrounds}

We have used the best existing map of 
Galactic dust emission, the $100 \mu$m emission 
map of  \citet{schlegelfd98}.
This map has 6$^\prime$ resolution and the method of 
extrapolating it to microwave frequencies was derived using comparison 
with FIRAS data \citep{finkbeinerds99}. 
We use a destriped, point-source-subtracted Haslam map at 408 MHz 
\citep{haslametal82}
to determine the synchrotron emission, with maps at 1.4 
\citep{reichr88}
and 2.3 GHz 
\citep{jonasbn98}
used 
to determine the synchrotron spectral index as a function of frequency 
(see \citealp{davieswg96} for cautionary notes).  
We do not add small-scale structure to this template, since 
we find that the Haslam maps are smooth enough on their smallest scales that 
adding simulated structure at higher resolution is unnecessary.
The typical method of doing so 
\citep{bouchetetal94}
involves adding Gaussian random $a_{\ell m}$ 
which obey the extrapolated power spectrum of synchrotron.   
This is a poor approximation to the highly correlated phases necessary 
to construct typical Galactic structure, which is filamentary 
and therefore highly non-Gaussian.
Free-free and  
spinning dust grain emission with the appropriate spectra were simulated 
using templates of the dust map and a model of the Local Bubble.

\subsection{Extragalactic foregrounds}

We have put known bright radio and IRAS galaxies and SZ clusters
 in their proper places on the sky, 
instead of relying on number counts alone and placing the 
sources using Poissonian clustering.  The real clustering of these sources 
is thus contained in our models, and it is possible to use 
knowledge of their locations to mask the pixels containing them.  This allows 
better point source subtraction than does a simple $5\sigma$ cutoff.  
Chapters \ref{chap:ir} and \ref{chap:radio} describe the process of 
simulating the microwave contribution from low-redshift infrared-bright 
galaxies and radio galaxies, respectively.  Chapter \ref{chap:sz} describes 
our predictions for the Sunyaev-Zeldovich effect from galaxy clusters and 
filaments.  

We have also included realistic simulations of the numerous high-redshift 
infrared-bright galaxies whose number counts at 353 GHz were recently 
determined by SCUBA 
\citep{smailib97}.  
We have simulated their non-Poissonian clustering 
as well.  
The population of high-redshift galaxies detected in the sub-mm by 
SCUBA appears to be the main source of the Far-Infrared Background radiation.  
Except for the source counts in the flux range seen by SCUBA, little 
is known about their overall $N(S)$ law, their redshift distribution, or 
the amplitude of their clustering. The model by 
\citet{tansb99}
 is consistent 
with observations across a wide range of frequencies and is therefore 
our current model for predicting source counts as a function of flux 
and redshift.  Alternative models from \citet{blainetal98}, 
\citet{guiderdonietal98}, and \citet{toffolattietal98} 
give a sense of the systematic uncertainties in these models.  
SCUBA observations constrain $N(S)$ around its break, and the low-flux end 
is constrained by the level of integrated background radiation.  However, 
the high-flux end is responsible for the vast majority of microwave 
anisotropy and it is not yet well constrained because SCUBA has 
surveyed a very small portion of the sky.  Thus there is no available 
spatial template for these sources, and there are tremendous systematic 
uncertainties in their number counts and clustering.  

An upper limit on 
the clustering of these sources and its implications for microwave 
foreground anisotropy is given by 
\citet{scottw98}.
They assume that high-redshift
 infrared-bright galaxies cluster like Lyman break 
galaxies, even though the Lyman break galaxies are selected using a technique 
that only allows them to have a narrow redshift distribution.  The redshift
distribution of high-redshift
 infrared-bright galaxies is undoubtedly broader, 
leading us to estimate that their clustering $w(\theta)$ will be between 
a factor of 4 and 10 less than that assumed by 
\citeauthor{scottw98}.  
As the 
amount of sky surveyed by SCUBA increases, an empirical determination of 
the clustering of these sources will narrow this range of uncertainty.
Section \ref{sect:conclusion_dusty} discusses the prospects for 
better determining the bright end of the number counts with forthcoming 
instruments.  

Another significant source of uncertainty in predicting these sources' 
contribution to anisotropy as a function of frequency is that of 
extrapolation.  The 353 GHz (850 $\mu$m) detections by SCUBA are 
almost definitely on the Rayleigh-Jeans side of the peak of their graybody 
emission, but not at low enough frequencies for the 
spectrum to be a pure power-law.  
Thus, although we expect the emissivity of these sources to be 
between 1 and 2, which leads to a Rayleigh-Jeans spectral index 
between 3 and 4, their effective spectral index above 100 GHz should 
be closer to 3, with an uncertainty of 0.5.  The spectral index of 
each source will vary somewhat around this average depending on the 
specific nature of its emitting dust, with a source-by-source uncertainty 
again around 0.5.  Note, however, that a typical CMB observation pixel 
receives roughly equal contributions from its several brightest high-redshift
infrared-bright sources, so the pixel-to-pixel variation will be less 
than this except for pixels dominated by a single unusually bright source.  
 Very little is known about the spatial location of 
high-redshift 
infrared-bright galaxies, hence 
the rough characteristics 
of these high-redshift foreground sources but not their simulated locations  
have  been revealed as part of our foreground predictions.

\begin{figure}[htb]
%\figurenum{1}
%\epsffile{lchdm00_ebb.ps}
\centerline{\psfig{file=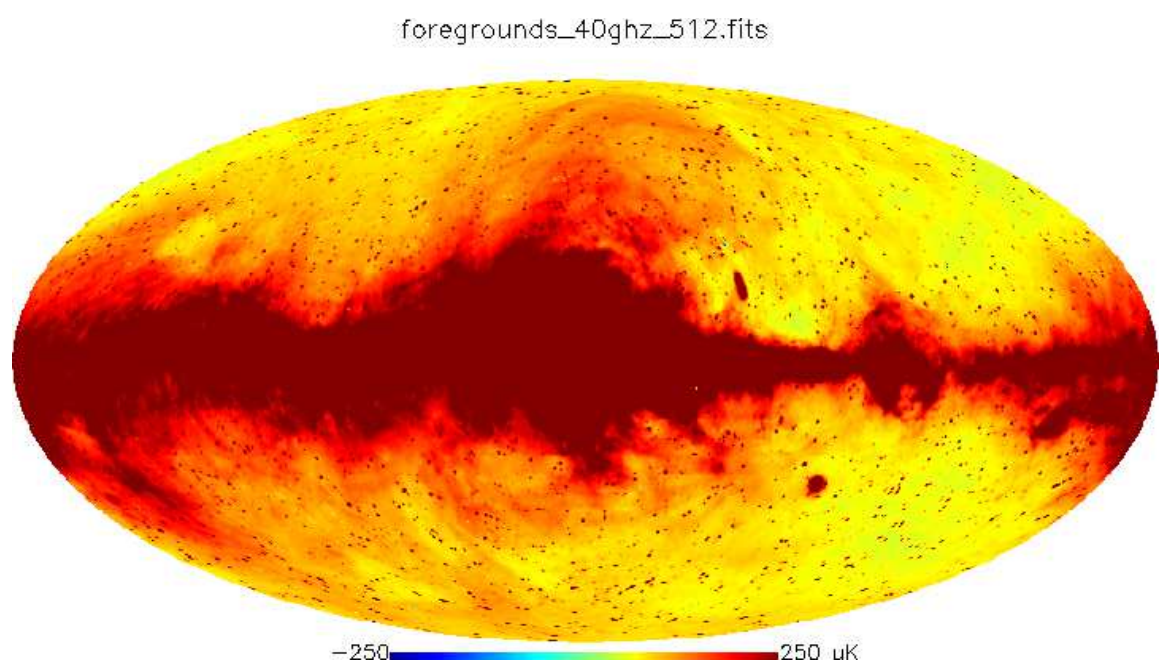,width=6in,angle=90}}
\caption{Simulated Galactic and extragalactic foregrounds for MAP 40 GHz 
channel with 28' FWHM.} 
 \label{fig:foregrounds}
\end{figure}

\begin{figure}[htb]
%\figurenum{1}
%\epsffile{lchdm00_ebb.ps}
\centerline{\psfig{file=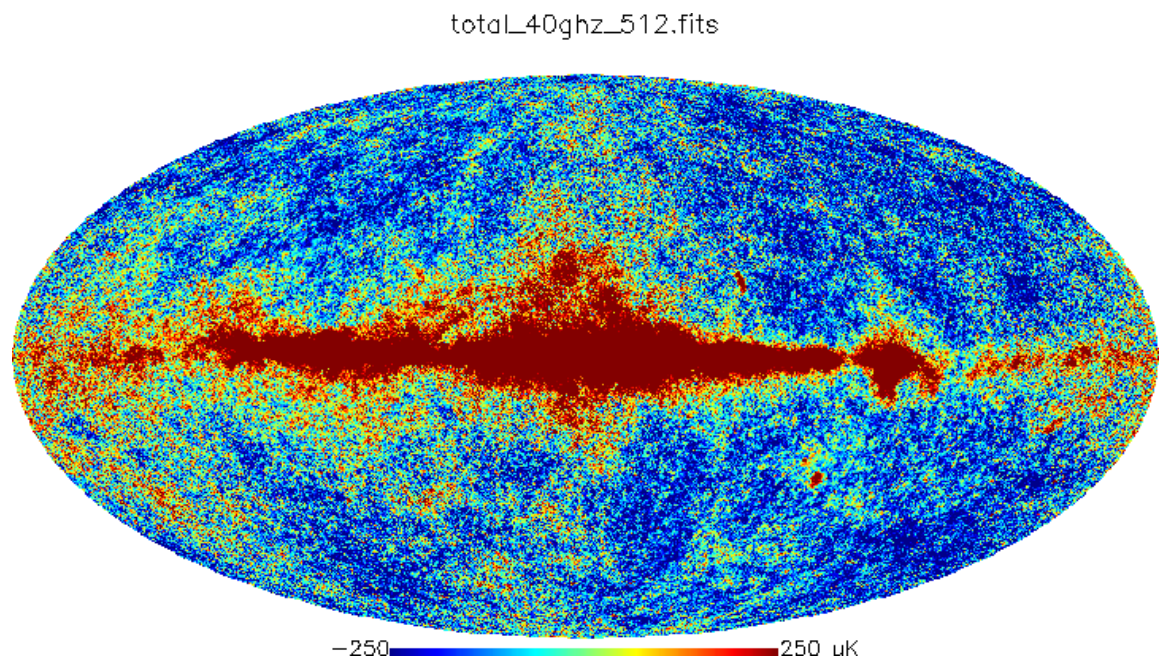,width=6in,angle=90}}
\caption{A realistic simulation of the microwave sky for MAP 40 GHz channel  
with 28' FWHM.  
 } 
\label{fig:total}
\end{figure}

\begin{figure}[htb]
%\figurenum{1}
%\epsffile{lchdm00_ebb.ps}
\centerline{\psfig{file=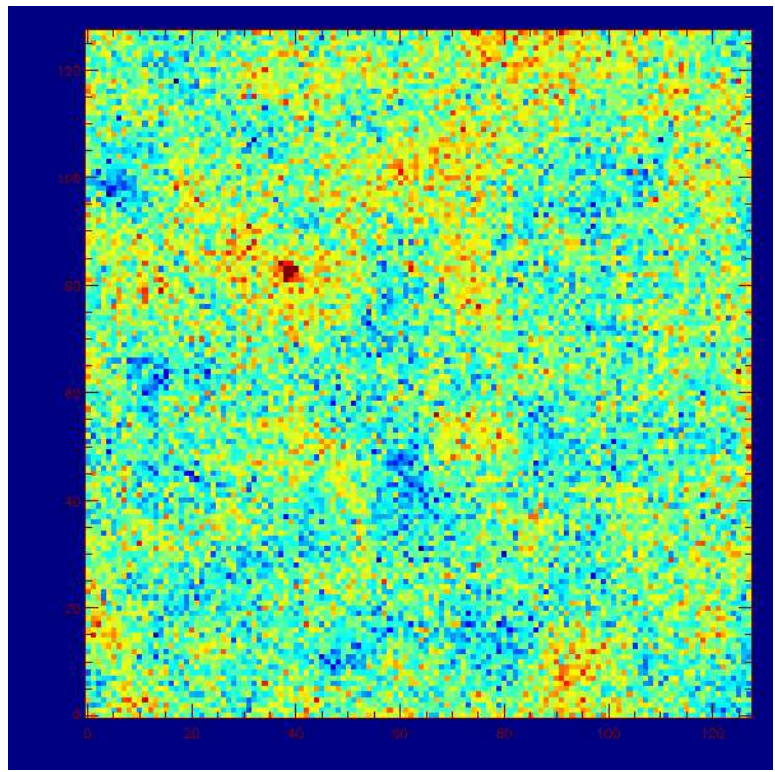,width=4in}}
\caption{
WOMBAT Challenge simulation of a 
 MAXIMA dataset at 150 GHz, 11' FWHM, $10^\circ \times 10^\circ$.}  
\label{fig:maxima}
\end{figure}

Figure \ref{fig:foregrounds} shows our simulation of Galactic 
and extragalactic foregrounds for the MAP 40 GHz channel at 28$^\prime$.  
Temperature fluctuations are given in thermodynamic $\mu$K, with the 
maximum and minimum at the expected level of $\pm 5 \sigma$ fluctuations.  
The Galaxy dominates at low latitude but strong extragalactic point 
sources (convolved with the beam) are visible.  
The SZ decrement is strong enough for 
a few clusters to turn the otherwise strong Galactic contamination into 
a temperature decrement.  
The WOMBAT Challenge consists of 5 simulated MAP datasets 
and 5 simulated MAXIMA datasets.  Each model has underlying 
CMB anisotropies corresponding to a particular cosmology, and each contains 
a different realization of the foreground contamination within the 
uncertainties of our best-fit foreground models.  
Figure \ref{fig:total} shows one of the WOMBAT simulations for the 
MAP 50 GHz channel at 28$^\prime$, including CMB anisotropy, instrument 
noise, and Galactic and extragalactic foreground emission.  
Figure \ref{fig:maxima} shows one of the $10^\circ \times 10^\circ$ 
MAXIMA simulations at 150 GHz, 11' FWHM.   
This also 
includes CMB anisotropy, instrument noise, 
and Galactic and extragalactic foreground emission.
The axis labels are pixel numbers within the 128$\times$128 
array.
Notice the bright point source in the upper left 
quadrant (40,80); 
this is the only one of the MAXIMA simulations with such a bright 
source, which means that it is either a simulated high-redshift galaxy or 
a known IR-bright or radio-bright galaxy where the realized systematic and statistical errors 
led to a much higher flux than its predicted value.

\section{A ``Hounds and Hares'' Exercise for Cosmology}
\label{sect:wombat_challenge}

Our purpose in conducting a ``hounds and hares'' exercise is to simulate the 
process of analyzing microwave skymaps 
as accurately as possible.  In real-world 
observations the underlying cosmological parameters 
and the exact amplitudes and spectral indices of the foregrounds are unknown, 
so Nature is the hare and cosmologists are the hounds.  
We provided a calibration 
map of CMB anisotropy with a disclosed angular power spectrum (SCDM) 
in January 1999 so that participants could test the download procedure 
and become familiar with HEALPIX.  
The WOMBAT Challenge 
began on March 15, 1999 with the release of our 
simulated microwave skymaps on the World Wide Web.  
 Participating groups have four 
months to analyze the skymaps, subtract the foregrounds and 
extract cosmological information. 
Options for reporting results 
include generating 
maps of the input CMB and/or foreground components, 
plotting $C_\ell$ spectra of 
these components, and specifying 
the input cosmological parameters that correspond 
to the CMB anisotropies in each simulated universe.

Two prizes will be given:  a case of champagne to the team or individual 
who best determines the CMB anisotropy power spectrum and cosmological 
parameters of each model, and a stuffed animal (guess which kind!) to the team 
or individual that 
can determine the most information about the 
Galactic and extragalactic foreground sources for each model.  
The fortnight between the 
end of the Challenge on July 15, 1999 and the announcement 
of input parameters on August 1, 1999 will be used to ensure that we 
understand the answers provided to us.  
Participants 
are encouraged to publish any research progress they make as a result of 
participation, such as algorithms and analysis methods.  
We ask that participants provide us with a brief 
description of the analysis techniques they have used.
  We will publish a summary 
of the results, indicating which methods of foreground 
subtraction and parameter estimation appear to be the most successful 
at present and how well they work.  
Participants may choose to remain anonymous in our presentation 
of the results.

\section{Discussion}

One of the biggest challenges in real-world observations is being prepared 
for surprises, both instrumental and astrophysical 
(see 
\citealt{scott98}
for an eloquent discussion).  
An exercise 
such as the WOMBAT Challenge 
is an excellent way to simulate these surprises, and we have  
included a few in our skymaps.  
We hope that the 
 results of the WOMBAT Challenge will provide estimates of the 
 effectiveness of  current 
techniques of foreground subtraction, power spectrum analysis, and 
parameter estimation.  

It is unclear 
how close the community 
is to being able to handle datasets as large as that of MAP (10$^6$ 
pixels at 13$'$ resolution for a full-sky map).  
Given current computing power, complex algorithms 
appear necessary for analyzing full-sky MAP datasets 
\citep{ohsh99}, 
although simpler approximations may 
be possible 
\citep[e.g.][]{wandelthg98}.  
Even the algorithm of \citeauthor{ohsh99} has yet 
to be tested on the slightly correlated instrument noise which 
will be present in the real MAP datasets (our simulations contain 
uncorrelated noise so they will not alleviate this concern).

Undoubtedly the most important scientific contribution of WOMBAT 
is the production of realistic full-sky maps of all major microwave 
foreground components with estimated uncertainties.  These maps are needed 
for foreground subtraction and estimation of residual foreground contamination 
in present and future CMB anisotropy observations.  They will allow 
instrumental teams to conduct realistic simulations of the observing and 
data analysis process without needing to assume overly idealized models 
for the foregrounds.  By combining various realizations of these 
foreground maps within the stated uncertainties with a simulation of 
the intrinsic CMB anisotropies, we have produced the best simulations 
so far of the microwave sky.  Using these simulations in a ``hounds and 
hares'' exercise should test how 
well the various foreground subtraction and parameter estimation techniques
work at present.  It is easy 
to question 
the existing tests of analysis methods which assume idealized foregrounds 
in analyzing similarly idealized simulations.

Data analysis techniques will undoubtedly improve with time, and we hope 
to reduce the current uncertainty in their efficacy 
such that follow-up simulations 
by the instrumental teams themselves can generate confidence in 
the results of real observations.  
We can test the resilience of CMB analysis methods to surprises such as 
unexpected foreground 
amplitude or spectral behavior, correlated instrument noise, and 
CMB fluctuations from non-gaussian or non-inflationary models.  
Cosmologists need to know if 
such surprises can lead to the misinterpretation of cosmological 
parameters.  
  In the future, we envision producing 
time-ordered data, simulating interferometer observations, and adding 
polarization to our microwave sky simulations.  

Perhaps the greatest advance we offer is the ability to evaluate the 
importance of studying the detailed locations of foreground sources.  
If techniques which ignore this phase information are still successful 
on our realistic sky maps, that is a significant vote of confidence.  
Alternatively, it may turn out that techniques which use 
phase information are needed in order to reduce foreground contamination 
to a level which does not seriously bias the estimation of cosmological 
parameters.  Combining various techniques may lead to improved 
foreground subtraction methods, and we hope that a wide variety of 
techniques will be tested by the participants in the WOMBAT Challenge.

\cleardoublepage

%\cleardoublepage
\nopagebreak

\cleardoublepage

%\cleardoublepage
\nopagebreak

\cleardoublepage
\chapter{Low-redshift Infrared Galaxies}
\label{chap:ir}
 %\documentstyle[12pt,aasms4]{article}

%\begin{document}

%\title{Contribution of Extragalactic Infrared Sources
%     to\\ CMB Foreground Anisotropy}

\section{Motivation}

	The COBE detection of large-angular scale anisotropy 
in the Cosmic Microwave Background radiation 
\citep{smootetal92} 
has sparked a drive 
to measure the anisotropy on smaller angular scales with the goal 
of determining crucial information about the density and 
expansion rate of the universe, the 
nature of dark matter, 
and the spectrum of primordial density perturbations.  
Current anisotropy observations look at  sub-degree angular scales 
which correspond to observable structures
in the present universe.  Improved instrumentation
and the MAP (Microwave Anisotropy Probe) and Max Planck Surveyor
(formerly COBRAS/SAMBA) satellite missions focus attention on 
angular scales between
one-half and one-sixth of a degree.

Due to its large beam size, COBE was basically unaffected by 
extragalactic foreground sources \citep{bandayetal96, kogutetal94}.
Because the antenna temperature 
contribution of a point source increases with the inverse 
of the solid angle of the beam,
observations at higher 
angular resolution 
are more sensitive to extragalactic foregrounds, including 
the low-redshift infrared-bright galaxies examined
here.  At frequencies above 200 GHz, these infrared galaxies are the 
dominant extragalactic foreground.  Predictions and simulations for 
high-redshift infrared galaxies were presented in 
Section \ref{sect:wombat_simulations} and limits on their 
anisotropy will 
be discussed in Chapter \ref{chap:sources}.  
   
	Previous work in this area 
\citep{toffolattietal95, franceschinietal89, wang91}
 used galactic evolution models with specific 
assumptions about dust
temperatures to predict the level of extragalactic foreground.  
We choose instead a phenomenological approach using 
the infrared-bright galaxies 
detected by the Infrared Astronomical Satellite (IRAS) and the Galactic 
emission detected by the COBE (Cosmic Background Explorer) satellite. 
Section \ref{sect:ir_results} 
 compares our results with those from galaxy-evolution models.  

	The FIRAS (Far-Infrared Absolute Spectrophotometer) instrument of COBE
gives evidence for the existence of 
Cold ($<15$K) Dust in the Galactic plane \citep{reachetal95}.  
If the Milky Way has Cold Dust, then it is likely present in 
other dusty spirals, which comprise the majority of bright 
low-redshift infrared sources.  Some observations 
\citep{chinietal95, blocketal94, devereauxy92}
indicate the presence of Cold Dust in other
galaxies.  Neither galactic evolution models nor pre-FIRAS observations 
\citep[see][]{ealeswd89}
were able to set tight constraints on emission from Cold 
Dust, but the FIRAS observations do.  
Emission from dust close in temperature to 
the 2.73 K background radiation
is difficult to separate from real CMB anisotropies. 
If Cold Dust in spiral galaxies 
is typically accompanied by  
the Warm Dust to which IRAS is sensitive, 
we can use
the FIRAS information about the total dust emission 
spectrum of the Galaxy to overcome
this spectral similarity and learn about the amount of Cold Dust in 
a galaxy by 
measuring its amount of Warm Dust.

\section{Extragalactic Infrared Sources}	

	The far-infrared discrete sources 
detected by IRAS are typically inactive spiral 
galaxies, although some are quasars, starburst galaxies, and Seyfert galaxies.
  The IRAS 1.2 Jy 
catalog \citep{fisheretal95}
provides flux measurements of 5319 galaxies at 12, 25, 60, and 100 $\mu$m, 
where interstellar dust emission is dominant.  
We compared the locations of these galaxies 
with those of a thousand of
the brightest radio sources, and only 7 possible coincidences resulted.  
This lack of coincidence shows that radio-loud galaxies can
be treated separately (see Chapter \ref{chap:radio}).  
The IRAS sources are roughly isotropic in distribution, except for
a clear pattern of the Supergalactic Plane.               
To reduce the
possibility of residual galactic contamination, we restrict our analysis
to  galactic latitude $|b| > 30^{\circ}$, which includes  
contributions from 2979 galaxies for a $0\fdg5$ beam.\footnote{The dependence
on beamsize is mild, but for a larger beam more sources centered outside 
of this area are convolved into it.}  

The nature of dust in spiral galaxies is still an open question.
It seems likely 
that there is dust at widely 
varying temperatures and possibly with different emissivities 
\citep{rowan-robinson92, franceschinia95}.
Attempts to fit observational data have yielded a variety of 
results; it is unclear if far-infrared luminous dust is well described by 
a one-component or a two-component model, and the emissivity power-law 
index is 
only known to be between 1 and 2.  We avoid specifying 
the nature of this dust by using the observed Galactic far-infrared emission
spectrum as a template for IRAS galaxies.
To check the accuracy of this template, we fit 
a two-component dust model to IRAS galaxies and
to the integrated 12, 25, 60, and 100 $\mu$m 
fluxes of the Milky Way measured by the DIRBE (Diffuse Infrared Background
Experiment) instrument of COBE.  This produces similar 
results for the Warm (15-40 K) Dust component to which IRAS and 
DIRBE are most 
sensitive; for an emissivity power-law index of 1.5, DIRBE gives
a Warm Dust temperature of 28K for the Milky Way, while the 425 
IRAS galaxies with highest-quality flux measurements are 
collectively fit to a Warm Dust temperature of 33K.  This Warm Dust 
accounts for the majority of the far-infrared emission of spiral galaxies.

\begin{figure}
\centerline{\psfig{file=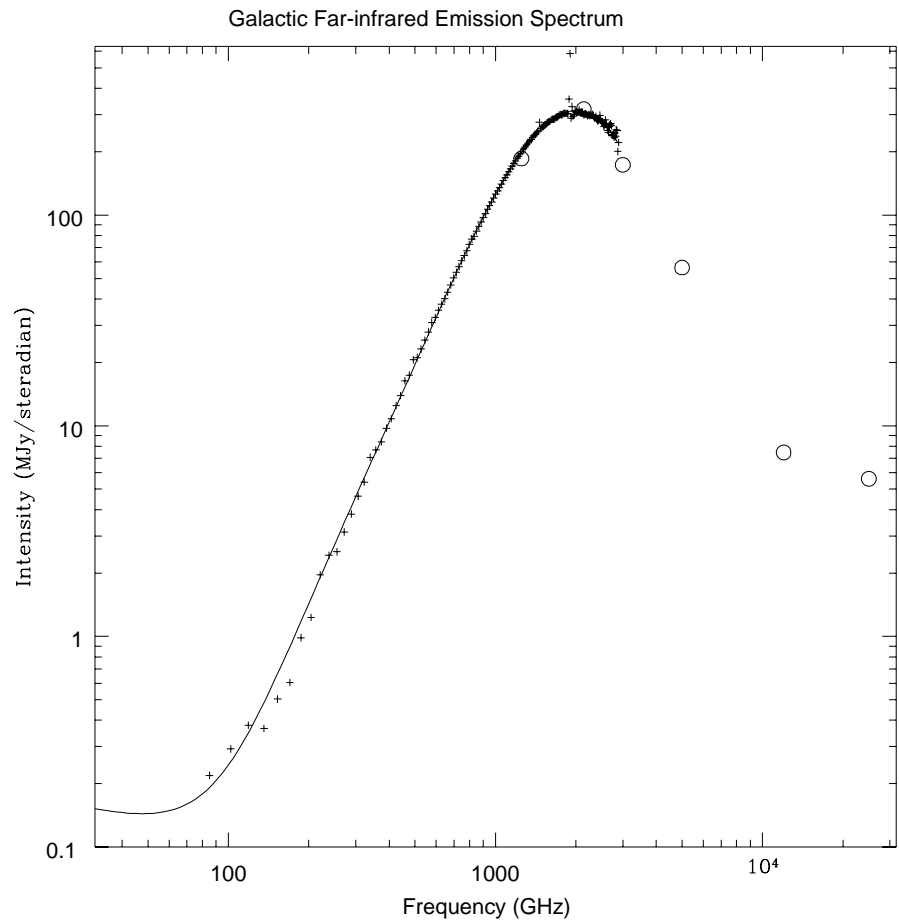,width=5in}} 
\caption{Galactic Far-IR Emission Spectrum} 
\mycaption{The FIRAS Galactic Dust spectrum, including emission lines, is 
shown by $+$ symbols.  
The smooth curve is a fit to this spectrum 
based upon a two-component dust model with synchrotron and free-free emission 
included using DMR results.  
The FIRAS error bars are not shown because they are
extremely small on this scale.  
The open circles are 
DIRBE integrated Galactic fluxes at 12, 25, 60, 100, 140, and 240 $\mu$m,
normalized to the FIRAS measurements.
}
\label{fig:iras_fig1}
\end{figure}

There is, however, observational evidence that the far-infrared emission
of inactive 
spirals is dominated by dust slightly colder than 20K 
\citep{neiningerg96, chinik93}.
Fitting the FIRAS 
spectrum of the Milky Way also leads to a Warm Dust temperature close to
20K.  These fits appear to conflict with the temperatures found above using
IRAS and DIRBE fluxes at $\lambda \leq 100 \mu$m.  
Using 
60, 100, 140, and 240 $\mu$m DIRBE fluxes, however, indicates a
Warm Dust temperature for the Galaxy of 24K.  This shows that temperature
fits to data on one side of 
the peak of a graybody (modified blackbody) spectrum can be 
inaccurate.  
Figure \ref{fig:iras_fig1} shows that
the spectra of the Milky Way found by DIRBE and FIRAS
are indeed
compatible.  It may be an oversimplification to represent the 
Warm Dust in a galaxy by a single temperature.

We recognize that 
not all IRAS galaxies have the same far-IR spectrum as the Milky Way.  
Active galaxies are warmer, 
with an average Warm Dust temperature of 33K (for emissivity index 2, 
\citealt{chinietal95}).
However, the
cirrus emission which dominates Galactic dust is consistent
with the emission from the majority of inactive spirals 
\citep{andreanif96, pearsonr96}.
Some observations indicate that our Galaxy is slightly warmer than 
the average inactive spiral \citep{chinietal95}.  None of these observations
includes enough frequencies to provide a 
template microwave emission spectrum, and
their results range by a factor of 3 depending on the choice of beam
corrections \citep{franceschinia95}.
The Milky Way is a 
good middle-of-the-road choice for a microwave template spectrum;
the 
DIRBE and IRAS dust 
temperature fits given above 
agree rather well.  For extrapolation to microwave frequencies, uncertainty 
in emissivity is of much greater importance than this level of temperature 
uncertainty, anyway.  

After removing Galactic emission lines (as in \citealt{reachetal95}), 
we fit a two-component dust model to the 
FIRAS dust spectrum.  The CO 1-0 emission line at 115 GHz is not clearly
detected by FIRAS but could be responsible for increased
emission at that frequency. 
With a $\nu^2$ emissivity law
assumed, the best fit is Warm Dust at $19.4$K and Cold Dust at $4.3$K with an 
optical depth 12.1 times that of the Warm Dust.  
It is possible to vary the parameters of the 
dust model significantly and still have an acceptable fit, so we refrain 
from assigning any physical importance to the parameters of the fit.  
We add synchrotron and free-free components with microwave-range
spectral indices of 
$-1.0$ and $-0.15$, 
respectively, so that these sources of microwave emission match COBE DMR
(Differential Microwave Radiometer)
observations below 100 GHz 
\citep{kogutetal96a, reachetal95, bennettetal92}.
Free-free emission is stronger than dust beyond the low-frequency end
of the FIRAS spectrum.     

We combine data from DIRBE, FIRAS, and DMR to form the  
broad Galactic spectrum shown in Figure \ref{fig:iras_fig1}.  
Each IRAS 1.2 Jy source is fit to 
the DIRBE end of the spectrum and extrapolated to the desired frequency using 
this template. 
In fitting each IRAS galaxy to the DIRBE fluxes of the Milky Way,
we give more weight to the 60 and 100 $\mu$m fluxes, which are most 
sensitive to Warm Dust, than to the 12 and 25 $\mu$m fluxes, 
which are also sensitive to Hot (100-300 K) Dust.  
The 1.2 Jy catalog gives redshifts for these galaxies.  
Most have $z < 0.05$ and all have $z < 0.3$. 
We take these redshifts into account while fitting 
and extrapolating.

It would be advantageous to 
fit each type of galaxy
to a specialized far-IR to microwave spectrum, but no other 
trustworthy template spectrum is currently available, 
so we use the Galactic far-infrared emission spectrum for all sources.
The Galactic spectrum agrees well with observed correlations between radio
and IR fluxes of IRAS galaxies 
\citep{condonb91, crawfordetal96}.
Our template spectrum is consistent 
with detections and 
upper limits for bright infrared galaxies
from DIRBE \citep{odenwaldns98}.  
This is helpful because DIRBE used 140 and 240 $\mu$m
channels, which IRAS lacks, allowing it to probe much cooler
dust temperatures than IRAS.
DIRBE rules out the 
possibility of extremely bright sources occurring in the 2\% of the high 
Galactic latitude sky not surveyed by IRAS and sees no evidence for 
sources whose emission comes predominantly from Cold Dust.

\section{Results}

\label{sect:ir_results}

We use the Galactic far-infrared emission spectrum to predict the 
microwave flux of each IRAS galaxy in Jy ($1$ Jy $= 10^{-26} W/m^2/$Hz).  
To convert from flux $S$ to antenna temperature $T_A$, we use

\begin{equation}
\label{eq:flux_to_temp}
 T_A = S \frac{\lambda^2}{2 k_B \Omega}\; \;,  
\end{equation}

\noindent 
where $k_B$ is Boltzmann's constant, $\lambda$ is the wavelength, and 
$\Omega$
is the effective beam size of the observing instrument.  Antenna temperature
is related to thermodynamic temperature by

\begin{equation}
\label{eq:t_to_ta}
 T_A = \frac {x}{e^x - 1} \; \; T,  
\end{equation}

\noindent
defining $x \equiv h \nu / k T$.  Small fluctuations in antenna 
temperature can be converted to effective thermodynamic 
temperature fluctuations using

\begin{equation}
\label{eq:dt_to_dta}
 \frac{dT_A}{dT} = \frac { x^2 e^x} {(e^x - 1)^2} \; \; .  
\end{equation}

Analysis of source counts indicates that the 1.2 Jy sample is complete down to 
an extrapolated flux of 3 mJy at 100 GHz.  We divide 
the sources logarithmically 
into groups of similar flux and 
find a gradual 
decrease in anisotropy as flux decreases, indicating that dimmer sources
will not generate significant anisotropy.  This may not hold true 
for high-redshift infrared galaxies, however, as the k-correction makes 
sources which are dim at 100$\mu$m quite bright in the sub-millimeter.  
\citet{toffolattietal95}
found a negligible contribution from non-Poissonian fluctuations.  
Poissonian fluctuations should
be dominated by those sources prevalent enough to have roughly one source
per pixel.  
If all sources have roughly the luminosity of the Milky Way, then 
for an instrument with a resolution of $10'$ to have one source
per beam, we must look at sources with $ z \simeq 0.24$.  
Assuming $(1+z)^3 $ luminosity evolution and including 
k-correction  
(see \citealt{pearsonr96, beichmanh91}), 
these sources will generate a 
temperature anisotropy only 2\% of that caused by IRAS 1.2 Jy galaxies.
We therefore expect the anisotropy generated 
by low-redshift 
sources too dim to make the 1.2 Jy catalog to be a small part of the total
anisotropy; the brightest sources are generating most of the fluctuations.

\begin{figure}
\centerline{\psfig{file=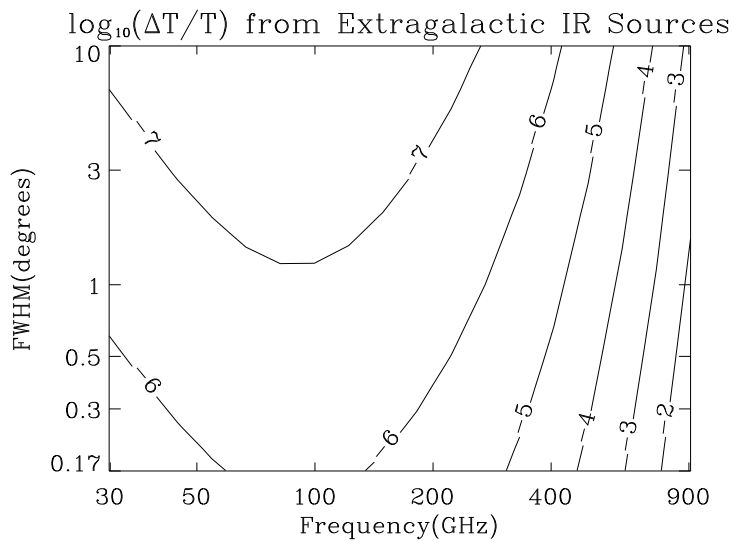,width=4in,angle=90}} 
\caption{Microwave anisotropy from low-redshift infrared galaxies.}
\mycaption{A log-log-log contour plot 
of equivalent thermodynamic temperature fluctuations  
due to extragalactic infrared sources as a function of frequency in GHz 
and angular resolution (FWHM) in degrees.  
The temperature anisotropy shown is $\log_{10} \frac{\Delta T}{T}$
where $\Delta T$ is the root mean square equivalent thermodynamic
temperature generated
by extragalactic infrared sources in Kelvin and $T$ is the 
temperature (2.73K) of the CMB.  The increase in anisotropy at low frequencies
occurs because synchrotron and free-free emission are included in our 
template spectrum. 
}
\label{fig:iras_fig2}
\end{figure}

	To simulate observations, we 
convolved all sources on pixelized skymaps (2X oversampled) of
resolution varying from $10'$ to $10^{\circ}$. 
The resulting maps, covering a range of frequencies from 30 to 900 GHz, 
were analyzed to 
determine the expected contribution of IRAS galaxies to foreground 
confusion of CMB temperature anisotropy.
The information contained in these skymaps can be used 
to choose regions of the sky in which 
to observe \citep{smoot95}.  
The contour plot in 
Figure \ref{fig:iras_fig2} shows the 
rms thermodynamic temperature anisotropy
produced by extragalactic infrared sources 
over the full range of frequencies and instrument resolution.  
The minimum value of $\frac{\Delta T}{T}$
is $1.3\times 10^{-8}$ at large FWHM and medium frequency 
and the maximum value is $0.092$ 
at small FWHM and high frequency.  For frequency in GHz and FWHM in degrees, 
our results for temperature
anisotropy are fit to within 10\% by
\begin{equation}
\log_{10}\frac {\Delta T}{T} = 2.0 (\log_{10}\nu)^3 - 8.6 (\log_{10}\nu)^2
+ 10.3 \log_{10}\nu - 0.98 \log_{10} (FWHM) - 9.2 \; . 
\end{equation}
The inverse linear relationship between anisotropy and
FWHM results from the combined effects of beam convolving 
and map pixelization (see Chapter \ref{chap:sources} for a derivation).  
Anisotropy from extragalactic infrared sources dominates 
expected CMB anisotropy at frequencies above 500 GHz.  This makes effective
foreground discrimination possible for instruments with a frequency range 
sufficiently wide to detect the extragalactic infrared 
foreground directly.

\begin{figure}
\centerline{\psfig{file=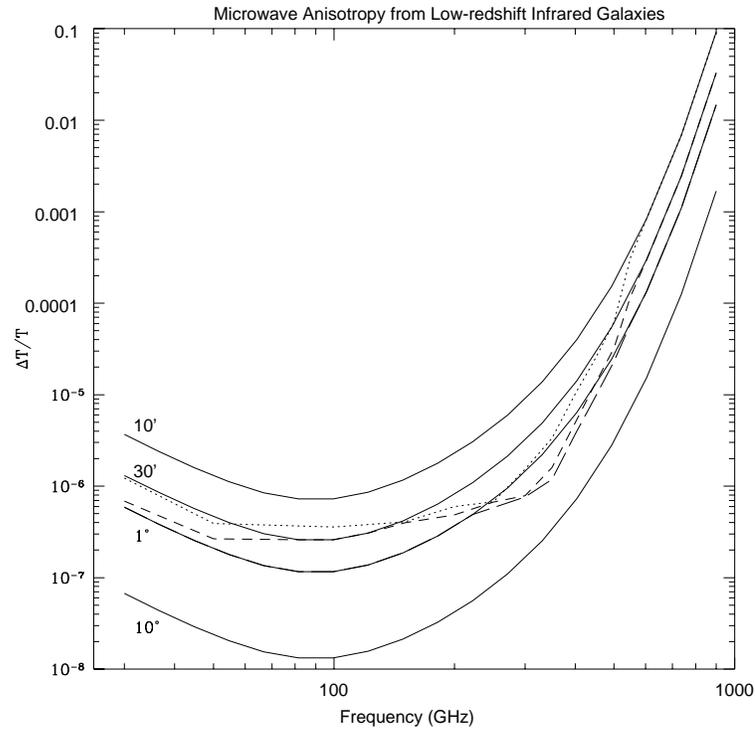,width=4in}} 
\caption{Confusion from low-redshift infrared galaxies}
\mycaption{Log-log plot of $\frac{\Delta T}{T}$
versus frequency for instrument resolutions of
$10'$, $30'$, $1^{\circ}$, and $10^{\circ}$,
showing window where foreground confusion should be
less than $10^{-6}$.  
Solid lines are for no pixel subtraction.  The 
dotted, dashed, and long-dashed lines show the results 
with pixels at a level of $5\sigma$ removed for resolutions of 
$10'$,$30'$, and $1^{\circ}$, respectively.  This $5\sigma$ subtraction 
makes no difference at any frequency for $10^{\circ}$.  
}
\label{fig:iras_fig3}
\end{figure}

Figure \ref{fig:iras_fig3} 
shows a summary of our results for several benchmark instrument
resolutions.  The dashed lines represent 
the results of subtracting pixels where
the fluctuations from extragalactic infrared sources are five times
times greater than the quadrature sum of the rms CMB anisotropy and the 
expected
instrument noise for the Planck Surveyor at that frequency.
These $5 \sigma$ pixels can be
assumed to contain bright point sources. 
Our results agree closely with those of \citet{toffolattietal98}
for their
model of moderate cosmological evolution of all galaxies.  
Our predictions for anisotropy are about a factor of three 
lower than those
of \citet{franceschinietal89},
who assume strong evolution of the 
brightest IR sources and include early galaxies with heavy starburst activity. 
\citet{wang91}
ignores the possibility of cold dust and uses galaxy evolution 
models to predict anisotropy levels somewhat lower than those found 
with our phenomenological approach.  

The $5 \sigma$ subtraction has a significant  
effect for small FWHM at frequencies below 500 GHz.   
The maximum effect is to subtract 0.002\% of the pixels, 
leading to a factor of 5 reduction in foreground 
temperature anisotropy.  
This is
further 
evidence that temperature anisotropy from extragalactic infrared sources 
is dominated by the brightest sources.
  The bright sources are a mixture of Local Group galaxies and 
more distant infrared-luminous galaxies such as starburst galaxies.  Optimal
subtraction of the extragalactic infrared foreground requires the contribution
from each bright source to be predicted accurately.

\begin{figure}
\centerline{\psfig{file=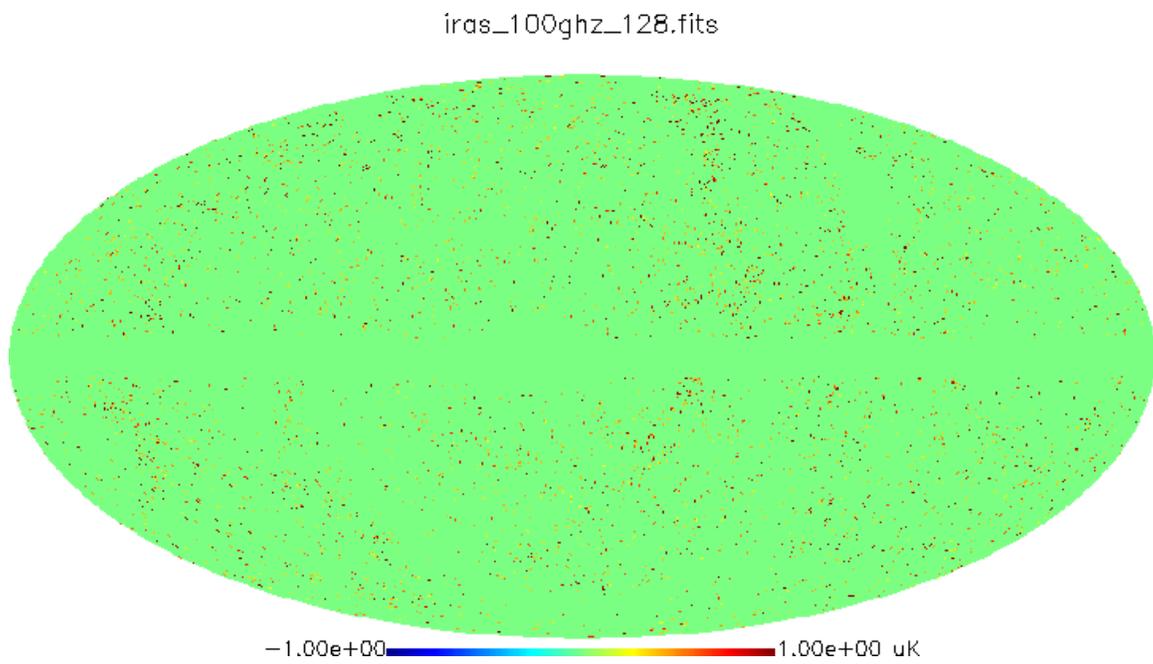,width=6in,angle=90}} 
\caption{
Skymap at 100 GHz of predicted contribution from IRAS 1.2 Jy sources.}
\label{fig:iras}
\end{figure}

Figure \ref{fig:iras} shows a  
skymap of our extrapolated IRAS 1.2 Jy catalog at 100 GHz.  
Very few of 
these sources are significant at frequencies below 200 GHz, although the 
entire 5319 sources should be detected by Planck HFI due to its high 
resolution at high frequencies.  
The skymap shows the structure of the Supergalactic Plane, 
as well as the regions of the sky not included 
due to Galactic contamination or the IRAS satellite's failure to observe them.
  Several 
sources are considerably brighter than the maximum of the color 
table, which has been set 
very low to show all of the catalog, even though most sources 
are quite dim at 100 GHz.
 
\section{Discussion}

Our usage of the Galactic far-infrared emission spectrum as a template
causes systematic errors on a galaxy-by-galaxy basis.  
Our method can be improved in the future 
to account for the spectral difference 
between Ultraluminous Infrared Galaxies and normal spirals.  
It is easy to place constraints on our results; if all
galaxies had only 33K dust as is typical for active galaxies, the resulting 
anisotropy would be a factor of 100 lower.  This is highly unlikely, because we
know that most IRAS galaxies are inactive spirals, and galaxies
with colder dust will dominate the anisotropy at mm-wavelengths because 
of the selection effect favoring sources with flatter spectra.  A robust upper
limit on microwave anisotropy from infrared galaxies can be set by assuming
that these IRAS 1.2 Jy galaxies cause the full cosmological far-infrared
background 
\citep{pugetetal96, buriganap98, fixsenetal98, schlegelfd98, 
hauseretal98}.  
In this case we have underestimated
the anisotropy by a factor of 100, but no predictions of the IR background
expect these nearby galaxies to produce more than a few percent of it.  
A more realistic check on our results comes from Andreani \& Franceschini
(1995), who measured a complete sample of IRAS galaxies at 1300 $\mu$m 
(240 GHz).  Their average flux ratio of 1300 $\mu$m over 100 $\mu$m is half
that of the Galaxy, but one of their beam correction methods brings their ratio
into agreement with the Milky Way.  They find that the 60 $\mu$m 
emission of spiral galaxies receives enough contribution from a starburst
dust component mostly absent in the Galaxy that including 60 $\mu$m fluxes
in our fits may have caused a factor of 2 overestimate.  
Combined, these corrections
give us a possible systematic overestimate of anisotropy by a factor of 4.  If
typical IR-bright galaxies have dust colder than the Milky Way, our
results could instead be an underestimate by a factor of a few, but this
appears less likely. 

We estimate an overall systematic uncertainty of a factor of 2 at 300 GHz, 
increasing to a factor of 3 at 100 GHz and to a factor of 5 at 30 GHz.  
As it is hard to predict the typical dust temperature and emissivity in 
a given galaxy as well as its relative amount of free-free and synchrotron 
emission, we estimate that our predictions for the microwave 
spectrum of each source have an overall factor of 5 uncertainty 
and an independent  
factor of 1.3 uncertainty at each frequency.  
The factor of 5 uncertainty preserves a source's spectral shape whereas the 
factor of 1.3 uncertainties allow for errors in the predicted 
spectral shape.  These uncertainties are quite large, but the extrapolation 
we have performed is over a factor of 10-100 in frequency.  We expect 
that forthcoming microwave observations will give us better information, 
especially about the brightest IRAS 1.2 Jy sources.

	The recently obtained spectral knowledge of our Galaxy has enabled
us to take into account the possible 
presence of Cold Dust.  
Our predicted level of temperature anisotropy makes the extragalactic
foreground from low-redshift infrared galaxies 
dominant over the Galactic foregrounds
of dust, free-free, and synchrotron for angular resolutions 
near $10'$ and frequencies above 100 GHz.  
Below 150 GHz, radio sources are
expected to be the dominant extragalactic
foreground.  
The extragalactic low-redshift 
infrared foreground will not be significant in comparison to
CMB anisotropies around 100 GHz but will be dominant above 500 GHz.  
  Despite the possible presence of Cold Dust in infrared-bright galaxies,
our results leave a window at intermediate frequencies
for the measurement of 
CMB anisotropies without significant confusion
from extragalactic infrared sources.

%\begin{figure}
%\centerline{\psfig{file=5sigma.ps}, width=5in} 
%\caption{
%A log-log surface plot of the number of pixels reaching the $5\sigma$ 
%level due to IRAS point source emission as a function of frequency 
%and instrument resolution.}  
%\label{fig:5sigma}
%\end{figure}

\cleardoublepage
  
%\end{document}

\cleardoublepage
\chapter{Radio Galaxies}
\label{chap:radio}
 %
%\documentstyle[12pt,aasms4,psfig]{article}

%\input epsf.sty

%\begin{document}

%Last modified by A. Sokasian around 1/29/98
%Last modified by E. Gawiser on 11/18/98

%\title{Contribution of Bright Extragalactic Radio Sources to
%      \\Microwave Anisotropy}

%\author{Aaron Sokasian
%\footnote{Current address:  Harvard-Smithsonian
%Center for Astrophysics, 60 Garden St., 
%Cambridge, MA  02138}$^,$\footnote{asokasia@cfa0.harvard.edu}
%}
%\affil{Department of Astronomy, Cornell University, 
%Ithaca, NY  14853
%}
%\author{Eric Gawiser\footnote{gawiser@astron.berkeley.edu} and 
%George F. Smoot\footnote{smoot@cosmos.lbl.gov}}
%\affil{Department of Physics, University of California, Berkeley, 
%and Lawrence Berkeley National Laboratory, Berkeley, CA  94720}
%\authoremail{asokasia@cfa0.harvard.edu}
%$^1$ Current address:  Harvard-Smithsonian Center for Astrophysics\\
%$^2$ asokasian@cfa.harvard.edu

%\begin{abstract}
%\end{abstract}

%\keywords{cosmic microwave background: data analysis, foregrounds -- 
%radio galaxies -- radio sources: extragalactic}

\section{Motivation}

High resolution CMB anisotropy observations are 
sensitive to extragalactic foregrounds, 
including the radio galaxies considered here, bright infrared 
galaxies (see Chapter \ref{chap:ir}), 
high-redshift infrared galaxies (see Chapter \ref{chap:sources}),
and the Sunyaev-Zeldovich effect 
from galaxy clusters 
(see Chapter \ref{chap:sz})
. 
Estimates of extragalactic foreground confusion 
are critical as many ground-based, balloon-borne, and 
satellite experiments (MAP, Planck Surveyor) 
plan to study CMB anisotropies at angular scales
from 5$'$ to $30'$, and preliminary anisotropy results are already
available (see Section \ref{sect:obs_cmb}).

To evaluate the impact of known 
radio sources on CMB anisotropy observations, 
we use flux data from a variety of catalogs 
(see Section \ref{sect:radio_catalog}),
including recent measurements, 
to construct models of source spectra as a function of frequency.  
We analyze simulated skymaps at frequencies from 
10 to 300 GHz to determine the expected contribution of 
radio galaxies to foreground 
confusion of CMB temperature anisotropy.  
This information will be useful when choosing frequencies and 
regions of the sky to observe CMB fluctuations on small angular scales.
This work represents a significant improvement over previous efforts 
\citep{toffolattietal98, toffolattietal95, franceschinietal89}
which depended upon galactic evolution models
to predict the contribution of simulated 
radio sources at microwave frequencies.  Our
catalog contains detailed observations of known sources 
and hence can be used to make a spatial template 
for masking out their emission, and we believe that this 
phenomenological approach will lead to greater accuracy in predicting source
counts and the overall level of foreground anisotropy.

\section{Our Catalog}
\label{sect:radio_catalog}

	The discrete radio sources used in this project were compiled 
from a large number of separate samples.  
Our current catalog includes flux 
measurements and their corresponding errors at multiple frequencies for 
2207 sources.  
We have focused our attention on obtaining all 
available radio observations at millimeter and sub-millimeter wavelengths, 
resulting in 5766 observations of 758 different sources at 90 GHz, 
890 observations of 229 different sources from 100-200 GHz, and 
2628 observations of 309 different sources at frequencies above 200 GHz.  
The sources are roughly isotropic in distribution, 
except for a significantly greater number of sources 
in the northern celestial hemisphere 
due to the anisotropic distribution of radio telescopes on Earth with 
high-frequency capability.  
In addition, there are noticeably fewer observations 
within $10^{\circ}$ of the 
galactic plane and the celestial north pole due to the difficulty of 
observing extragalactic radio sources in those locations.  
  
Our catalog includes the full-sky 5 GHz-selected 
1 Jy sample of K\"{u}hr et al. (1981).  We add  
high-frequency ($>90$ GHz) 
measurements 
\citep{steppeetal88, steppeetal92, steppeetal95, 
tornikoskietal96,
kreysa98,
antonucciba90,
beichmanetal81,
chinietal89,
edelson87,
gearetal94,
holdawayor94,
knappp91,
landauer80, landauetal83, landauetal86,
lawrenceetal91,
nartalloetal98,
owenetal78, owensc80,
stevensrh96,
chandler95,
vla95}
and centimeter-wavelength observations  
\citep{herbigr92, 
patnaiketal92,
wirenetal92,
stanghellinietal97,
perley82,
alleretal85,
vla97} .
An updated version of the catalog 
will be described in detail by
\citet[][hereafter GSS]{gawiserss99}
. 

\section{Spectral Fitting}
Some 
extragalactic radio sources have complex spectra which cannot be 
approximated by simple functional forms due to emission from 
both compact and extended structures which dominate at different 
frequencies.  
In most radio galaxies, 
the emission comes from radio lobes located symmetrically around the core.  
The dominant emission mechanism, synchrotron, 
can be well approximated by a simple power law,
\begin {equation}
S\propto \nu^{-\alpha}
\end {equation}

\noindent with a flux spectral index, $\alpha$, typically between 0.5 and 
1.0 
\citep{plataniaetal98}
.  
Some radio sources have 
compact active nuclei which generate flat-spectrum radio emission.   
%These sources are often associated with quasars but are frequently found in 
%the centers of normal galaxies.  
The spectra of these sources can be  
inverted ($\alpha > 0$) for most of the 
radio frequency range due to self-absorption of the lower frequency emission.
Attempts to explain the observational 
data have yielded a variety of results.
The central engine of a typical active 
galaxy may consist of a supermassive black hole surrounded by an accretion  
disk and accelerating a jet of relativistic particles perpendicular to the 
disk plane 
\citep[e.g.][]{urryp95}
. 
\citet{boettcherrl97}
proposed 
a  model in which the inverted spectrum  of NGC 3031 is assumed to be the 
emission of a jet component, becoming optically thin to the radio emission 
of a monoenergetic pair plasma at decreasing frequencies as it moves outward 
and expands.

\begin{figure}[htb]
%\figurenum{1}
\centerline{\psfig{file=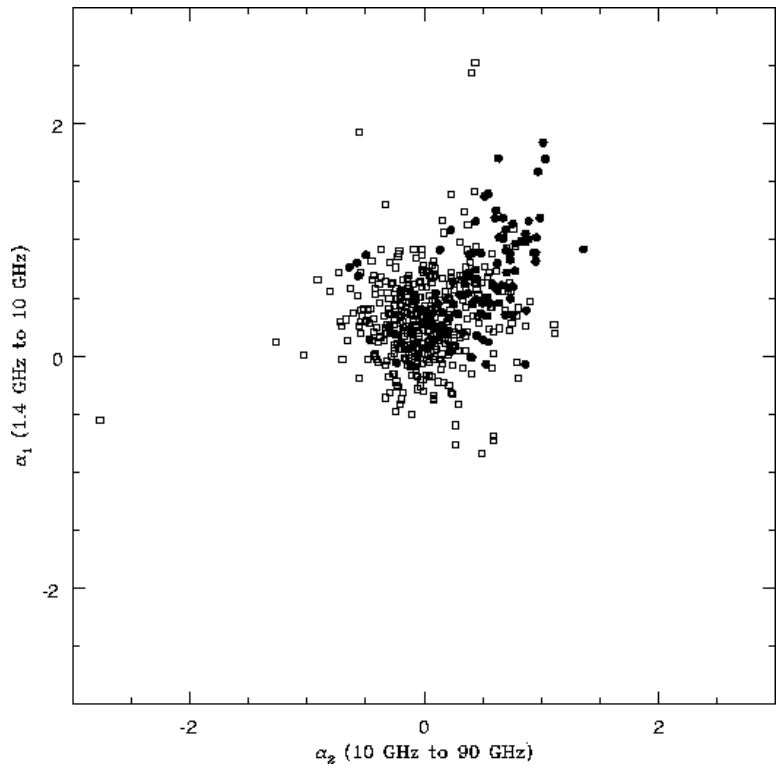,width=4in}}
\caption{
	Spectral indices $\alpha_1$ from 1.4 to 10 GHz and 
$\alpha_2$ from 10 to 90 GHz.   
Solid circles represent the brightest sources at 1.4 GHz; 
open squares represent dimmer sources at 1.4 GHz.  Note the lack
of clustering into distinct archetypal spectra.
}
\label{fig:radio_fig1}
\end{figure}

For sources which lack direct high-frequency observations, we 
avoid trying to determine the nature of the emission mechanism.  Instead, 
we use a phenomenological approach based on the expectation that the 
spectra of most radio sources approach power law behavior at 
frequencies higher than $\simeq 5$ GHz 
\citep{verschuurk88}
.
This power law may then be used to extrapolate the spectrum  
to typical CMB observation frequencies.  
To determine if template spectra could be used 
we use sources that have been measured near 1.4, 10, and 90 GHz and 
plot each source's spectral index from 1.4 GHz to 10 GHz versus its index
from 10 GHz to 90 GHz in Figure \ref{fig:radio_fig1}.  
There is a vague clustering of bright sources (circles)  
consistent with the notion that 
the brightest sources selected at low frequencies  
tend to have steep spectra.  The overall scatter of source spectra 
in Figure \ref{fig:radio_fig1} 
shows
that it is wrong to categorize radio sources into template spectra
or a narrow spectral index range.  This motivates us to  
 fit the spectra of each source individually.  
A previous phenomenological approach 
\citep{tegmarke96}
extrapolated 1.4 GHz source counts by assuming flat-spectrum emission for 
all sources. 
Our method has the advantages of using 
the actual source locations,
which can be turned into a template 
for masking the brightest pixels on the sky, and of choosing the 
spectrum for extrapolation on a case-by-case basis.

To determine the frequency beyond which a power law (a line on a 
log-log plot) can be fitted to the 
spectrum of a given source, we use an iterative model 
which starts with the best-fit line to the three highest frequency data points
and repeatedly includes the next highest frequency data point to the set to
which it fits a line.  
The fitting stops when the reduced $\chi^{2}$ 
starts to get worse or becomes acceptable ($\simeq 1$).
 There is little evidence that inverted spectra are common
past 30 GHz 
\citep{steppeetal95, stanghellinietal97}, 
so we set the handful of inverted ($\alpha \leq 0$)
 high-frequency spectra in the 
catalog to flat ($\alpha = 0$) spectra.  Upon 
closer inspection, these inverted spectra appear 
to result from variable sources being observed at different epochs 
at different frequencies, and we find that most of the sources 
with $\alpha_2 \leq 0$ in Figure \ref{fig:radio_fig1} 
based on their mean 10 and 90 
GHz fluxes are better fit by an $\alpha \geq 0$ power-law when 
all observations are taken into account.  
The average high-frequency spectral index was 0.5
with 27\%  of the sources in our catalog 
having steep spectra ($\alpha > 0.75$), 
and 37\% having flat spectra ($\alpha < 0.25$).

\begin{table}
\label{tab:radio_tab1}
\caption{
%Table 1: 
Average Errors from Extrapolation.  The average 
extrapolation error is the mean of $\mid (S_P - S_O) / S_P \mid$ where 
S$_P$ is the predicted flux and S$_O$ is the observation.  
The average error factor is the mean of $S_P/S_0$.  
}

\begin{center}
\begin{tabular} {l|l|l|l}

\hline
$\nu$ tested & $\nu$ ignored & 
Avg. Extrapolation Error & 
Avg. Error Factor \\
%Add median factor, use reciprocal of this factor
\hline
90 GHz & $\geq$ 2 GHz & 209 \% & 2.5  \\
90 & $\geq$ 10 & 148 & 2.3  \\
90 & $\geq$ 20 & 133 & 2.0  \\
90 & $\geq$ 90 & 92 & 1.6  \\
150 & $\geq$ 90 & 94 & 1.5 \\
230 & $\geq$ 90 & 250 & 3.2 \\
\hline
\end{tabular}
\end{center}
\end{table}

To check the accuracy of this technique, we ran our extrapolation 
method on sources with observations at 90, 150, and/or 230 GHz  
while ignoring the observations above  
certain frequencies and then compared the measured fluxes  
with the extrapolated fluxes.  
The results (Table \ref{tab:radio_tab1}) show that the 
extrapolation method works best when there is at least one measurement 
at 20 GHz or greater, as expected since many spectra become power-law 
past 5 GHz.  Table \ref{tab:radio_tab1} 
shows that on average we overpredict the 
flux at 90 GHz by a factor of 1.6, even when measurements above 20 GHz 
are used.  However, the median such error factor is only a factor 
of 1.1 overestimate, so we have roughly an equal number of over- and 
under-estimates.  This is no longer the case at 230 GHz, where even 
the median error factor is 1.9; our extrapolation method is overestimating 
the typical flux due to flat spectra 
falling off to more typical synchrotron spectra at frequencies around 100 GHz
\citep{gearetal94}
.
It is difficult to predict how far this fall-off will last, as thermal 
emission from low levels of dust in these radio-bright galaxies are 
expected to dominate their spectra by 500 GHz, except for the BL Lacs 
which have flat spectra up to infrared wavelengths 
\citep{knappp91, chinietal89, landauetal86}
.
  We therefore 
only trust our extrapolation in the range that has been tested, up to 
a maximum frequency of 300 GHz.  As the radio sources that have been 
observed at 30-300 GHz were selected at lower frequencies for 
brightness and flat spectra, 
our errors are only good estimates for this type of radio sources.  
This selection effect is not a great concern, however, as 
those are exactly the type of radio sources that threaten CMB 
anisotropy observations.
When interpolation is required, we use a cubic spline which passes 
through the mean fluxes at the observed frequencies.  
We visually inspected all 2207 sources to check the algorithm and 
eliminate any serious errors or outliers.  

For planned CMB anisotropy experiments, an additional concern is that 
the flat-spectrum radio sources can vary by up to a factor of ten 
in flux since their 
emission comes from a compact, active core.  Typical variations occur 
on timescales of one month to one year.  Outbursts 
are seen first at the highest (most transparent) frequencies 
and gradually shift to lower frequency, but except for this 
effect the overall 
spectrum shape is often preserved for a decade or longer 
\citep{tornikoskietal93}.
We use the scatter in 
the observed fluxes of a source at each frequency to estimate the typical 
range of variability, which yields an error bar on the source's flux 
at that frequency about the mean of all observations.  Because the 
variations are not periodic, there is little more that can be done, unless 
sources are observed nearly simultaneously at higher resolution and nearby 
frequencies.  GSS looks at the issue of variability 
in detail, including the possibility of extrapolating long-term drifts 
in source flux to the next epoch of observation.  Radio sources 
are typically 4-7\% polarized,
and this polarization is variable 
\citep{nartalloetal98}
, so 
radio-source foreground subtraction will be an important consideration
 for CMB polarization 
observations as well.

\section{Results}

We use the fitted spectra to predict the 
microwave flux of each radio galaxy.  
Equations \ref{eq:flux_to_temp} and \ref{eq:dt_to_dta} give 
the conversion from flux to antenna temperature to thermodynamic 
temperature fluctuations.  
The intrinsic $\Delta T / T$ of the CMB found by COBE is $\simeq 10^{-5}$
and is expected to vary between that and $3 \times 10^{-5}$ due 
to the acoustic oscillations discussed in Chapter \ref{chap:overview} 
at the 
angular resolutions considered here.

An analysis of source counts shows no indication 
of incompleteness for the northern celestial 
hemisphere subset of our catalog 
down to an extrapolated flux of  1.0 Jy at 
90 GHz while the southern hemisphere is incomplete 
below 2.0 Jy at 90 GHz.  
For the purposes of statistical analysis we have concentrated on the 
northern hemisphere where we appear to have measurements of the 
200 brightest sources in the hemisphere.  
We cannot rule out the existence of an unrelated
population of sources peaking around 90 GHz which are not bright at lower 
frequencies, as 90 GHz observations have only been made for 
sources selected at frequencies below 10 GHz (this hypothetical
source population is limited in Chapter \ref{chap:sources}).  The brightest
sources will dominate the anisotropy unless they are masked, 
because uncertainty in their exact fluxes makes subtraction highly 
inaccurate.  After masking, 
the brightest remaining sources will dominate unless non-Poissonian 
clustering becomes appreciable.  
\citet{toffolattietal98}
have shown that
non-Poissonian clustering is not expected to make an important contribution
to the foreground anisotropy from radio sources.  

To simulate observations, we convolve  all sources on pixelized sky maps 
(twice oversampled) of resolution varying from 
$10'$ to $10^{\circ}$ at frequencies between 
10 and 300 GHz.  
To avoid underestimating the anisotropy and to reduce the 
possibility of residual galactic contamination, we use only the 
portion of each skymap which 
covers galactic latitudes $|b| > 30^{\circ}$ and corresponds to the northern
celestial hemisphere to produce estimates of $\Delta T / T$.

\begin{figure}[htb]
\centerline{\psfig{file=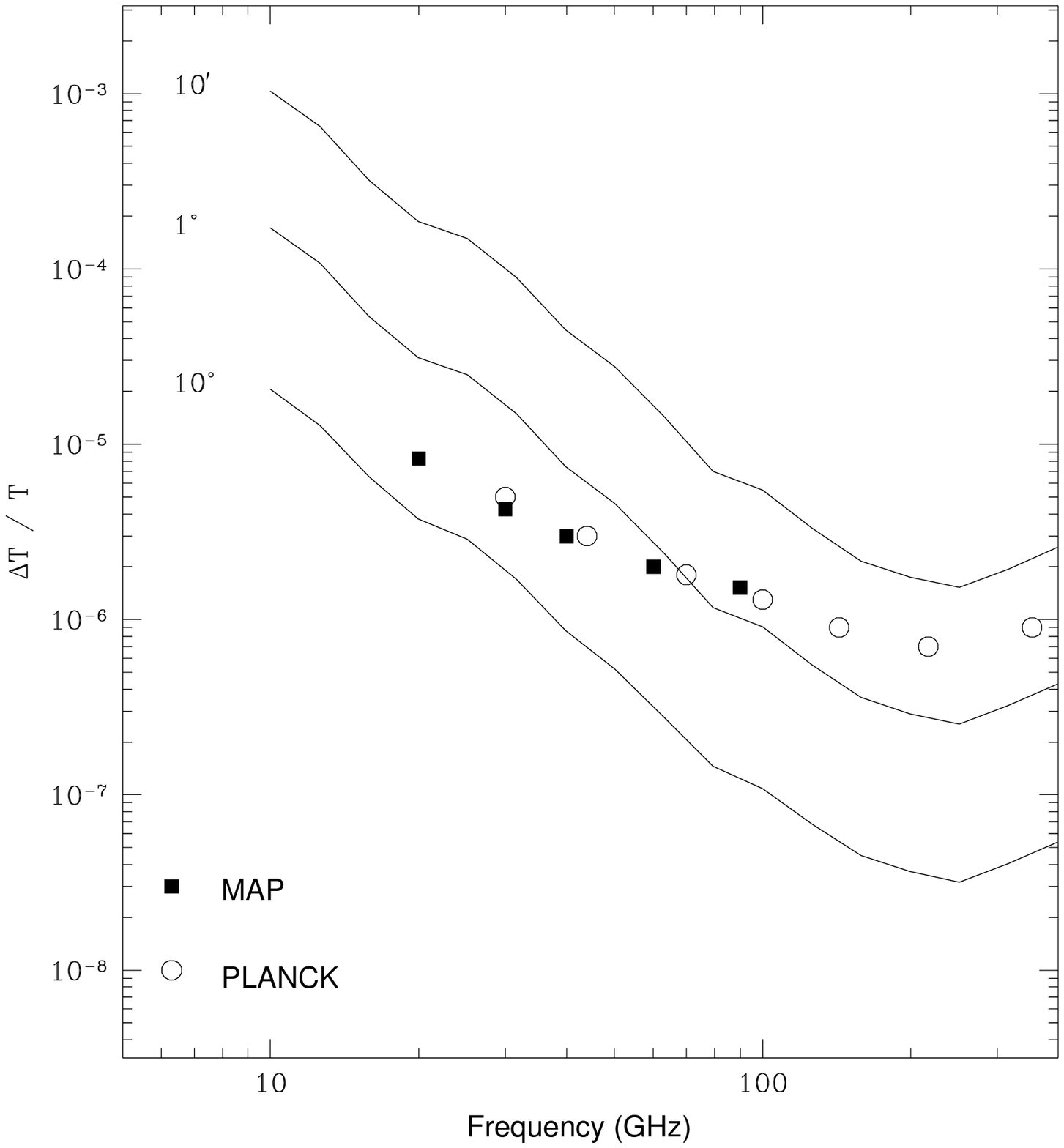,width=4in}}
\caption{Microwave anisotropy from radio sources}
\mycaption{
Plot of $\log_{10} \frac{\Delta T}{T}$ in the northern celestial 
hemisphere
(without pixel subtraction) versus frequency for instrument resolutions of
$10'$, $1^{\circ}$, and $10^{\circ}$,
showing window where foreground confusion should be
$\simeq 10^{-6}$.  
The rise beyond 200 GHz is 
caused by the exponential falloff in CMB antenna temperature beyond 100 GHz.  
The $5 \sigma$-source 
subtracted predictions for MAP (solid squares) and Planck (open 
circles) from Table \ref{tab:radio_tab2} are also shown.  
The $1/FWHM$ scaling will extend to other instrument resolutions but 
the $5\sigma$ source detection threshold is mildly instrument dependent.
}
\label{fig:radio_fig3}
\end{figure}

Figure \ref{fig:radio_fig3} 
shows a summary of our results for several relevant 
instrument resolutions.  
The inverse relationship between anisotropy and 
FWHM arises due to the combined effects of beam convolving and pixelization.  
The exact level of oversampling causes a small change
in the measured anisotropy, but the 1/FWHM behavior should hold for
extrapolation to smaller resolutions (see Chapter \ref{chap:sources}).  

We also analyze $\frac{\Delta T}{T}$ in the northern hemisphere 
based on only the 758 sources with 90 GHz measurements.  
The resulting rms 
$\frac{\Delta T}{T}$ at 90 GHz with a FWHM of $30'$ 
is $2\times 10^{-6}$ which dominates the anisotropy since the  
rms $\frac{\Delta T}{T}$ from extrapolating the spectra of 
the other 1449 sources 
amounts to only  $7\times 10^{-7}$.  
This indicates 
that we have flux measurements for the vast majority of bright 90
GHz radio sources, and 
there is typically less uncertainty in predicting the fluxes of 
sources which have already been observed at 90 GHz than in 
extrapolation.   
\citet{refregiersh98}
find that the $5 \sigma$ source detection limit for 
$0\fdg3$ MAP pixels will be 2 Jy at 90 GHz.  We have 108 sources 
in our catalog which have been observed to be brighter than 2 Jy at 90 GHz 
at least once, but only 42 sources have a weighted average flux that high, 
and a total of 52 sources are predicted to be brighter than 2 Jy at 90 GHz.  
We estimate that there will be 40-50 radio sources on the sky 
brighter than 2 Jy at 90 GHz.  At the 0.4 Jy level, 
\citet{toffolattietal98}
predict roughly twice as many sources as we do, but our 
prediction falls within their range of uncertainty.  
As the source counts we predict at this level based on the 
northern celestial hemisphere should be nearly complete, 
we recommend a slight recalibration of the galaxy evolution models 
used by \citet{toffolattietal98}, 
although a factor of two represents remarkable agreement 
for such different approaches.

\begin{table}
\label{tab:radio_tab2}
\caption{
Foreground Anisotropy in MAP \& Planck channels after source removal.  
Sources which contribute to the anisotropies at the $5 \sigma$ level 
or higher are considered detected and are  
removed by masking the pixels containing them.  No attempt 
has been made to use multi-frequency information or further prior 
information to detect and remove dimmer sources.  
}
\begin{center}
\begin{tabular} {r|c|l|c|l}

\hline
Frequency (GHz) & FWHM  & 
Source detection limit  &
\# detected &  
$\Delta T/T$ \\
\hline
MAP 20 & 56$'$ & 1.4 Jy & 186 & $8\times10^{-6}$     \\
30 & 41 & 1.2 & 216 & $4\times10^{-6}$   \\
40 & 28 & 0.9 & 265 & $3\times10^{-6}$  \\
60 & 21 & 1.1 & 168  & $2\times10^{-6}$  \\
90 & 13 & 1.0 & 161  &  $1.5\times10^{-6}$  \\
Planck 30 & 33 & 0.9 & 290 & $5\times10^{-6}$     \\
44 & 23 & 0.8 & 285 & $3\times10^{-6}$   \\
70 & 14 & 0.6 & 360 & $2\times10^{-6}$   \\
100 & 10 & 0.6 & 304  &$1.3\times10^{-6}$   \\
143 & 7 & 0.6 &  323  &$9\times10^{-7}$    \\
217 & 5  & 0.3 & 533 & $7\times10^{-7}$     \\
353 & 4.5 & 0.2 & 644 &$9\times10^{-7}$    \\
545 & 4.5 & 0.4 & 289 & $8\times10^{-6}$   \\
857 & 4.5 & 0.7 & 125  & $4\times10^{-4}$  \\
\hline
\end{tabular}
\end{center}
\end{table}

Table \ref{tab:radio_tab2} lists the expected level of anisotropy and the 
number of detected radio sources in MAP and Planck channels 
if this type of straightforward $5 \sigma$ source detection and subtraction 
is performed.  
Since the 90 GHz MAP channel will have a resolution close to $0\fdg2$ we 
expect a $5 \sigma$ source detection limit of 1 Jy at 
90 GHz.  
These detected sources represent a list of the few hundred 
brightest radio sources in the sky at each frequency.  
The anisotropy levels are shown in Figure \ref{fig:radio_fig3}.  
The level of source confusion drops if the 
brightest sources ($\geq 5 \sigma$) are subtracted. 
Table \ref{tab:radio_tab3}
 shows how the expected level of temperature 
anisotropy from radio sources varies with cutoff level, where we use 
our catalog to obtain prior information on which pixels are expected 
to contain sources at a given flux level and then mask those pixels.  
While all $5 \sigma$ pixels can be masked without such prior information 
if the CMB anisotropies are assumed to follow a Gaussian distribution, 
it is impossible to remove all $1 \sigma$ pixels without 
crippling the 
analysis.  
The actual improvements 
from masking all pixels expected to contain $1 \sigma$ sources may be 
less than indicated, unfortunately, due to the effect of incompleteness 
in our catalog.  GSS will attempt to fill in 
this incompleteness using full-sky PMN and GB6 catalogs at 5 GHz.  
If we settle
for making the southern celestial hemisphere as complete as the northern 
is now, we could create a mask for all sources expected to contribute 
at the 3$\sigma$ level, which is enough to make a significant reduction 
in radio source contamination versus 5$\sigma$ subtraction alone.

\begin{table}
\label{tab:radio_tab3}
\caption{Foreground Contamination in 13$'$ MAP channel at 90 GHz.   
This analysis 
assumes that our catalog is used to identify 
sources whose fluxes will be above the threshold and that 
the pixels
containing those sources are masked.
The results given are 
for the northern celestial hemisphere, where our catalog is 
estimated to be complete for the brightest few hundred sources, so the 
final line is likely an underestimate of anisotropy.    
}
\begin{center}
\begin{tabular} {c|c|c}

\hline
Threshold (Jy)& 
\# Sources above Threshold & 
$\Delta T / T$ \\
\hline
None & 0 &  $4.4\times10^{-6}$     \\
2 (10 $\sigma$) & 49 &  $2.1\times10^{-6}$   \\
1 ($5 \sigma$) & 161 &  $1.7\times10^{-6}$   \\
0.6 ($3 \sigma$) & 346 &  $1.2\times10^{-6}$  \\
0.2 ($1 \sigma$) & 940 &  $3.8\times10^{-7}$  \\
\hline
\end{tabular}
\end{center}
\end{table}

\section{Improving Sky Coverage with the PMN Survey}
\label{sect:radio_pmn}

In order to have a roughly isotropic distribution of radio sources, we 
have added the 1770 brightest sources from the Southern sky Parkes-MIT-NRAO 
catalog 
\citep{griffithw93, griffithetal94, griffithetal95, wrightetal94, 
wrightetal96}  
which were not already included in the SGS catalog.  These 
sources have fluxes measured at 4.85 GHz, and some have also been observed
at 2.7 GHz allowing for a rough measurement of their radio-frequency 
spectral index.   

\begin{figure}[htb]
\centerline{\psfig{file=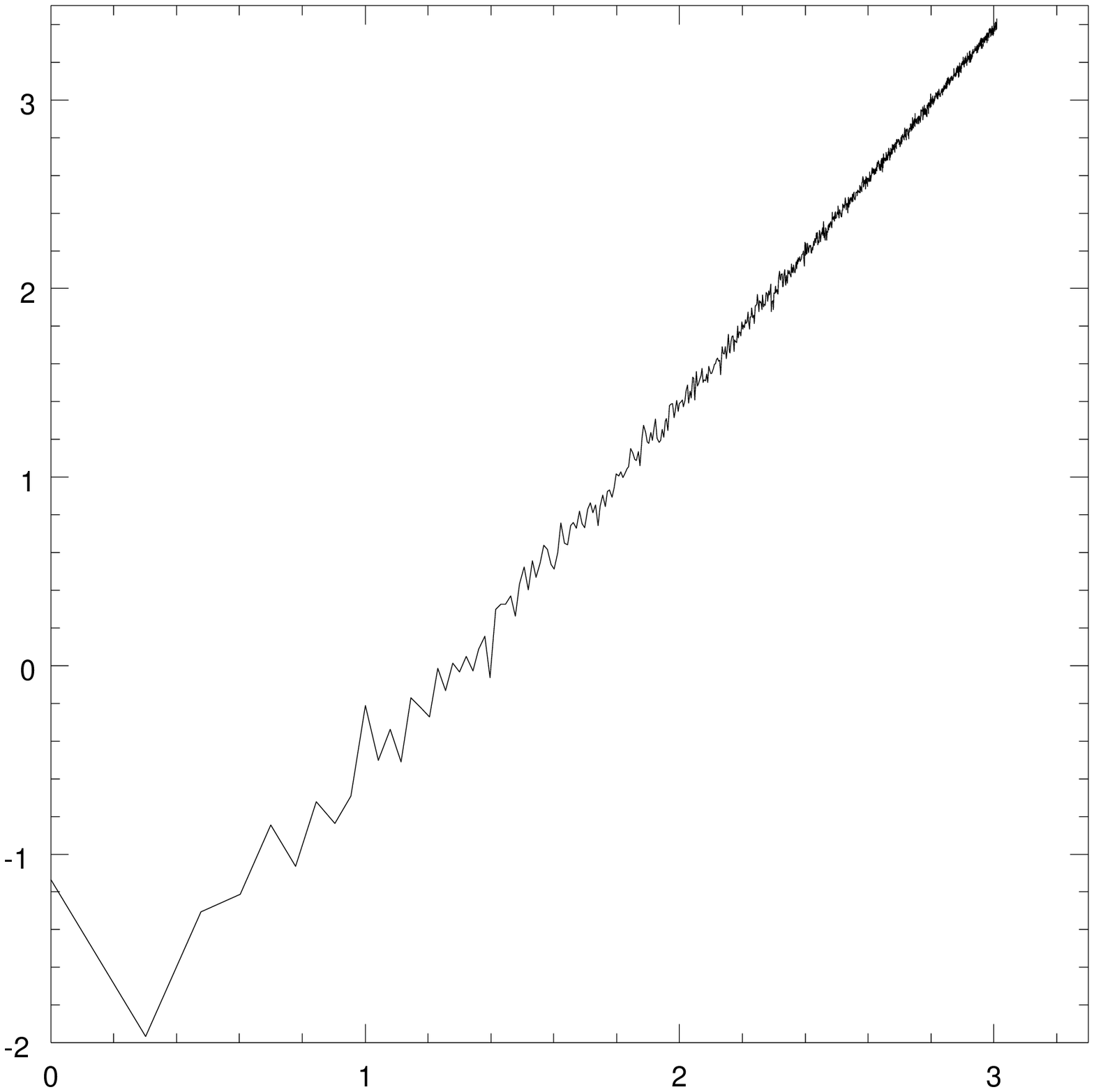,width=4in}}
\caption{Plot of predicted log$(\ell(\ell + 1)C_\ell)$ vs. log$_{10} \ell$
for the radio catalog extrapolated to 100 GHz and 
pixelized at HEALPIX level 1024 (3.4' pixels). 
}
\label{fig:radio_cl}
\end{figure}

\begin{figure}[htb]
%\figurenum{2}
\centerline{\psfig{file=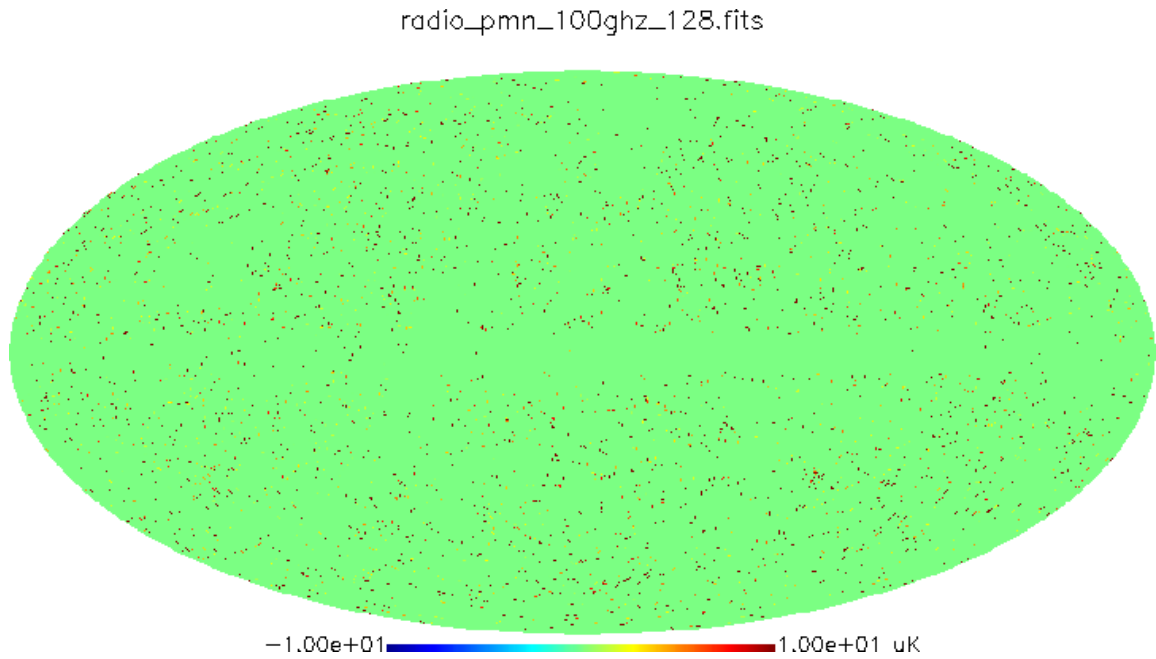,width=6.0in,angle=90}}
%\centerline{\psfig{file=radio_fig2.ps,width=7.0in,angle=90}}
\caption{
Sky map in Galactic coordinates 
of our catalog of radio sources (including PMN)  
extrapolated to 100 GHz and pixelized
at HEALPIX level 128 (0\fdg5).  
The color table (in units of thermodynamic 
temperature fluctuations) 
reaches a maximum
for all sources which will be directly detectable by future satellites.
}
\label{fig:radio_pmn}
\end{figure}

Figure \ref{fig:radio_pmn} shows a 
skymap (in Galactic coordinates) 
of our extrapolated/interpolated radio source
predictions at 100 GHz.  
MAP should detect about 200 of these sources at the $5\sigma$ level.  
The map shows the 
$|b|>5$ cut we made in the PMN catalog
to eliminate Galactic contamination and the dearth of sources 
at the celestial North pole, where radio source 
observations are difficult.  
Figure \ref{fig:radio_cl} shows confirmation of the expected power-law 
power spectrum for radio galaxies at 100 GHz.  

We estimate an overall systematic uncertainty of a factor of 1.5 for our 
predictions at frequencies less than 100 GHz, increasing to a factor 
of 3 at 200 GHz and a factor of 5 at frequencies above 300 GHz.  Two 
major sources of uncertainty affect these predictions at high frequency:
the likelihood that the spectral index will fall off around 100 GHz due 
to a falloff in each galaxy's electron energy spectrum, and the possibility 
of appreciable thermal emission from dust in each galaxy contributing 
at frequencies above 100 GHz.  
We expect an overall error for each source which 
preserves the source's predicted spectrum of a factor of 1.5 for 
sources from the SGS catalog.  This overall error is a function of 
the 4.85 GHz flux error and the spectral index uncertainty for 
sources from the PMN catalog.  The uncertainties for each source  
are based on an estimated 
spectral index uncertainty of 0.2 for sources with measured 2.7 GHz 
fluxes and of 0.5 for others.  PMN sources with apparently rising spectra 
were set to flat spectra and assigned a spectral index uncertainty of 
0.5 since radio spectra that rise from 2.7 to 5 GHz typically turn over 
and fall thereafter.  
For sources from the original catalog, we expect there to be an additional 
independent uncertainty at each frequency of a factor of 1.3 
at frequencies below 150 GHz, a factor of 2 at frequencies between 
150 and 300 GHz, and a factor of 3 at frequencies above 300 GHz.  
For sources from the PMN catalog, this frequency-by-frequency error 
is a function of the spectral index uncertainty.

\section{Discussion}

Our results indicate that the brightest radio sources will dominate microwave
anisotropy for a wide range of resolutions and frequencies.  
Our skymaps predict the 
location and flux of the brightest radio sources at each frequency, 
making it straightforward to develop a template for masking 
the pixels containing them.  This masking should be sufficient to protect
high resolution CMB anisotropy observations from unacceptable radio 
source confusion.

Spectral analysis of bright radio sources 
indicates that their spectra are complex and cannot in general 
be categorized into template spectra or single power-laws.
The results from our analysis of extrapolation errors suggest that our 
phenomenological approach of fitting a power law to the 
high-frequency end of each spectrum is a reasonable model to use to 
extrapolate radio sources to microwave frequencies.  
Although subject to systematic errors on a galaxy-by-galaxy basis, 
we expect our overall extrapolation results to be accurate to within 
a factor of two at 90 GHz.

Our analysis of foreground 
confusion from extragalactic radio sources 
indicates that they contribute negligibly to COBE resolution 
observations of the CMB, consistent with the conclusion of 
\citet{bandayetal96}.  
However, they do become problematic at higher resolution.  
Our results set a lower limit on the anisotropy and provide a 
list of the brightest sources in the sky which can be used to mask pixels in 
future high-resolution CMB observations.
The contribution of extragalactic radio sources to CMB anisotropy is comparable
at 200 GHz to that of bright extragalactic infrared sources 
\citep{gawisers97, toffolattietal98}.  
Our current
results indicate a valley at around 200 GHz where the anisotropy from
radio sources is a minimum; adding in the contribution from infrared-bright
galaxies should move that valley towards 150 GHz.  

The results of this investigation motivate an expansion of our catalog 
so that sources which will 
contribute to 
anisotropies on the 
1$\sigma$ level can be masked.  
It is clear that the current generation of 
CMB anisotropy experiments must pay close attention to the possibility of 
radio point source contamination at all frequencies.  Masking
pixels which contain bright radio galaxies should reduce this foreground
to a manageable level.

%\clearpage
%\begin{figure}[htb]
%\figurenum{2}
%\centerline{\psfig{file=radio_fig2_998.ps,width=7.0in}}
%\caption{
%Sample radio sources from our catalog with fits used for extrapolation.  
%Notice how varied the source spectra are, defying the usage of simple 
%templates.
%}
%\end{figure}

\cleardoublepage

%\end{document}

\cleardoublepage
\chapter{Observational Limits on Anisotropy from Point Sources}
\label{chap:sources}

%\documentstyle[12pt,aasms4]{article}

%\begin{document}

%\title{Observational Constraints on Microwave Anisotropy from Point Sources}

%\author{Eric Gawiser$^{1,4}$, Andrew Jaffe$^{2,5}$, \& Joseph Silk$^{1,2,3,6}$}
%\affil{
%Departments of Physics and Astronomy and Center for Particle 
%Astrophysics, 
%   University of California, Berkeley, CA 94720}
%\authoremail{gawiser@astron.berkeley.edu}

%\vspace{2 in}

%$^1$ Physics Department

%$^2$ Center for Particle Astrophysics

%$^3$ Astronomy Department

%$^4$ gawiser@astron.berkeley.edu

%$^5$ jaffe@cfpa.berkeley.edu

%$^6$ silk@astron.berkeley.edu 

\section{Motivation}

Improved instrumentation
and the upcoming MAP (Microwave Anisotropy Probe) and Planck 
Surveyor
satellite missions focus current attention on 
angular scales between one-half and one-tenth of a degree, and there is 
theoretical motivation for undertaking future observations 
at even smaller scales 
\citep{huw97a, metcalfs98, jaffek98}.  
Because the antenna temperature 
contribution of a point source is inversely proportional to  
the solid angle of the beam,
observations at higher 
angular resolution 
are more sensitive to foreground contamination from 
point sources, including the  
radio sources and low- and high-redshift 
infrared-bright galaxies which have already been 
discussed and the Sunyaev-Zeldovich 
effect from galaxy clusters (see Chapter \ref{chap:sz}).  
The dominant contribution of the Galaxy to microwave 
anisotropy is from diffuse emission 
\citep{toffolattietal98, finkbeiner99}.

Until recent SCUBA observations, almost all sources observed from 
10-1000 GHz were selected at higher or lower frequencies, so there 
was little direct knowledge of point source populations with emission 
peaking in this wide frequency range.  \citet{blainis98} 
(see also \citealt{blainetal98} and \citealt{scottw98}) 
use models for high-redshift galaxies
normalized to SCUBA counts at 353 GHz to predict anisotropy 
from extragalactic point sources 
down to 100 GHz, but this extrapolation is highly 
model-dependent.  Previous predictions of the total point source
contribution 
\citep{toffolattietal98, toffolattietal95, franceschinietal89, wang91}
 used galactic evolution models with specific 
assumptions about dust
temperatures and luminosity evolution 
to predict the level of point source contamination.  More
phenomenological approaches 
(see Chapters \ref{chap:ir} and \ref{chap:radio} and \citealp{tegmarke96})
lack information on infrared galaxies at high redshift and on dim but 
numerous radio sources.    
 
	Cosmic microwave background observations contain contributions
to anisotropy from two groups of point sources.  The bright sources
at a level of at least $ 5 \sigma$ 
($\sigma$ is the quadrature sum of instrument
noise, CMB fluctuations, diffuse Galactic emission,
 and underlying point source
fluctuations) can be detected individually and eliminated by 
masking the pixels containing them.  This detection limit can 
be lowered by using prior 
knowledge of the locations of bright sources obtained from 
extrapolating far-infrared and radio frequency observations as
described in Chapter \ref{chap:ir} and \ref{chap:radio} 
(as well as filtering, fourier transform, 
 and wavelet techniques; see 
\citealt{tegmarkd98, ferreiram97, tenorioetal99}).  
Numerous dimmer sources 
will add to anisotropy but cannot
be detected without performing further observations at higher resolution 
at nearby frequencies.
For most planned CMB observations, these simultaneous observations will be 
difficult due to large sky coverage at high resolution of
the primary instrument (although Planck's wide frequency coverage 
will help with foreground subtraction.)

We utilize recent sub-arcminute resolution
observations to constrain the contribution to anisotropy from this
second group of point sources that will inevitably contaminate
measurements of CMB anisotropy.    
Recent observations using BIMA by 
\citet{wilnerw97}
detected no sources brighter than 3.5 mJy in the 
15 arcmin$^2$ of the Hubble Deep Field (HDF) at
4.7$''$ resolution at 107 GHz.  We combine this constraint with the 
counts of sources detected in blank fields at 353 GHz by SCUBA 
\citep{hughesetal98, bargeretal98, ealesetal98}
and at 8.4 GHz with the VLA 
\citep{richardsetal98, fomalontetal97}
, 
with blank field upper limits from BIMA/OVRO 
at 28.5 GHz 
\citep{holzapfel98, carlstrom98}
, 
SuZIE at 142 GHz 
\citep{churchetal97}
, 
IRAM at 250 GHz 
\citep{kreysa98, grewing97}
, SCUBA at 667 GHz 
\citep{hughesetal98}
,  
and CSO at 850 GHz 
\citep{phillips97}
, and with the 
detection of Far-Infrared Background radiation 
in FIRAS data 
\citep{pugetetal96, buriganap98, fixsenetal98}
.  

It has long been feared that a population of 
sources with spectra peaking near 100 GHz, due to self-absorbed radio 
emission or thermal emission at very high redshift, might remain
undetected by radio and far-infrared observations while contributing 
significantly to measurements of CMB
anisotropy.  
Now that high-resolution observations are available in the 
frequency range relevant to CMB anisotropy observation, we
set upper and lower limits on point source confusion between 
10 and 1000 GHz 
by assuming 
that the emission of point sources originates from 
the well-understood physical processes of 
synchrotron, free-free, thermal dust, and spinning dust grain emission.  

\section{Extragalactic Point Sources}	

The main emission mechanism 
of bright far-infrared 
 sources is graybody reradiation of starlight and/or Active Galactic Nuclei 
(AGN) radiation absorbed by dust. 
Chapter \ref{chap:ir} predicted the 
level of microwave anisotropy from the 5319 low-redshift
infrared-bright galaxies in the IRAS 1.2 Jy survey. 
We expect there to be numerous higher-redshift starburst 
galaxies like the prototypes Arp 220, F 10214+4724, SMM 02399-1236
\citep{ivisonetal98, frayeretal98}, 
and APM 08279+5255 
\citep{lewisetal98}
which generate
similar dust emission, and with their spectra redshifted considerably 
these sources could easily be missed by far-infrared surveys and yet 
make significant contributions to the microwave sky.  There may  
well exist a population of ultraluminous proto-elliptical galaxies which 
cannot be described by models using smooth evolution of the IRAS 
luminosity function.  Recent detections of the 
Far-Infrared Background radiation 
and of submillimeter sources by SCUBA 
\citep{smailib97, smailetal98}
 give us the first clues about
the nature and abundance of these high-redshift objects.

	A separate population of extragalactic point sources 
are radio-loud, typically elliptical galaxies or AGN.    
Radio sources which have nearly flat spectra up through microwave 
frequencies are called blazars, a class which includes 
radio-loud 
quasars and BL Lacertae objects.
Chapter \ref{chap:radio} examined 2207 bright
radio sources in detail, but there are over ten thousand 
of these sources which are bright enough to have some impact on 
arcminute-resolution microwave observations.  

For instruments of resolution $\geq 10'$, galaxy clusters will be  
unresolved and will provide an additional family of point sources 
via the Sunyaev-Zeldovich
 effect 
\citep{sunyaevz72}.
The observations used here are basically insensitive to SZ clusters 
as the fields have been chosen to avoid known clusters and are 
typically observed at sub-arcminute resolution. 
However, 
anisotropy from SZ sources is not expected to seriously impair CMB anisotropy 
observations 
\citep{refregiersh98, aghanimetal97}.

\section{Analysis}

We assume for these calculations that observations use pixels
of width equal to the FWHM of their beam.  A chosen 
amount of overpixelization will 
lead to a small, calculable correction in the level of anisotropy 
and makes it easier to distinguish point sources,
which contribute to several pixels, from instrument noise, which is often 
uncorrelated between neighboring pixels.  

We can rigorously predict the fluctuations due to
sources using the techniques of $P(D)$ analysis 
\citep{scheuer57, scheuer74, condon74, franceschinietal89, toffolattietal98}
. 
To
begin, we must estimate the cumulative flux distribution of sources, 
$N(>S)=\int_S^\infty N(S)\; dS$.  
The SCUBA results give a list of sources and their
fluxes with error bars; they also provide a limit on the
low-flux tail of the distribution from their measured residual
fluctuations 
\citep{hughesetal98}
. 
We estimate $N(>S)$ directly, using a
Gaussian of width given by the error on the observed flux for each
source.  
We calculate $2\sigma$ error bars on $N(>S)$ 
for this estimated distribution by having the fluctuations in number be 
consistent with Poissonian fluctuations for each cumulative distribution.
We use a top-hat experimental beam to 
convert $N(>S)$ to an observed flux distribution. We then convert
this to the probability distribution, $P(D)$, of getting a total flux,
$D$, in the beam
\citep{scheuer57, scheuer74, condon74}.  
Whereas the integrated background is determined by 
the slope and low-flux cutoff of $N(>S)$, the anisotropy is dominated 
by the brightest sources seen by SCUBA.  

We consider both the detected sources and the rms noise in 
the instrument.  The observed instrument noise is usually roughly 
Gaussian with mean near zero.  This provides a good upper limit on 
confusion from undetected sources 
because one can bury only about half that noise in anisotropy from 
dim sources without increasing the mean noise level by much or making the 
noise distribution noticeably non-Gaussian.
\citeauthor{hughesetal98}
report a noise level of 0.45 mJy
per 8.5$''$ beam. To allow for the possibility that a large
fraction of this may actually be from sources, we define the total flux,
$y$, the sum of $D$ and this noise contribution.  We take
the noise distribution to
be a zero-mean Gaussian with the reported variance, scaled by 
the desired beam area.  The distribution of $y$
is just the convolution of $P(D)$ and the noise distribution.  
From $P(D)$ or $P(y)$ we determine the impact on CMB measurements by
estimating the variance 
($\sigma_D$ and $\sigma_y$).  Our $2\sigma$ upper and lower limits from 
SCUBA at 353 GHz are 
67 mJy and 8 mJy respectively.  
Such careful calculations are not strictly necessary; 
the following easily-reproduced back-of-the envelope
calculation is accurate to within 
a factor of two, adequate for present purposes.

Our upper and lower limits correspond to $2\sigma$ 
 confidence levels from the reported observations.  
If an observational field contains 
$N_{obs}$ sources, we estimate the upper/lower 
limit 
on the number of sources $N$ in a typical such field on the sky using 
$N_{obs} = N \pm 2 \sqrt{N}$, which leads to limits on the fluctuation of the 
number of sources in a typical field on the sky of $\sqrt{N} = 
\sqrt{N_{obs}+1} \pm 1$.  
 
For $N$ sources with flux $S$ per beam, the rms flux anisotropy on the sky 
is 
\begin{equation}
 \Delta S = S \sqrt{N} \; \;  
\end{equation}
in 
the Poisonnian limit of large N.  
\citeauthor{toffolattietal98}
predict a negligible contribution from
non-Poissonian clustering of sources for beams of 10$'$ and larger.  
\citet{scottw98}, 
however, suggest that clustering will lead to fluctuations 
twice as large as Poissonian fluctuations for a 10$'$ beam, with less 
enhancement at higher resolution.  
For N$<1$ (one source per several
beams), we have only a few pixels receiving flux, the mean flux is $NS$, and

\begin{equation}
  \Delta S = \sqrt { N (S-NS)^2 + (1-N)(NS)^2 } = S \sqrt{ N - N^2} \; \; , 
\end{equation}

\noindent
which also tends towards $S \sqrt{N}$ 
as N becomes small.  

We extrapolate our upper and lower limits 
from an observed frequency by using the 
most extreme known physical emission mechanisms in that frequency 
range; the fastest the flux should fall is as 
very steep spectrum synchrotron emission, i.e. $\nu ^{-2}$ 
\citep{steppeetal95}
, or above 300 GHz as a Wien tail with $\nu^1$ emissivity,
i.e. 
$\nu^3 / (\exp(h \nu / k T_{CMB}) - 1)$, 
since it is unreasonable 
for a cosmological object to have an effective temperature less than 
T$_{CMB}$.  
Conversely, the fastest a spectrum should be able to rise is as 
Rayleigh-Jeans thermal emission 
with $\nu^2$ emissivity, i.e. as $\nu^4$.   Free-free and spinning dust grain
emission 
\citep{drainel98a, drainel98b}
produce less conservative extrapolations.

The way to maximize anisotropy for the observed integrated 
Far-Infrared Background
is to make individual sources as bright as possible. 
The low emissivity ($\nu^{0.6}$) fit by 
\citet{fixsenetal98}
means that 
no one graybody spectrum (emissivity between $\nu^1$ and $\nu^2$) can be 
responsible for the FIRB.  Therefore, we set upper limits on the 
anisotropy at a given frequency by making hypothetical sources whose spectra
peak at that frequency be as bright as possible.  
The brightness of these high-z IR sources 
is constrained by requiring their dust to have temperature 
greater than 20K (since low-z inactive spirals have 20K dust) and 
greater than 3K(1+z) (so that the dust is never 
colder than the CMB at that redshift), and to have a  
bolometric luminosity 
no greater than that of a quasar ($10^{39}$ W).  We also examine 
a second model where the luminosity constraint is raised to $10^{41}$W, the 
likely luminosity of APM 08279+5255 once lensing is accounted for 
\citep{lewisetal98}
.
Using these constraints, we predict an upper limit of 
$\Delta T / T = 10^{-6} (10^{-5})$ 
for a 10$'$ beam at 200 GHz for a high-z IR population of 
luminosity $10^{39}$W ($10^{41}$W)
 whose total emission generates the FIRB.  However, this upper limit 
is less robust than those from direct observations, because there 
could be separate source populations, one which yields the integrated
background but small fluctuations on the relevant angular scales, 
and another which dominates the flux
anisotropy but produces only a small fraction of the FIRB.

Flux variation
is converted to antenna temperature and thermodynamic equivalent 
temperature fluctuations using Equations \ref{eq:flux_to_temp} and 
\ref{eq:dt_to_dta}.  
This yields an equivalent thermodynamic temperature variation which
 scales as fwhm$^{-1}$ for a given flux anisotropy on $10'$ scales:
\begin{equation}
\frac{\Delta T}{T_{CMB}} = \Delta S_{10'}(Jy) 
\left ( \frac{fwhm}{10'} \right )^{-1} (5\times10^{-4})
 \left ( \frac {(e^x - 1 )^2}{x^4e^x} \right ) \;  \;  .
\end{equation}
The fwhm$^{-1}$ behavior occurs because the number of sources in 
a beam, $N$, is proportional to fwhm$^2$, the flux variation $\Delta S$
due to Poissonian clustering 
is proportional to $\sqrt N$ and the temperature fluctuations 
after beam convolution are given by 
$\Delta T \propto (\Delta S)$fwhm$^{-2} \propto$ fwhm$^{-1}$.

\begin{table}[h] 
\label{tab:sources_tab1}
\caption{Noise levels in 
high-resolution microwave observations.}
\mycaption{We list the frequency, resolution, noise per 
beam, and the upper limit 
for $\Delta S_{10'}$ that results from assuming that half of this
noise is really produced by unresolved point sources.}
\begin{center}
\begin{tabular}{|r|c|c|c|c|}
\hline
Instrument &  $\nu$ (GHz) & FWHM  & 
   Noise/beam & $\Delta S^{upper}_{10'}$ \\

\hline

VLA	& 8.4	&6$''$ 	&0.0028 mJy &0.14 mJy  \\

BIMA	& 28.5 	&90$''$ 	& 0.12 mJy &0.4 mJy  	\\

BIMA	& 107	&4.7$''$	&0.7 mJy &45 mJy  	\\

SuZIE   & 142   & 100$''$       & 10 mJy & 30 mJy       \\

IRAM	& 250	&11$''$ 	&0.5 mJy &14 mJy  	\\

SCUBA	& 353	&15$''$ 	&0.45 mJy &16 mJy 	\\

SCUBA	& 667	&7.5$''$ 	&7 mJy	&280 mJy 	\\
	
CSO	& 857  	&10$''$ 	&100 mJy &3000 mJy  	\\

\hline

\end{tabular}
\end{center}
\end{table}

%\newpage

\begin{table}[bht] 
\label{tab:sources_tab2}
\caption{Microwave source detections.}
\mycaption{Upper and lower limits correspond 
to the observed fields being 2 $\sigma$ Poissonian fluctuations above or below
the typical source density on the sky (see text).  
The totals include the noise 
totals given in Table 1 added in quadrature with the limits from each source 
population.  The range of sources in the $>3$mJy SCUBA bin allows for 
the incompleteness correction suggested by Eales et al. and the approximate 
number in the $1-2$mJy bin is based on the P(D) analysis of Hughes et al.}
%\label{tab:models}
\begin{center}
\begin{tabular}{|r|c|c|c|c|c|c|}
\hline
Instrument &  $\nu$ (GHz) & Field size & 
S$_{source}$ & N$_{sources}$ & $\Delta S^{upper}_{10'}$ & 
$\Delta S^{lower}_{10'}$ \\

\hline

VLA	& 8.4	&40 sq. $'$ &$>0.5$ mJy  &3 &2.6mJy &0.9 mJy \\

	& 	& 	&0.05-0.5 mJy & 8 &0.4 mJy  &0.2 mJy \\ 

	& 	& 	&0.009-0.05 mJy & 18 	& 0.1 mJy & 0.08 mJy\\

	& 	& 	&0.006-0.009 mJy & 19 & 0.04 mJy & 0.03 mJy\\

	& 	& 	& TOTAL 	& 	& 2.7 mJy & 0.9 mJy \\

SCUBA	& 353	&46 sq. $'$ & $>3$ mJy & 15-20 &29 mJy &16 mJy	\\

	& 	&9 sq. $'$  & 2-3 mJy & 2    & 12 mJy & 7 mJy   \\

	&       &9 sq. $'$  & 1-2 mJy & $\simeq 18$  & 18 mJy & 15 mJy \\ 

	& 	& 	  & TOTAL   & 		& 40 mJy  & 23 mJy \\

\hline

\end{tabular}
\end{center}
\end{table}

\section{Results}

\begin{figure}
\centerline{\psfig{file=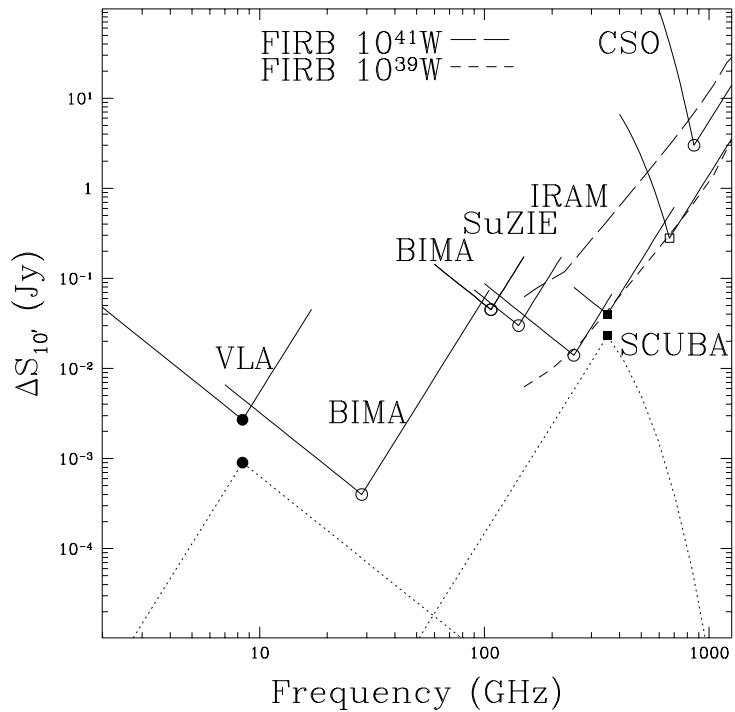,width=4in}} 
\caption{Limits on flux anisotropy.}
\mycaption{Upper (solid) and lower (dotted) 
limits on flux anisotropy (in Jy) for a 10$'$ beam  
from VLA, BIMA/OVRO, BIMA, SuZIE, IRAM, SCUBA (squares), and CSO.  
Filled points indicate detections, open points are non-detections, 
and the extrapolations are based on steep-spectrum radio emission, 
Rayleigh-Jeans thermal emission with $\nu^2$ emissivity, and 
Wien tail thermal emission with $\nu^1$ emissivity.  The long (short) 
dashed lines
indicate the upper limit on flux anisotropy derived from extreme 
models of the  
Far-Infrared Background radiation with a maximum 
source luminosity of $10^{41}$W ($10^{39}$W).  
}
\label{fig:sources_fig1}
\end{figure}

Table \ref{tab:sources_tab1} shows our upper limits for the possibility 
of dim sources buried in the instrument noise of non-detections, and 
Table \ref{tab:sources_tab2} 
lists source detections and the resulting limits on 
$\Delta S_{10'}$.
Figure \ref{fig:sources_fig1} shows our upper and lower limits for
flux anisotropy 
from point sources from 10-1000 GHz, as well 
as the results for the extreme models of the Far Infrared Background
radiation.

%\nopagebreak
            
\begin{figure}
%\figurenum{2}
\centerline{\psfig{file=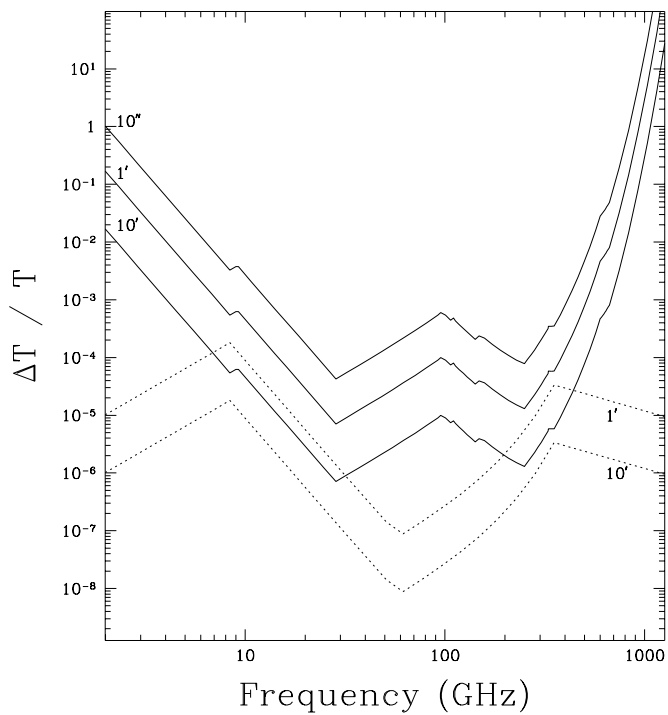,width=4in}} 
\caption{Limits on temperature anisotropy.}
\mycaption{
Net upper and lower limits on 
$\Delta T / T$ for $10'$, $1'$, and $10''$ based on the observations
and extrapolated limits shown in Figure \ref{fig:sources_fig1}.  
The lower limit for $10''$ is zero because all sources detected by 
SCUBA and the VLA should also be detected and subtracted 
by future observations at that 
resolution.  The upper limits are based on assuming that the 
combination of instrument noise and CMB fluctuations 
is too high to subtract any of the SCUBA or VLA sources.
  }
\label{fig:sources_fig2}
\end{figure}

We plot
the resulting limits on temperature anisotropy in Figure 
\ref{fig:sources_fig2} for a
range of angular scales and frequencies.  The less robust nature of 
the FIRB constraint 
prevents us from using this as an upper limit in Figure 
\ref{fig:sources_fig2}.  
Because the angular power spectrum $C_\ell$ 
of 
Poissonian distributed point
sources increases with multipole $\ell$ 
relative to the expected CMB angular power spectrum 
\citep{scottw98} and the rms temperature fluctuation 
is an integral over this power spectrum,
an rms $\Delta T / T$ from point sources close to $10^{-5}$
will seriously impair
the measurement of the CMB angular
power spectrum on the smallest angular scales.  However, 
a value less than 
$10^{-6}$ means that foreground contamination is not a major concern.
The 10$'$ lower limit shows that $\Delta T / T < 10^{-6}$ is only possible 
from 20-300 GHz, and the upper limit for 10$'$ shows that $\Delta T / T < 
10^{-6}$ at 30 GHz and $\Delta T / T \simeq 10^{-6}$ at 250 GHz.  The limits
are much less stringent near 100 GHz, where a pathological population of 
point sources could lead to anisotropy closer 
to $10^{-5}$.  Typical radio 
and far-IR sources that fall within these limits at 30 and 250 GHz will 
end up much closer to the lower limit near 100 GHz, however.  
Our upper limit constrains all types of point sources, including any 
hypothetical high-latitude or halo Galactic point sources.  
Our limits diverge considerably near 100 GHz, so while they 
are compatible with the model-dependent extrapolations of 
\citeauthor{blainetal98} and 
\citeauthor{toffolattietal98}
,
they would also be compatible with significantly different extrapolations.

\citeauthor{scottw98}
 indicate that clustering 
may lead to a factor of two amplification of our predictions for a 10$'$ beam; 
the correction is less for higher resolution.  However, they 
adopted the angular correlation function of Lyman-break galaxies observed
in a narrow redshift range, whereas the SCUBA sources are expected and 
observed to follow a much broader redshift distribution 
\citep{bargeretal99, richards99}, leading to a factor of 
4-10 reduction in the angular correlation function (as discussed in 
Section \ref{sect:wombat_simulations}).

\begin{figure}
\centerline{\psfig{file=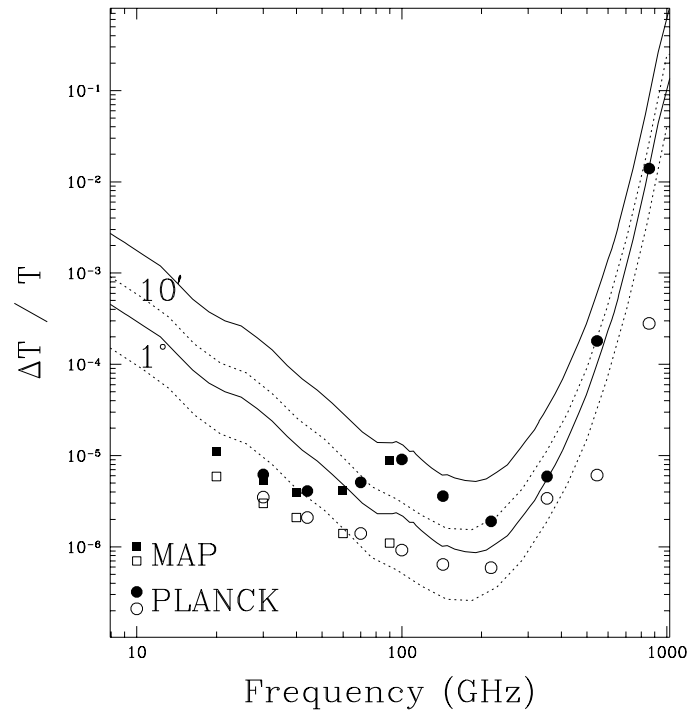,width=4in}} 
\caption{
Net upper (solid line) and lower (dotted line) limits for $10'$ and $1^\circ$.}
\mycaption{
This includes anisotropy for known bright sources from 
Figures \ref{fig:iras_fig3} and \ref{fig:radio_fig3} 
with a 
factor of three uncertainty shown.     
5 $\sigma$ subtracted upper (solid) and lower (open) 
limits for MAP (squares) and Planck (circles) are also 
shown.  The channels are treated independently here, although in practice 
they could be combined to produce somewhat lower anisotropy levels.  The 
combination of pixelization and beam-size effects discussed in the 
text leads to 1/fwhm scaling for all of these point source populations.    
}
\label{fig:sources_fig3}
\end{figure}

The blank fields observed by VLA, BIMA, IRAM, SCUBA, and CSO  
were chosen 
to avoid known bright point sources.  
Therefore, for observations which avoid known bright sources or 
mask the pixels containing them, Figure \ref{fig:sources_fig2} 
gives full upper and lower 
limits on point source anisotropy.  
Chapters \ref{chap:ir} and \ref{chap:radio} have 
analyzed the contribution of 
bright infrared and radio point sources, 
respectively, 
so for a randomly chosen location on the sky 
the expected anisotropy is the quadrature sum of the 
anisotropies from those types of bright sources and our result
 in Figure \ref{fig:sources_fig2}.  
Figure \ref{fig:sources_fig3} adds in results for known bright
sources from Chapters \ref{chap:ir} and \ref{chap:radio}
for a 10$'$ beam and 
shows the results for MAP and Planck 
after subtracting sources detected at $5 \sigma$. 
Since 
$ \Delta T / T$ is strongly influenced
by a few bright pixels due to the highly non-Gaussian distribution, 
the values are significantly lower after bright source subtraction.    
Figure \ref{fig:sources_fig3}
 shows that for a 10$'$ beam without source subtraction, the 
point source anisotropy will be $\geq 10^{-6}$ at all frequencies.  
From 70-200 GHz, the 
upper limit from Figure \ref{fig:sources_fig2}
 dominates the anisotropy from known bright 
radio and IRAS sources.  MAP and Planck can detect the brightest few 
hundred sources at each frequency (see Chapter \ref{chap:radio}) 
so the upper and lower limits 
for the satellites diverge over a wider frequency range, making the impact 
of our uncertainty about the level of anisotropy from dim but numerous 
point sources a significant problem in predicting foreground contamination.  
The highest-frequency Planck channels can detect nearly all 5319 IRAS 1.2 Jy 
sources, so it is the dimmer 
high-redshift IR galaxies constrained by SCUBA that 
dominate their source confusion.  
Our limits here treat each channel independently, but it will be possible
to detect bright sources at particular frequencies and mask the corresponding
pixels in all channels.  This will enhance the importance of dim but 
numerous sources relative to known bright sources but will reduce the 
overall level of foreground contamination.

\section{Discussion}

We find impressive agreement between the SCUBA observations, the IRAM and 
SCUBA upper limits, and the upper limit for flux anisotropy produced by 
$10^{39}$W sources which generate the integrated Far-Infrared Background 
shown in Figure \ref{fig:sources_fig1}.  
This is  
consistent with the FIRB being produced by the SCUBA sources (since the upper 
limits and the detections differ by only a factor of two), and this indicates 
that the FIRB sources must be close to maximizing their anisotropy i.e. they 
are highly luminous but not too numerous.  
The $10^{41}$W model, however, 
predicts more anisotropy than is consistent with the observed SCUBA 
source counts and IRAM and SCUBA upper limits, suggesting that starburst 
galaxies 
like APM 08279+5255 are more luminous than 
typical FIRB sources.  This conclusion is also 
supported by the near-blackbody spectrum of APM 08279+5255 in the 
sub-millimeter; there is no way to sum such spectra at various redshifts 
and produce a graybody of emissivity 0.6 as is seen for the FIRB 
\citep{fixsenetal98}
.  
The IRAM upper limit is low enough 
to show that the far-IR sources detected by SCUBA have rising spectra, so 
this is further evidence that their emission is thermal in origin.  

The CMB anisotropy damping tail on arcminute scales is a sensitive probe 
of cosmological parameters and has the potential to 
break degeneracies between models which explain the larger-scale anisotropies
\citep{huw97a,metcalfs98}.  
The expected level of temperature 
anisotropy 
is $\Delta T / T \simeq 10^{-6}$, 
which Figure \ref{fig:sources_fig2} indicates may be enough to 
dominate the point source confusion from 30-200 GHz.  The upper limit on 
point source confusion, however, would completely swamp the fluctuations 
of the damping tail, so more knowledge of dim sources is needed before 
we can expect such observations to be feasible.   A high resolution instrument
will probably need to 
 use its highest resolution for point source detection and 
subtraction.

The greatest promise for seeing CMB anisotropies through the 
obscuration of point source confusion occurs near 100 GHz, but 
this is also the frequency range where we know the least about 
the true level of foreground anisotropy on the sky.  
Our upper limits for 10$'$ near 100 GHz give us confidence that useful 
information will be obtained from CMB anisotropy 
observations, but it remains possible that point sources will cause 
thermodynamic fluctuations 
almost equal to 
the intrinsic CMB fluctuations.  
Probably the true upper limit near 100 GHz 
is a factor of a few lower than shown, 
because the sources responsible for the current upper limit 
would need to have spectra rising like a graybody but falling off 
like a steep spectrum radio source; this is either unphysical 
or requires an effective 
temperature less than that of the CMB.   
Since the point source 
fluctuations come from the highest multipoles, this could seriously impair
attempts to measure cosmological parameters from the 
CMB angular power spectrum.
Thus, 
further high-resolution observations 
of blank fields at frequencies near 100 GHz are critical in order 
to determine the actual level of point source confusion,
and CMB anisotropy 
analysis methods must account carefully for contamination from 
point sources.   

%\clearpage

%\newpage

%\begin{figure}
%\figurenum{4}
%\centerline{\psfig{file=sources_fig1.ps}} 
%\end{figure}

%\begin{figure}
%\figurenum{5}
%\centerline{\psfig{file=sources_fig2.ps}} 
%\end{figure}

%\begin{figure}
%\figurenum{6}
%\centerline{\psfig{file=sources_fig3.ps}} 
%\end{figure}

%\end{document}

\cleardoublepage
\chapter{The Sunyaev-Zeldovich Effect}
\label{chap:sz}
\section{Theory}

Rich clusters of galaxies contain hot electron gas which scatters 
the Cosmic Microwave Background photons, leading to a temperature 
decrement at frequencies less than 217 GHz and a temperature increment at 
higher frequencies (relativistic corrections lead to a slight modification 
of this spectral signature).  This 
is called the thermal Sunyaev-Zeldovich effect
(see \citealp{holderc99} for a review).  
  There is also 
a kinetic SZ effect due to a cluster's peculiar velocity; this Doppler 
shift of the apparent CMB temperature 
is independent of frequency and is very difficult to 
detect.  
The SZ effect may also be produced by the lobes of radio galaxies 
\citep[see][]{yamadass99}
or by patchy early reionization around quasars 
\citep[see][]{aghanimetal96, gruzinovh98, knoxsd98} and 
in the Lyman $\alpha$ forest \citep{loeb96}.

For long-wavelength observations, the Sunyaev-Zeldovich effect in 
galaxy clusters will produce a temperature shift of 
\begin{equation}
 \frac {\delta T}{T} = - 2 y = 
- 2 \tau_{0} \frac {k T_{x}} {m_{e}c^{2}} \sim
-5 \times 10^{-5} h^{-\frac{1}{2}}
\end{equation}
where $T_{x}$ is the plasma temperature and $\tau_{0}$ is the optical
depth of the cluster plasma \citep{peebles93}.  This is comparable to the  
temperature fluctuations in the CMB observed by COBE and can therefore 
mask the true CMB temperature anisotropies on scales of several arcminutes 
corresponding to clusters.  The SZ effect is a localized spectral distortion
and appears as a temperature anisotropy.  One way to distinguish
it from the true primordial fluctuations in the CMB is to use several
different frequencies, including some in the Wien tail where the SZ effect
will increase the apparent temperature.  

The frequency dependence of SZ thermodynamic temperature fluctuations  
is given by 
\begin{equation}
	j(x) = 2\frac{\exp(x)}{(\exp(x)-1)^2}  
	      \left(x \frac{\exp(x) + 1}{\exp(x)-1} - 4\right) 
		(\cosh(x) - 1),  
\end{equation}
where we have defined 
\begin{equation}
        x \equiv \frac{h \nu}{k_b T_{cmb}}.
\end{equation}
This allows a temperature decrement at a known frequency $\nu_0$ to 
be extrapolated to a different frequency using 
\begin{equation}
	dT(\nu) = dT(\nu_0) \frac{ j(x)}{ j(x_0)}.
\end{equation}

\section{Simulations}

We use 
the XBACS cluster catalog
\citep{ebelingetal96} 
to predict the location on the high-Galactic latitude 
sky ($|b|>30$) of the 
226 brightest SZ sources.  Clusters at higher redshift will have less 
flux, despite comparable temperature decrements, because they are decreasing
in angular size (XBACS only covers up to $z=0.2$).  
The MAP satellite can 
hope to detect about 10 SZ clusters at the $5\sigma$ level, and to barely 
resolve 
Coma and Virgo 
\citep{refregiersh98}.  
Once the minimum of the angular diameter distance is 
neared, the SZ flux from clusters of a given X-ray 
temperature becomes effectively independent of redshift.
Hence, for beams smaller 
than 10$'$, as will be used in the future with Planck HFI and microwave 
interferometers, clusters at all redshifts will have an impact.
Although Planck will also see nearby clusters
as brighter than more distant ones, its high sensitivity and improved
resolution present the prospect of detecting numerous 
high-redshift clusters through 
their SZ effect.

Previous simulations have ignored the known positions and characteristics 
of nearby galaxy clusters 
but offer an illustration of the expected number counts 
of high-redshift clusters, which is strongly dependent on cosmology 
\citep{aghanimetal97}.
For simulations at Planck resolution, we 
will need to add a simulation of the clusters not included in XBACS using 
the $N(S)$ predictions of the Press-Schechter method and adopted temperature
and density profiles for clusters of a given mass.  
A complete 
realistic 
simulation of the SZ effect from galaxy clusters 
on the full-sky requires three components:  
\begin{itemize}
\item
 XBACS clusters ($z < 0.2$)
\item 
Low-galactic latitude clusters at $z < 0.2$.  These must be simulated 
since they are not present in the XBACS sample.  
\item
 Clusters at $z > 0.2$, which also must be simulated because full-sky 
X-ray surveys to determine their actual locations and fluxes do not exist.  
\end{itemize}

\citet{refregiersh98} used an 
isothermal beta-model to fit 
the X-ray data and predict the flux at 90 GHz due to the thermal SZ effect in 
 XBACS clusters.  
There are two major sources of systematic errors for the predicted 
90 GHz fluxes, the   
overall normalization, which depends on the gas fraction and 
on the virialization state, and 
the error in the X-ray temperatures in the XBACS catalog, which 
were derived from ROSAT spectra.  
Data from MAP will be used to improve the overall normalization, which 
can also be tested against N-body simulations.
Pending more detailed analysis, we believe that our overall normalization 
of the predicted 90 GHz fluxes is uncertain by a factor of 1.5, and that 
the prediction for each cluster is also uncertain by a factor of 1.5.  The 
spectral dependence
of the SZ effect is well-understood and should not 
be a significant source of error.  

\begin{figure}[htb]
\centerline{\psfig{file=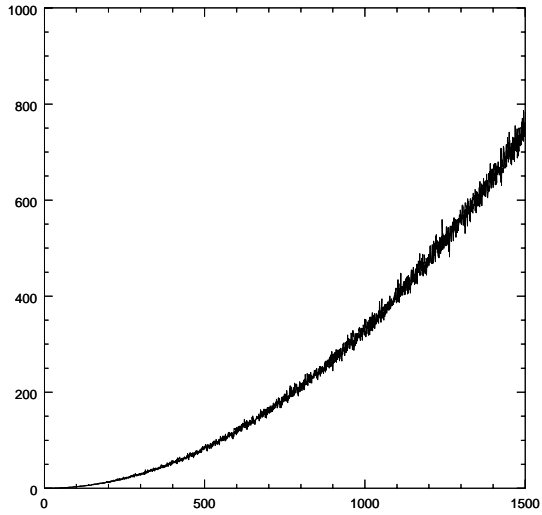,width=3in}}
\caption{Predicted SZ angular power spectra.}
\mycaption{$\ell (\ell + 1) C_\ell$ in $\mu$K$^2$ versus $\ell$ 
for both the XBACS catalog with clusters at nominal positions and 
with clusters at random positions.  The two curves overlap, illustrating that 
the visually distinguishable non-Poissonian clustering of the XBACS clusters 
is insufficient to affect the angular power spectrum of their SZ contribution.
}
\label{fig:xbacs_cl}
\end{figure}

One possible advantage of having the actual positions of these clusters 
from the XBACS catalog is that their non-Poissonian clustering is included.  
However, when we calculate the angular power spectrum at 90 GHz for the 
XBACS clusters with positions randomized within $|b|>30$ we find no 
noticeable difference in the $C_\ell$ (see Figure \ref{fig:xbacs_cl}).  
There is non-Poissonian clustering noticeable to the human eye in the 
XBACS catalog, and this randomized version has many fewer cases 
of multiple clusters located near each other on the sky.  This indicates 
that even mildly non-Poissonian clustering of point sources 
distinguishable by eye is 
sometimes not sufficient to alter the $C_\ell$ spectrum from its 
Poissonian behavior (constant).  This is different for high-redshift 
infrared galaxies because there are many of them per pixel and the 
fluctuations themselves are therefore magnified by non-Poissonian clustering
rather than just redistributing the fluctuations in a more clustered way.  

\begin{figure}[th]
%\figurenum{3}
\centerline{\psfig{file=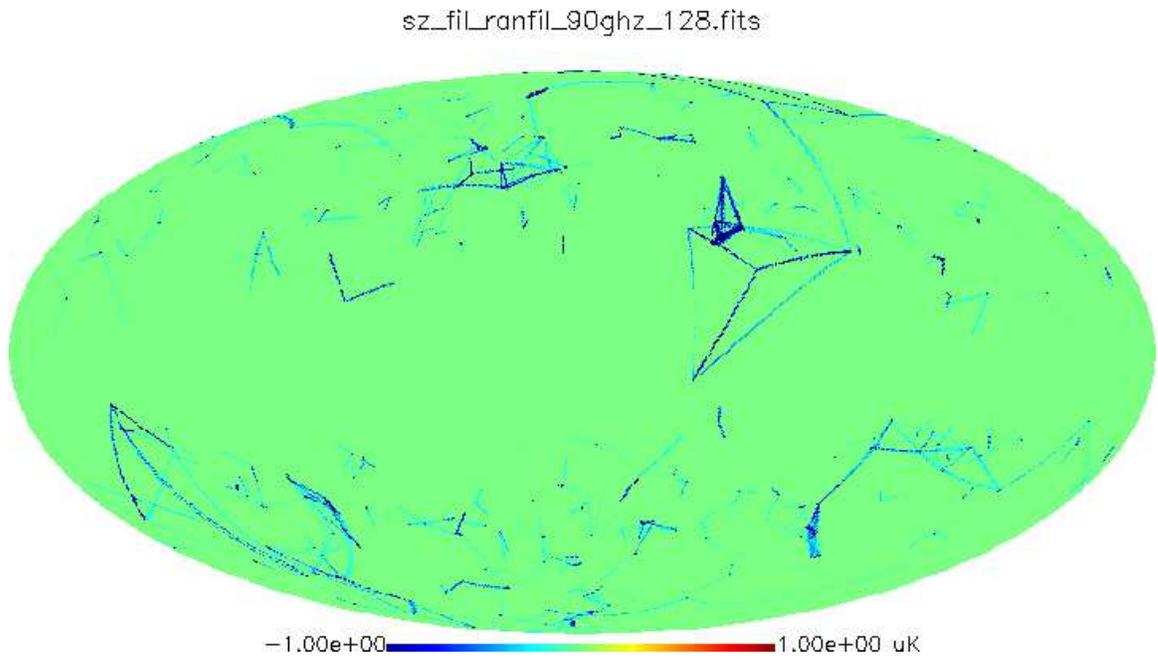,width=6in,angle=90}}
\caption{
Predicted temperature fluctuations at 90 GHz from filaments extending from 
and connecting XBACS clusters using the toy model described in the text.
}
\label{fig:sz_filaments}
\end{figure}

We also explore whether filaments of hot gas connecting and extending from 
clusters can have an impact on CMB anisotropy
\citep{persietal95, boughn98}.  
We use a toy model for the filaments where XBACS clusters that are 
within 50$h^{-1}$Mpc of each other are connected by a filament and clusters
located too far apart to have neighbors in the catalog are 
assigned two filaments heading in random directions.  
Each filament is assigned a total number of baryons equal to its 
cluster endpoint (or the average of its endpoints).  Hence if the gas were 
equally hot in filaments and clusters the total SZ flux from a filament 
would equal the average of that of its endpoints.  
This reproduces the rough structure seen in the simulations of 
\citet{ceno99}\footnote{
See http://www.astro.princeton.edu/\~{}cen/PROJECTS/p1/p1.html for 
nice visualizations of their simulations.}. 
The results of our simulation of the SZ effect from 
filaments are shown in 
Figure \ref{fig:sz_filaments}.  
While these filaments look 
great visually\footnote{and 
contain the type of patterns that might get significant  
media attention, especially from the Weekly World News},
 they are nearly insignificant in comparison 
to the XBACS clusters themselves in terms of SZ effect 
because their 
electron gas temperature is expected to be lower than that 
of rich clusters by about a factor of 10.   
Thus even if half of the baryons 
are in filaments and groups, as predicted, 
their impact on anisotropy is small.  Detecting
these filaments by cross-correlating CMB and large-scale structure maps
has been discussed by \citet{refregiersh98}, and that method can be 
tested using the WOMBAT Challenge simulations, which contain 
a variety of models for the filaments as well as systematic 
variations of the cluster fluxes.

\begin{figure}[htb]
%\figurenum{3}
\centerline{\psfig{file=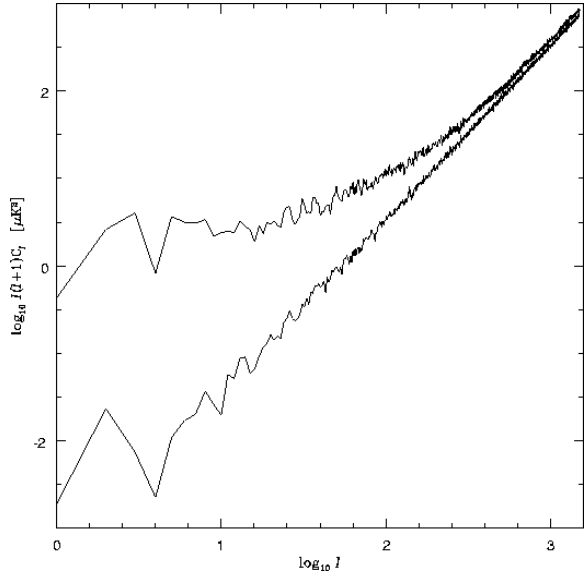,width=3in}}
\caption{Angular power spectra from SZ effect in filaments and clusters.}
\mycaption{
Log-log plot of $\ell (\ell + 1) C_\ell$ versus $\ell$
for SZ simulation based on XBACS only (lower curve) compared to 
simulation including filaments as well (upper curve).
The high-$\ell$ behavior is Poissonian, and the 
filaments have not significantly increased the total power.  The filaments 
lead to significant non-Poissonian clustering on large angular scales, but 
this effect will be entirely swamped by large-angular CMB fluctuations and 
Galactic foregrounds.
}
\label{fig:fil_cl}
\end{figure}

\begin{figure}[bh]
%\figurenum{3}
\centerline{\psfig{file=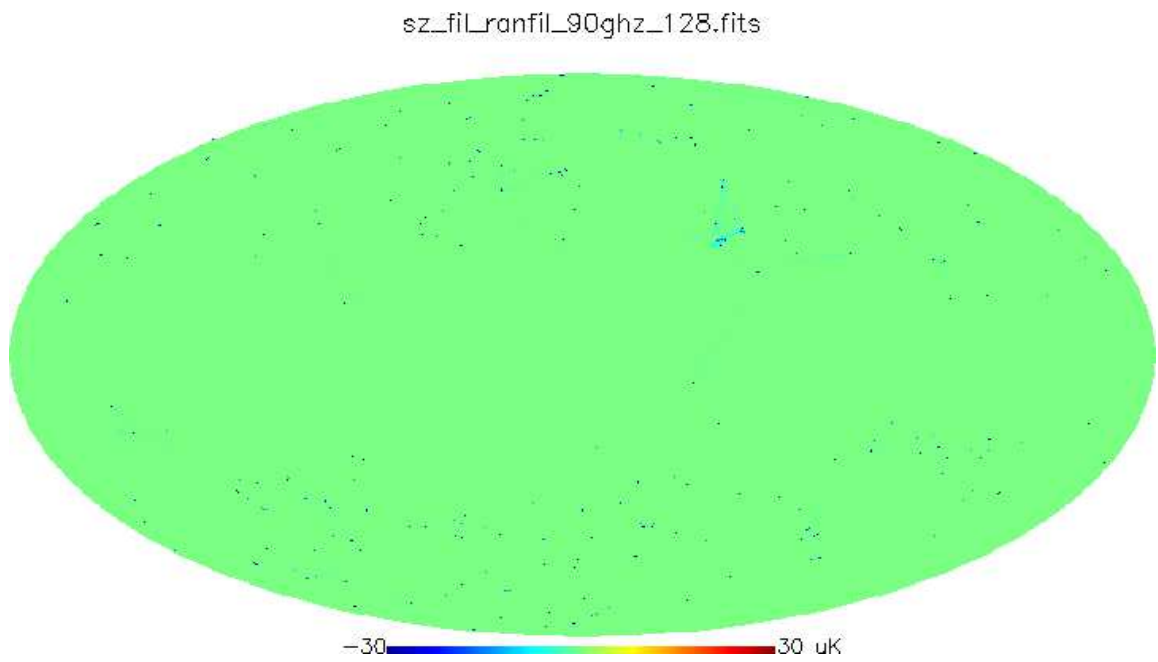,width=6in,angle=90}}
\caption{
Combined prediction of the SZ effect from XBACS clusters and a simulation 
of the filamentary structure connecting them.  
}
\label{fig:sz}
\end{figure}

Combining the SZ effect from clusters and filaments leads to only 
a small increase in the angular power spectrum, as shown in Figure 
\ref{fig:fil_cl}.    
Figure \ref{fig:sz} 
shows all of the sources with a wider 
temperature scale than the plot of filaments alone.  
  The catalog is incomplete 
at $|b|<30$ and reveals large-scale structure at $z < 0.2$.  However, 
almost none of the filaments are visible, illustrating that the 
expected reduction of temperature in the filaments is sufficient 
to make their SZ effect negligible.  
The filaments are not detectable by forthcoming CMB instruments, 
but the XBACS sources are strong.  The prediction of 
\citet{refregiersh98} that MAP can detect about a dozen XBACS clusters 
is consistent with the results of our simulations.

%\begin{figure}[htb]
%\figurenum{3}
%\centerline{\psfig{file=sz_nofil.ps,width=6in}}
%\caption{
%Predicted temperature fluctuations at 90 GHz from the Sunyaev-Zeldovich 
%effect in XBACS clusters.  [RESOLUTION, T SCALE] }
%\label{fig:sz_nofil}
%\end{figure}

\cleardoublepage

\nopagebreak

\cleardoublepage
\chapter{A Pixel-space Method for Foreground Subtraction}
\label{chap:pixel}

\section{Motivation}

\begin{figure}[htb]
\centerline{\psfig{file=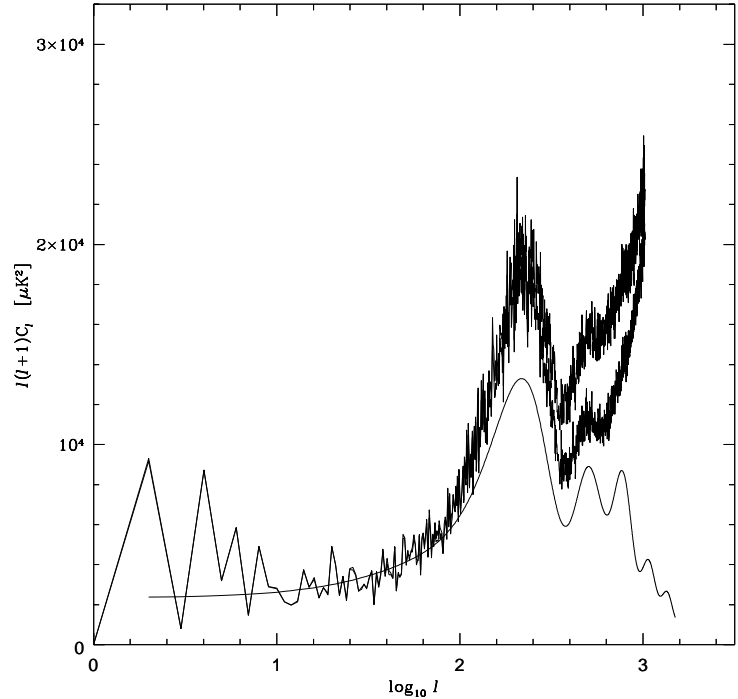,width=4in}}
\caption{
$C_\ell$ for one of the WOMBAT Challenge simulations measured at $|b|>30$ compared to the expected $C_\ell$ for SCDM. 
}
\label{fig:wombat_cl}
\end{figure}

\begin{figure}[htb]
%\figurenum{3}
\centerline{\psfig{file=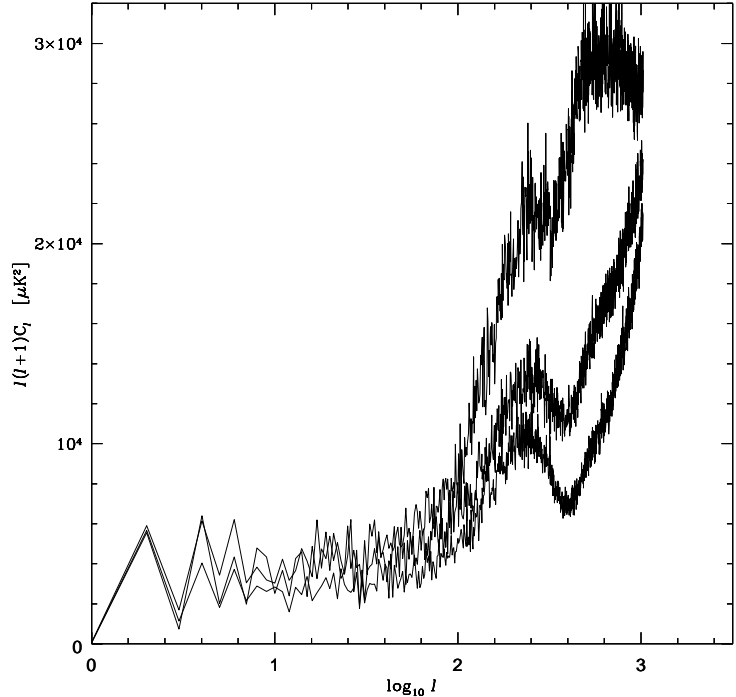,width=4in}}
\caption{
$C_\ell$ for 3 different WOMBAT Challenge simulations, measured at 
$|b|>30$ with $5 \sigma$ pixels subtracted.}
\label{fig:wombat_cl2}
\end{figure}

Figure \ref{fig:wombat_cl} shows the anisotropy power spectrum for one 
of the WOMBAT Challenge simulations at 90 GHz analyzed using 
the HEALPIX routine anafast90.  The upper curve is for $|b|>30$ and 
the curve below this corresponds to the same region of sky with 
pixels at the level of $5 \sigma$ subtracted.  As these 
$5 \sigma$ pixels are most likely due to extragalactic point sources, 
this  illustrates that
extragalactic foreground contamination is considerable but can 
be reduced significantly by this simple method.  The 
angular power spectrum for SCDM is shown 
for reference (smooth curve).  It seems likely that the first acoustic peak 
of the CMB fluctuations in this simulation is simply higher than that of SCDM, 
as the shape and location of the peak are consistent with a host of models 
preferred by the current Saskatoon observations, and Galactic and 
extragalactic foregrounds are unlikely to preserve this shape and amplify it 
by so much.  However, at smaller scales where the second and third acoustic 
peaks are being sought, it is not certain if the second peak is 
being seen and no third peak is evident.  
It is clear that more sophisticated methods are needed to 
reconstruct the second and third acoustic oscillations to high precision.
Figure \ref{fig:wombat_cl2} shows these 
high-Galactic latitude, $5 \sigma$ pixel-subtracted 
versions of the anisotropy power spectrum for three additional 
WOMBAT Challenge simulations.   The simulations display 
a wide range of power at small scales, due to some combination of 
differences in 
intrinsic 
CMB anisotropy, foreground contamination, and instrument noise.

In practice, foreground contamination of CMB anisotropy observations 
is typically estimated by cross-correlating 
template maps at radio or far-IR frequencies with the observations.  Either  
the correlated component proves to be quite small or an attempt is then 
made to subtract it.  On the other hand, most formalized
methods of foreground subtraction are performed in Fourier or spherical 
harmonic space, where phase information about the known spatial distribution 
of Galactic emission
 and bright extragalactic point sources is ignored.  A fiducial
CMB power spectrum must often be assumed, and the risk of aliasing power 
into or out of the CMB in $\ell$-space is great.

The goal here is to design a method for CMB foreground subraction that is 
conceptually simple and computationally fast.  This is not a detailed 
method but rather a framework recommending a series of steps and 
outlining their basic equations; the exact method for a given instrument 
will require tuning some of the parameters and thresholds of this framework 
with simulated observations.  
We will formalize the process of cross-correlating 
and subtracting spatial foreground templates which have been extrapolated 
to the observing frequency with known systematic uncertainties.  Unrealistic
idealization of foregrounds can be prevented by keeping a sense of 
the uncertainties in foreground models and their extrapolations and 
then admitting that these uncertainties are a source of correlated 
``noise'' in the cleaned map.  Where available, we recommend using 
spatial templates for the foregrounds.

Where spatial foreground templates are not available, one must rely on 
the thermodynamic nature of CMB fluctuations to distinguish them 
from foreground contaminants.  It is also possible to use the expected 
difference in angular power spectra but we do not recommend filtering 
in $\ell$-space, as the methods available for that, multi-frequency 
Wiener filtering and Maximum Entropy 
\citep[e.g.][]{tegmarke96,hobsonetal98a}, 
require a priori assumptions 
about the CMB power spectrum and can 
alias power out of true CMB fluctuations.  
We will be subtracting foregrounds in pixel space; when we cross-correlate
with spatial templates this is the same as filtering in 
$a_{\ell m}$ space 
\citep[see][]{white98}
as the $a_{\ell m}$ of those components are 
known up to the same correlation coefficient we are determining.  An 
additional phase of $\ell$-space filtering could be added, but we prefer 
working in pixel space where uncorrelated physical processes add 
and therefore any errors we make in foreground subtraction are 
uncorrelated with the underlying CMB fluctuations.  This means that we can 
only add power to the CMB map and should never have intrinsic CMB 
anisotropy power aliased into other components.  Since the CMB should 
dominate the foregrounds at the frequencies and angular resolutions being 
considered, it makes more sense to allow some foreground power to remain 
added to the CMB than to let an uncertain fraction of CMB power alias 
into detected ``foregrounds.''  Simulations can estimate the residual 
effect on the $C_\ell$, and one should check for the constant-$C_\ell$ 
contamination that is characteristic of point sources at high-$\ell$.

\section{Cross-correlation of Foreground Templates}

Consider a set of observed CMB skymaps with $N_\nu$ channels 
each at frequency $\nu$.  A given map has $N_{p\nu}$ pixels.  
All maps are assumed to be in units of thermodynamic equivalent 
temperature fluctuations.  Then $T^\nu_i$ is the temperature 
of pixel $i$ at frequency $\nu$.  $N^\nu_i$ is the expected level 
of noise in pixel $i$, with covariance matrix given by $C^\nu_{ij,obs}$.  
For differential instruments like COBE or MAP, $C^{\nu}$ is nearly diagonal 
to start, whereas for instruments with signficant correlated noise 
it is already a dense matrix \citep{dodelsonkm95}.  
As we subtract foregrounds we will 
add uncertainties into the updated covariance matrix $C^\nu_{ij}$ but 
these uncertainties are systematic in nature so the errors 
are no longer truly Gaussian.  Only simulations can tell us how much this matters, because these systematic errors often reflect subjective 
judgments and do not follow a quantifiable distribution.  

Foreground templates for various components $\mu$ are $F^\mu_i$ and 
should come with error maps $\Delta F^\mu_i$.  We envision here using 
templates for Galactic dust, synchrotron, and free-free emission (from 
forthcoming $H \alpha$ observations), IRAS galaxies, radio galaxies, 
and the brightest X-ray clusters.  
Whereever information is available on the variation of spectral index 
with position, these templates should be extrapolated to the observational
frequencies with the errors increased to reflect uncertainties of the 
spatial variations.

We can cross-correlate the extrapolated foreground templates $F^{\mu \nu}_i$
which have error maps $\Delta F^{\mu \nu}_i$ and some uncertain 
overall amplitude of fluctuations, which is acceptable because we will 
mean-subtract all maps for the purposes of cross-correlation.  The 
value of $\sum_i F_i^{\mu \nu} T_i^\nu$ or the 
best-fit line on a basic 
scatter plot determines the cross-correlation coefficient, 
$\alpha^{\mu \nu}$, 
and its uncertainty, $\Delta \alpha^{\mu \nu}$, can be determined from the 
$1 \sigma$ uncertainties in the slope of the best-fit line of the 
scatter plot or from varying the cross-correlation over regions likely 
to be systematically different such as different Galactic latitudes 
or opposite celestial hemispheres.  Performing this cross-correlation for 
various patches of sky one at a time 
\citep[see][]{jewellll99}
will improve the eventual foreground 
subtraction but introduces low-$\ell$ structure in the cleaned map.  Some 
amount of low-$\ell$ contamination is allowable, so this tradeoff will 
have to be optimized for each experiment individually.  Of course, it is 
possible to use an observational channel as a foreground template 
\citep{dodelsonk95} 
but care is needed if that channel contains significant contributions from 
multiple 
foregrounds or CMB fluctuations.  

Now we can subtract the correlated foreground component 
\citep[see][]{jaffefb99},
\begin{equation}
T^\nu_i = T_{i,obs}^\nu - \sum_\mu \alpha^{\mu \nu} F_i^{\mu \nu}, 
\end{equation}
and account for the added noise using 
\begin{equation}
C^\nu_{ij} = C_{ij,obs}^\nu + \sum_\mu (\alpha^{\mu\nu})^2 
	\Delta F_i^{\mu \nu} \Delta F_j^{\mu \nu} 
	+ \sum_\mu (\Delta \alpha^{\mu \nu})^2 F_i^{\mu \nu} F_j^{\mu \nu}.
\end{equation}
One can imagine using multiple templates for the same foreground 
component sequentially by first subtracting the correlated component of the 
best template, then forming an orthogonal template using the two best 
templates and subtracting the component that correlates with 
 this orthogonal template, etc.
  If many good foreground templates were available, 
the set of orthogonal templates could be designed from the start, but 
this is not generally the case, as can be seen by trying to use 
multiple synchrotron templates to gain more information than is available 
from the best of them alone.

\section{Testing for a Thermodynamic Spectrum}
 
Other foreground components (high-redshift 
IR galaxies, high-redshift clusters, inhomogeneous reionization, and 
perhaps spinning dust grains) cannot be accounted for with templates.  
  We must therefore scan the dataset to look for non-thermodynamic emission, 
which is the clearest signal of foreground contamination, as long as the 
deviation from a thermodynamic spectrum is inconsistent with fluctuations 
due to the expected instrument noise \citep[see][]{brandtetal94}.  
Pixels typically 
oversample the beam so we can add up to a beam size to reduce the noise  
and mask suspicious areas.  True features on the sky will be convolved 
by the beam function $B(\theta)$ so we turn that into a symmetric 
matrix $B_{ij}$ which 
indicates how much of a signal which fills only 
pixel $i$ has been convolved into pixel $j$ and vice versa.  
We will use $B_{ij}$ to choose weights for the pixels around pixel $i$ and 
form a weighted
beam-size average at each frequency:
\begin{equation}
WT_i^\nu = \frac{\sum_j T^\nu_j B_{ij}}{\sum_j B_{ij}}
\end{equation}
with uncertainty given by 
\begin{equation}
\Delta WT^\nu_i=\left(\sum_j \frac{B_{ij}}{C^\nu_{jj}}\right)^{-\frac{1}{2}}.
\end{equation}
This formula for the weighted average assumes that the instrument noise 
is well-approximated as uncorrelated between pixels; if not, the noise 
covariance matrix can be used to form a more sophisticated set of 
weights and uncertainties.  One could identify $5 \sigma$ pixels at this 
point; since sky signals should be as large as the beam this is now the 
optimal temperature map to use to look for point sources.  In fact, this 
is roughly the method used by available software packages such as DAOPhot, 
and it may be superior to the wavelet methods which are now being 
explored \citep[e.g.][]{tenorioetal99}.  

Now we can test at location $i$ for a thermodynamic temperature fluctuation, 
which has the same value of $T_i$ in every channel, by forming the weighted 
sum over all frequencies of the beam-size weighted temperature
at each frequency, 
\begin{equation}
WT_i = \frac{\sum_\nu \frac{WT_i^\nu}{(\Delta WT_i^\nu)^2}}
	     {\sum_\nu \frac{1}{(\Delta WT_i^\nu)^2}}
\end{equation}
which we can now test versus the hypothesis that the beam-size weighted 
temperatures $WT_i^\nu$ are all the same by forming 
\begin{equation}
\chi^2 = \sum_\nu \frac{(WT_i^\nu-WT_i)^2}{(\Delta WT_i^\nu)^2}
\end{equation}
which should obey a $\chi^2$ distribution with $N_\nu - 1$ degrees of 
freedom since the mean has been fit to the $N_\nu$ data channels.  

One decision to be made by each instrumental team is what 
threshold to use for deciding if a given beam-size area is 
thermodynamic.  We recommend a $3 \sigma$ detection of non-thermodynamic 
fluctuations as the threshold, as this should only yield a false 
positive in one percent 
of the beam-size areas of the map, and the cost of losing that much 
area is small.  False negatives are caused by instrument noise conspiring 
to hide the non-thermodynamic signal, so they 
become increasingly probable as the amplitude of the non-thermodynamic 
foreground decreases, which is tolerable.  
Prior information about point source locations 
can be used to mask contaminated pixels as well, even if there is great 
uncertainty in the extrapolated spectrum of the point source.  Specifically, 
any areas where the foreground subtraction uncertainty is greater than 
the expected level of intrinsic CMB fluctuations should be masked, e.g. 
on the Galactic plane or at the locations of known $1 \sigma$ point sources 
whose fluxes are uncertain by a factor of two.  

 Typical experiments (like MAP) often have 
different beam sizes at different frequencies, requiring the smoothing of 
higher-resolution maps to make the weighted average and test for 
thermodynamic fluctuations.  We advocate an iterative 
approach where the areas of the high-resolution maps that survive the 
thermodynamic test at all frequencies are then re-examined with less 
smoothing.  For instruments with broad frequency coverage such as Planck, 
it may be possible to form a few distinct spectral indices at each beam-size 
area centered on a given pixel, allowing one to make a color-color plot 
\citep{spergel98}
.  
The advantage of this versus the single $\chi^2$ test is that a beam-size 
pixel may look thermodynamic at most frequencies and simply have contamination 
at one extreme of the frequency range.  This form of contamination obviously 
does not require masking at all frequencies, and can even be allowed by 
setting the ``noise'' large at the extreme frequencies where the 
foreground contaminant appears to dominate.

Now we have produced what most methods of power spectrum analysis require
as input, a foreground-subtracted map of thermodynamic temperature 
fluctuations, with all frequency channels combined (although for the 
highest-resolution analyses it may be preferable to only use the 
highest-resolution channels).  
However, our map has a large number of pixels that have been 
masked.  
If a method
of determining the anisotropy power spectrum requires continous-sky coverage, 
these masked areas can be replaced by either zero or the median 
value of a ring around them with a radius of a few beam widths.  
This will lose a small amount of high-$\ell$ structure but that effect 
can be tested with simulations.  
However, it is better to 
utilize a method that will simply ignore the masked pixels 
\citep[e.g.][]{ohsh99}.    

The final ``noise'' covariance matrix of this weighted average map 
will have full off-diagonal structure  due to uncertainty in 
foreground subtraction.  If an analysis method can handle large 
variances, there is no need to fully mask pixels, i.e. their variance 
can simply be made large.  But some methods, such as Fast Fourier Transforms, 
require assigning equal weight to all pixels, so the ring-replace method 
or something similar must be used.

This method of foreground subtraction can be tested on the WOMBAT Challenge 
simulations.  This will indicate how feasible it is to subtract foregrounds 
without appealing to their expected angular power spectra and how much 
advantage is gained by using known spatial information on the various 
components.

\cleardoublepage
\chapter{Conclusion}
\label{chap:conclusion}

\section{Prospects for Detecting Dusty High-Redshift Galaxies}
\label{sect:conclusion_dusty}

\begin{figure}[hbt]
\centerline{\psfig{file=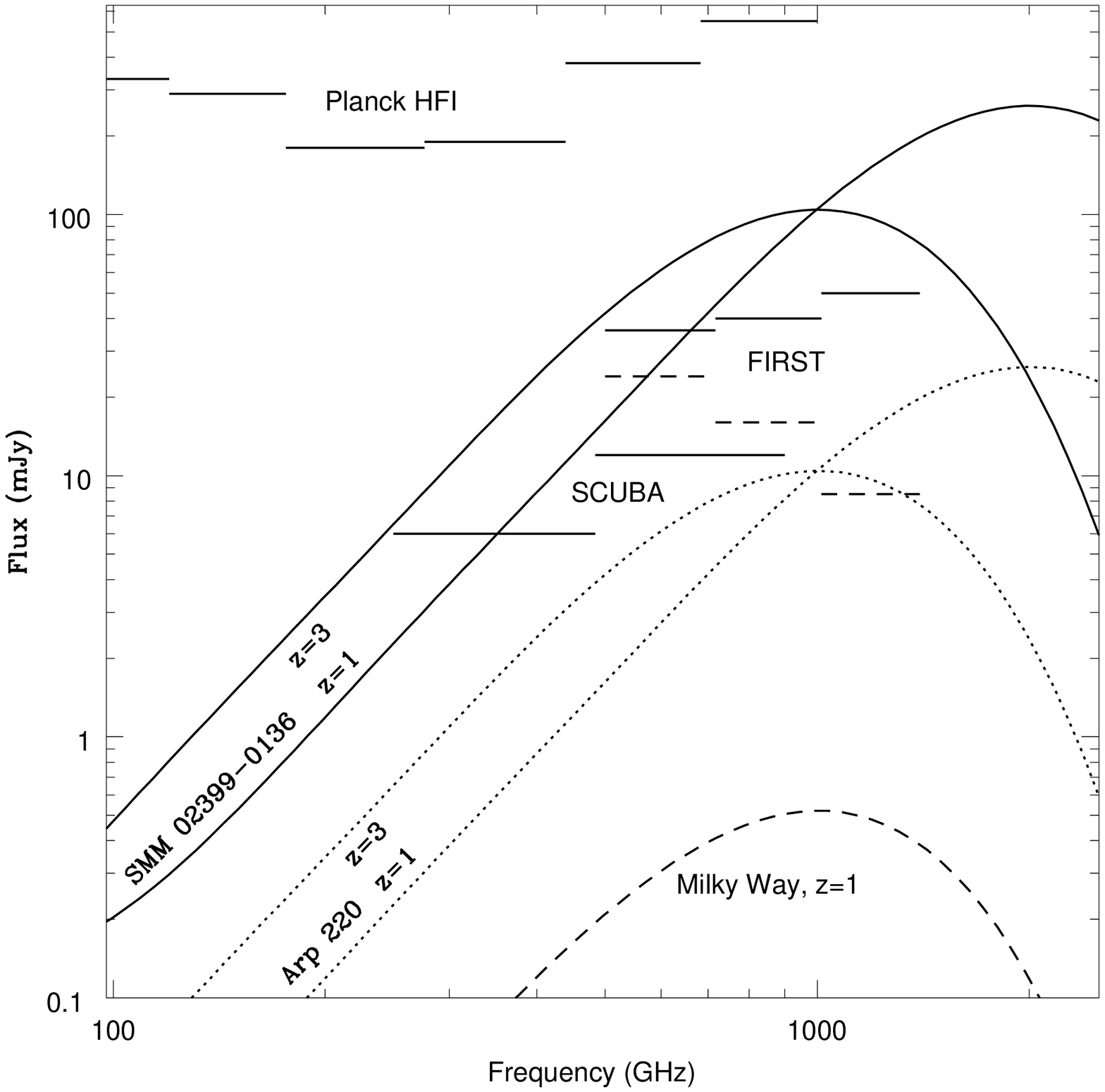,width=4in}}
\caption{Prospects for submillimeter detection of high-redshift galaxies.}
\mycaption{The far-IR/sub-mm spectrum of ultraluminous IR galaxies such
as SMM 02399-0136 (solid) and Arp 220 (dotted) are shown at z=1 and z=3, 
along with a z=1 L$_*$ galaxy with the spectrum of the Milky Way (dashed).  
The
expected intensities are compared with the  
$5\sigma$ detection limits of FIRST channels centered at 
600, 857, and 1200 GHz, 
SCUBA channels at 350 and 670 GHz, and Planck HFI
channels at 100, 143, 217, 353, 545, and 857 GHz.  The dashed version
of the FIRST detection limits assumes that the detection limits will 
be determined by point source confusion rather than Galactic cirrus.  
}
\label{fig:galaxy}
\end{figure}

Figure \ref{fig:galaxy} 
shows that the frequency range covered by FIRST is 
favorable for the detection of high-redshift star-forming galaxies such 
as SMM 02399-0136, which was recently detected by 
\citet{ivisonetal98} 
using SCUBA.  
SCUBA has good sensitivity but a small field of view, so it will see 
many dim sources but detect only a few as bright as SMM 02399-0136.  
FIRST, on the other hand, will be limited by source confusion but can 
cover a much larger area of sky, leading to a much bigger sample of 
these ultraluminous objects.  
The spectrum of the 
Cosmic Far-Infrared Background detected by \citet{pugetetal96}
and \citet{schlegelfd98} has 
a peak indicating dominant contributions from 
starbursting galaxies near $z=1-3$, 
and the redshift distribution of sources detected 
with SCUBA is consistent with this so far \citep{bargeretal99, hughesetal98,
richards99}.  
  FIRST source counts and 
measurements of the spectrum of point source confusion should considerably
restrict the set of models of galaxy evolution which are consistent with
the FIRB and preliminary SCUBA detections at present.

The 5$\sigma$ source detection limits plotted in the figure are based on 
a quadratic
sum of instrument sensitivity achievable within a reasonable (10 hour) 
integration time, confusion from CMB anisotropy,
confusion from point sources \citep[see][]{blainis98},
and confusion from Galactic cirrus extrapolated 
as in \citet{toffolattietal98}.  
The Planck HFI instrument
will see thousands of low-redshift galaxies, including 
all 5319 IRAS 1.2 Jy sources 
 but Planck has too much noise and 
foreground confusion to see the sources responsible for the far-infrared
background radiation.  FIRST is the only one of these instruments whose
detection sensitivity limit is dominated by expected Galactic confusion; 
if that could be further minimized by choosing lower dust-contrast 
regions, the FIRST detection limits 
could be up to five times lower (see Figure \ref{fig:galaxy}), 
making galaxies like Arp 220 with luminosities around 100 $L_*$ 
which are likely to be missed by SCUBA detectable with FIRST.
These limits, combined with FIRST's suitability for wide-area surveys, lead
\citet{ivisonetal98}
to conclude that FIRST could detect about 100 sources 
similar to SMM 02399-0136 per hour of integration, which would provide 
us with an invaluable window into the optically-obscured 
era of star formation which 
generated the far-infrared background radiation.  
Because the FIRST frequencies cover
the peak of the emission spectrum for galaxies at $z=3$ (assuming 
40K dust), a combination of FIRST photometry with redshifts
determined using spectroscopy from optical follow-ups or FIRST itself 
will allow an examination  of the intrinsic luminosity, 
dust emissivity, and dust mass of these fascinating objects.

\section{Future Constraints on Structure Formation Models}

\begin{figure}
\centerline{\psfig{file=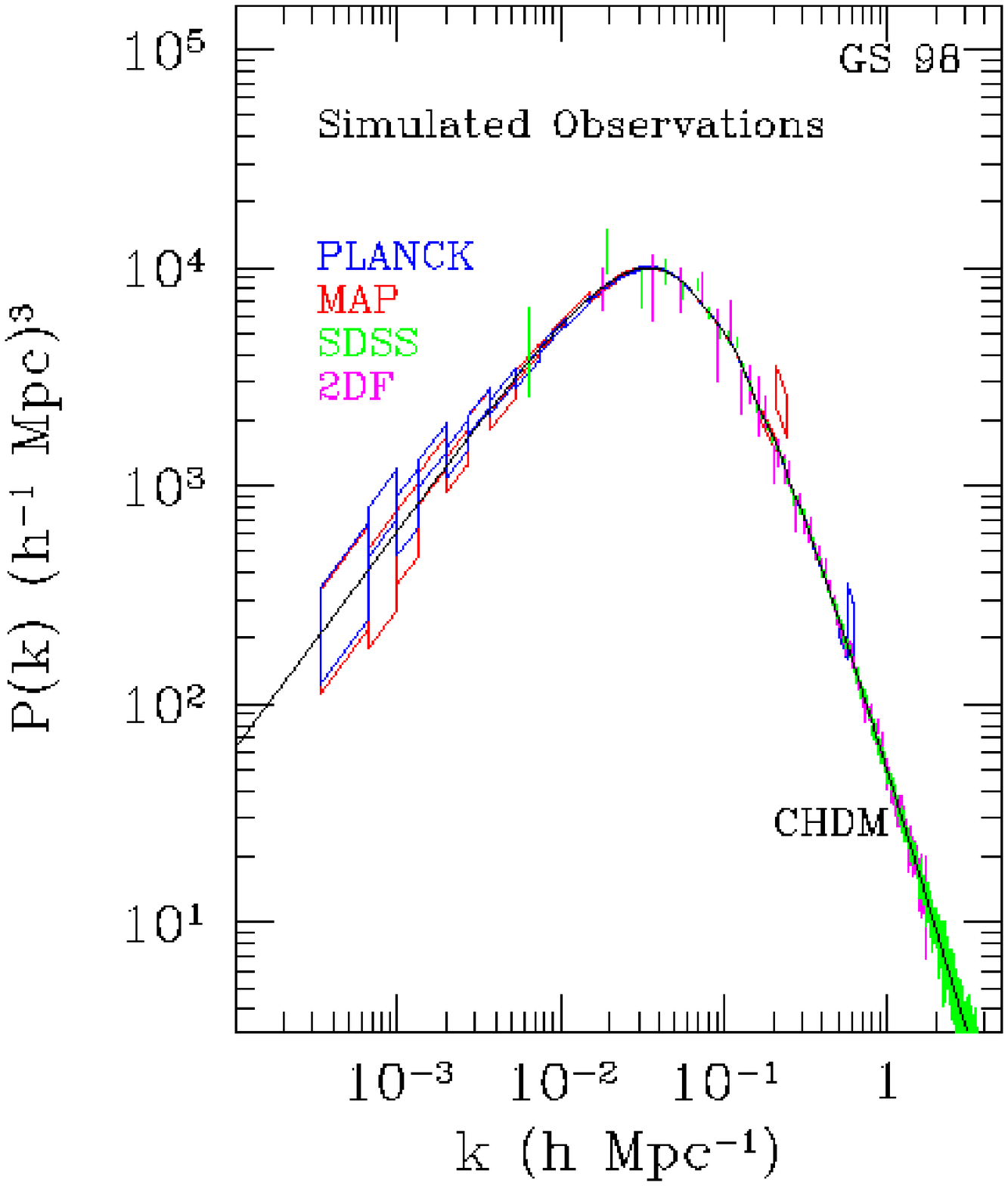,width=4in}}
\caption{Simulation of high-precision future structure formation constraints.}
\mycaption{Simulated observations  
of CMB anisotropy are shown for the MAP(red boxes) and Planck 
Surveyor (blue boxes) satellites, under the assumption that CHDM is 
in fact the correct model of structure formation.    
Green 
error bars show accuracy of the Sloan Digital Sky Survey and 
magenta are for the 2 Degree Field Survey.     
All four data sets are indistinguishable from the underlying CHDM model 
for a wide range of $k$.  No attempt has been made in this figure to account
for redshift distortions or non-linear evolution, 
which will complicate observations at $k>0.2$.   
}
\label{fig:chdm_future}
\end{figure}

Figure \ref{fig:chdm_future} shows the impressive precision
with which future observations by the MAP and Planck Surveyor
satellites and the Sloan Digital Sky Survey and 2 Degree Field survey
should confirm 
the correct model of structure formation (in this case 
assumed to be CHDM).  Instrumental parameters from \citet{bondet97}
were used to predict the precision of these constraints for 
CMB anisotropy observations.  
\citet{huet98} look 
at how well $\Omega_\nu$ can be determined by SDSS observations,
and \citet{wangss99}
examine the ability of combined MAP and SDSS observations to 
constrain cosmological parameters.  

  The overlap in scale between microwave background anisotropy detections and
large-scale structure observations will increase tremendously in the
next several years, and the errors in these measurements will decrease
significantly as well.  The CMB promises to measure several 
cosmological parameters to great accuracy, but some parameters such as 
the fraction of Hot Dark Matter, $\Omega_\nu$, are more directly probed
by large-scale structure.   All of this, of course, assumes 
that the correct model is one that we will consider, i.e. that we 
are searching the right parameter space.  Instead of searching the 
adiabatic CDM parameter space to determine the best-fit model (which due 
to the dangers of relative likelihood analysis might not really 
be a very good 
fit), we advocate an 
empirical approach to structure formation \citep{gawiseraas98}, 
where high-precision observations of CMB anisotropy and large-scale 
structure are first used to determine the validity of the adiabatic CDM 
paradigm.

\section{Conclusions}

We have extracted
the spectrum of primordial density fluctuations from the current 
observations of structure formation in the universe and  
find that it is close to that of the Cold + Hot Dark Matter
model and significantly different from  
that of any other model we have considered. 
However, a wide range of models agree on a qualitative level 
with both the 
high-redshift density variations imprinted in the CMB and 
the low-redshift inhomogeneities in the galaxy distribution, 
giving strong 
evidence for the success of the gravitational instability paradigm of cosmological 
structure 
formation.
Increasingly precise measurements of the mass density in the universe 
appear to rule out $\Omega_m=1$ and the Type Ia supernovae measurements 
favor a positive cosmological constant.  
The agreement of $\Lambda$CDM with the 
structure formation 
data, however, is not particularly 
good, and it is harmed by adding even a small fraction of Hot Dark Matter.  
This allows us to set upper limits on the mass of the most massive neutrino of 
2 eV,
 assuming a scale-invariant primordial power spectrum, and of 4 eV, 
assuming 
only a scale-free primordial power spectrum (consistent with the 
vast majority of 
inflationary models).   
We have reconciled the $\Lambda$CDM
model with the existing structure formation observations by appealing to 
a radically non-scale-invariant primordial power spectrum which 
features a smooth enhancement on 100-200 Mpc scales.  Only further 
observational data will reveal whether such a feature is actually necessary to 
explain the origin and nature of structure in the universe.  
Forthcoming high-precision CMB anisotropy and large scale structure data 
will provide us with independent probes of the primordial density 
fluctuations which 
can be used to test models of structure formation without assuming a 
particular primordial power spectrum.

We have developed predictions for microwave emission from radio and  
infrared-bright galaxies and the SZ effect from clusters.  These predictions 
form part of the input information for the WOMBAT Challenge simulations, 
which are the most realistic simulations to date of the microwave sky. 
 Analysis 
of these skymaps shows 
that a reasonable observational window is available for 
CMB anisotropy measurements.
These simulations can be used to test various foreground 
subtraction methods and to explore systematic errors in cosmological 
parameter determination.  The pixel-space 
method for foreground subtraction that we have 
 presented may offer an improved method for 
performing this important task.

\cleardoublepage

%\begin{table}
%\begin{tabular}{|l|r|}
%  \hline 
%Title & Author \\
%\hline
%War And Peace & Leo Tolstoy \\
%The Great Gatsby & F. Scott Fitzgerald \\ \hline
%\end{tabular}
%\caption{A normalsize table.  There has been a complaint that table
%captions are not single-spaced.  This is odd because the code
%indicates that they should be.}
%\end{table}

%\begin{table}
%\caption{A small table.}
%\begin{scriptsizetabular}{|l|r|}
%  \hline 
%Title & Author \\
%\hline
%War And Peace & Leo Tolstoy \\
%The Great Gatsby & F. Scott Fitzgerald \\ \hline
%\end{scriptsizetabular}
%\end{table}

%\chapter{Previous Work}

%Some other research was once performed.

%\begin{figure}
%\caption{A first figure.}
%\end{figure}

%\begin{figure}
%\caption{A second figure.}
%\end{figure}

%\chapter{Conclusion}

%try to shrink bibliography some...

\ssp
\small

%\nocite{*}
%\bibliographystyle{plain}
\bibliographystyle{apj}
%\bibliography{uctest}
\bibliography{apj-thesis,refs}  %apj-thesis 

%\appendix
%\chapter{Color figures}

%Ancillary material should be put in appendices, which appear after the
%bibliography. 

\end{document}